\documentclass{jfm}
\usepackage{graphicx}
\usepackage{epstopdf, epsfig}
\usepackage{enumerate}
\usepackage[table,usenames,dvipsnames]{xcolor}
\usepackage{multirow}
\usepackage{hyperref}

\usepackage{amsmath}
\usepackage{color}

\usepackage[ruled,vlined]{algorithm2e}

\def\m1line{\vrule width3pt height2.5pt depth -2pt}

\def\bdot{\raise.2em\hbox to .15em{.}}

\allowdisplaybreaks[1]

\shorttitle{Most unstable route to boundary-layer turbulence}
\shortauthor{R. Jahanbakhshi and T.A. Zaki}

\title{Nonlinearly most unstable route to high-speed boundary-layer turbulence}

\author{Reza Jahanbakhshi\aff{1}
  \and Tamer A.  Zaki\aff{1}
  \corresp{\email{t.zaki@jhu.edu}}}

\affiliation{\aff{1} Department of Mechanical Engineering, Johns Hopkins University \\
Baltimore, MD 21218-2682, USA}

\begin{document}

\maketitle

\begin{abstract}

Laminar-to-turbulence transition in zero-pressure-gradient boundary layer at Mach 4.5 is studied using direct numerical simulations.
For a given level of total disturbance energy, the inflow spectra was designed to correspond to the nonlinearly most dangerous condition that leads to the earliest possible transition Reynolds number.  
The synthesis of the inlet disturbance is formulated as a constrained optimization, where the control vector is comprised of the amplitudes and relative phases of the inlet modes;
the constraints are the prescribed total energy and that the flow evolution satisfies the full nonlinear compressible Navier-Stokes equations;
the cost function is the average wall friction which, once maximized, corresponds to the earliest possible transition location.  
An ensemble variational (EnVar) technique is developed to solve the optimization problem.  
EnVar updates the estimate of the control vector at the end of each iteration using the gradient of the cost function, which is computed from the outcome of an ensemble of possible solutions.
Two inflow conditions are computed, each corresponding to a different level of energy, and their spectra are contrasted:    
the lower energy case includes two normal acoustic waves and one oblique vorticity perturbation, whereas the higher energy condition consists of oblique acoustic and vorticity waves.
The focus is placed on the former case because it can not be categorized as any of the classical breakdown scenarios, while the higher energy condition undergoes a typical second-mode oblique transition.  
At the lower energy level, the instability wave that initiates the rapid breakdown to turbulence is not present at the inlet plane.
Instead, it appears at a downstream location after a series of nonlinear interactions that spur the fastest onset of turbulence.
The results from the nonlinearly most potent inflow condition are also compared to other inlet disturbances that are selected solely based on linear theory, and which all yield relatively delayed transition onset.

\end{abstract}

\section{Introduction}
\label{sec:Introduction}

The location and mechanism of laminar-to-turbulence transition in hypersonic boundary layers are sensitive to the flow environment.  
Linear theory provides criteria for the onset of exponential instability, and these primary waves subsequently undergo secondary instability and breakdown to turbulence.  
However, similar criteria for transition onset can not be established theoretically because breakdown to turbulence is nonlinear.  
Nonlinear simulations can predict the evolution of any inflow perturbation, although starting with the linearly most unstable mode does not guarantee the earliest possible transition location. 
And when the inflow perturbation is composed of multiple waves, the choice of their relative amplitudes and phases can appreciably influence transition onset.  
These parameters are herein optimized using an ensemble variational (EnVar) algorithm in order to determine the earliest possible transition Reynolds number in zero-pressure-gradient high-speed boundary layer, for a specified level of energy of the inlet disturbance.

\subsection{Transition in high-speed boundary layers}
\label{sec:instability}

The precursors that ultimately lead to transition to turbulence in high-speed boundary layers can be traced upstream to the early amplification of small-amplitude acoustic, entropic or vortical fluctuations \citep{Leyva2017}.
When their amplitudes are infinitesimal, these waves and their growth rates are accurately modelled by linear stability theory  (LST) \citep{Mack1984}.  
According to the theory, oblique first-mode instabilities, which are similar to the Tollmien-Schlichting waves at lower speeds, are most amplified in supersonic conditions.
With increasing Mach number, however, two-dimensional acoustic instability waves, which are commonly referred to as Mack's (second-) modes \citep{Mack1984}, play an increasingly important role.
Secondary instability theory was also developed \citep{Herbert1988} in order to examine how the state comprised of the mean-flow profile and the primary instability will itself become unstable to new perturbations. 
In order to account for non-parallel effects on the amplification of instability waves, the linear parabolized stability equations (PSE) were introduced by \citet{Bertolotti1992}. 
In addition, when the disturbance amplitudes become appreciable, the nonlinear parabolized stability equations \citep{Bertolotti1992,Chang1994,Herbert1998} can account for nonlinear modal interactions and the base-state distortion.
The final stage of transition is marked by the appearance of localized turbulence patches, or spots, which spread as they are advected downstream.  
Intermittency, which is defined as the fraction of time that the flow at a given streamwise location is turbulent \citep{Narasimha1985}, thus rises from its initial value of zero in the laminar regime to unity where the spot merge to form a fully turbulent boundary layer. 
Recent numerical and experimental studies have focused on advancing our fundamental understanding of transition \citep[see the reviews by][]{Fedorov2011, Zhong2012, Schneider2015}.    
A summary of select efforts is provided here, with the objective of highlighting transition mechanisms that are relevant to boundary layers at free-stream Mach numbers larger than four.

A key determining factor of the transition mechanism, be that in simulations or in experiments, is the spectral makeup of the upstream disturbance.   
For example, in numerical simulations, this inflow disturbance must be specified, and changes in the relative amplitudes and phases of inflow waves can lead to quantitative and qualitative changes in the transition process.
\cite{Franko2013,Franko2014} performed direct numerical simulations of three transition mechanisms at Mach six, in zero- and adverse-pressure-gradient (ZPG and APG) boundary layers: 
first-mode oblique breakdown, second-mode oblique breakdown, and second-mode fundamental resonance.  
In all cases, the disturbance frequencies and spanwise wavenumbers were chosen based on linear stability theory and $e^{N}$ method.
They concluded that, for all of three mechanisms, APG increases the linear growth rates and promotes earlier breakdown to turbulence.
Nonetheless, the nonlinear breakdown of each of the three mechanisms was qualitatively similar for ZPG and APG conditions.
\cite{Sivasubramanian2015,Sivasubramanian2016} also preformed DNS of fundamental resonance and of oblique breakdown in Mach 6 boundary layers on a sharp cone, with a half vertex angle equal to seven degree.
A series of low-resolution simulations, informed by linear stability theory, were used to identify the instability waves that promote early transition.
These waves are then used to performed fully resolved simulations.
They reported that both second-mode fundamental and oblique breakdown lead to strong nonlinear interactions, and are viable candidates to affect transition on the cone geometry.  
\cite{Novikov2016} performed a numerical study of laminar-turbulent transition over a flat plate at Mach 5.37. 
In that effort, perturbations were introduced into the boundary layer through a small round hole on the surface of the plate by forced pulsations of the vertical velocity component.
The authors reported that the linear stages of transition are dominated by first-mode oblique waves, while second-mode plain waves become dominant in nonlinear regions.
Most recently, \cite{Zhao2018} studied the effect of local wall heating and cooling on the stability of a flat-plate boundary layer at Mach 6.
Their results revealed that the location of thermal treatment relative to the synchronization point significantly affect the stability of a second-mode instability.
These studies highlight the variety of transition mechanisms that can be explored using nonlinear computations, and by varying the inflow condition.

In a Mach 6 wind tunnel, \cite{Zhang2013,Zhang2015} and \cite{Zhu2016,Zhu2018} investigated transition using Rayleigh-scattering visualization, fast-response pressure measurements, and particle image velocimetry.
They found that the second-mode instability is a key modulator of the transition process, through the formation of high-frequency vortical waves and triggering a fast breakdown to turbulence.
They also reported that the second-mode waves, accompanied by high-frequency alternating fluid compression and expansion, produce intense aerodynamic heating in a small region.
\cite{Casper2016} studied boundary-layer transition on a sharp seven-degree cone in hypersonic wind tunnels at Mach five, six, eight, and fourteen. 
At the lowest Mach number, transition was initiated by a combination of first- and second-mode instabilities, while at higher values the boundary layer was dominated by second-mode instabilities.
\cite{Laurence2016} and \cite{Kennedy2018} also investigated transition on a slender cone at Mach $6$-$7$ and Mach 14, respectively, using high-speed Schlieren visualizations.
They specifically focused on formation and evolution of second-mode instabilities, and reported that the wave-packets form rope-like structures that become progressively more interwoven as transition develops.
The experiments were performed for both low and high-enthalpy conditions, and highlighted the difference in the wall-normal distribution of the disturbances in each condition.
Despite laboratory experiments accounting for all stages of transition fully, starting from receptivity through the development of the turbulent boundary layer, they do not necessarily reproduce the environmental disturbances of high-altitude flight.  
As a result, flight experiments have been performed, and more than 20 such tests were surveyed by \citet{Schneider1999}.
The author noted the high cost of each experiment which often precludes repeating the tests.
More recently, the Hypersonic International Flight Research Experimentation (HIFiRE) program performed a series of tests to study transition scenarios on a cone in hypersonic flight \citep[see e.g.]{Juliano2015,Stanfield2015,Kimmel2018}. 
The HIFiRE data show that, depending on the flow condition and angle-of-attack, second-mode, crossflow, and separation induced transition are all potential causes of turbulence on different parts of the vehicle.
It is important to note that only a limited amount of flight data on transition are available and that in-flight environmental conditions are variable and uncertain.

\subsection{The nonlinearly most dangerous disturbance}
\label{sec:motivation}

With a wealth of instability waves at hypersonic speeds, transition mechanisms in numerical simulations can vary quantitatively and qualitatively depending on the inflow disturbance spectrum.
Existing approaches synthesize the inflow as a superposition of instability waves that are often selected based on their growth rates or appearance in experiments. 
The former approach does not guarantee that these waves are the most dangerous with respect to the onset of nonlinear breakdown, and relative modal amplitudes and phases are not prescribed in a manner that exposes the most dangerous transition scenario.  
When motivated by experimental observations, the choice of the computational inflow condition attempts to reproduce the instability waves that were experimentally measurable; precursor interactions that may have led to the generation of these modes may not be included.  
In addition, inflow conditions that are relevant to in-flight environments are often uncertain or unknown.  
These challenges motivate the present work and the question: how can we guarantee robust flow design in high-speed flows?
The adopted approach is to determine, for a given level of disturbance energy, the nonlinearly most unstable inflow condition that will cause transition at the lowest possible Reynolds number.
No other inflow disturbance with the same energy can cause earlier breakdown to turbulence.

Compared to theory and nonlinear simulations of boundary-layer stability, studies of nonlinearly most unstable disturbances in transitional flows are relatively recent \citep{Pringle2010,Cherubini2011,Rabin2012}. 
The limited number of previous efforts all focused on incompressible flows; the present study is the first to consider high-speed flow.  
Another common feature in earlier efforts is the use of adjoint-variational techniques, which are commonly adopted in flow control \citep{Bewley2001,Cherubini2013,Luchini2014,Xiao2017}, in order to optimize a control vector that is either the initial or inflow disturbance\textemdash the optimal value is the nonlinearly most dangerous disturbance.
The two main ingredients of the algorithm are a cost function whose optimality corresponds to the earliest transition location, and iterative numerical solution of the forward and adjoint equations to evaluate the local gradient of this cost function with respect to the control vector. 
One key advantage of adjoint methods is that a forward-adjoint loop directly yields the local sensitivity of the cost function to the control vector.  
However, the adjoint approach has a number of limitations that are relevant for the present purposes: 
Firstly, an accurate adjoint model is required and is not always available, for example in the case of nonlinear parabolized stability equations.  
Secondly, when the forward equations are nonlinear, the adjoint model depends on the time history of the forward state variables which must therefore be stored with full or high spatial and temporal resolution.  
Lastly, in chaotic systems it is very difficult to accurately satisfy the forward-adjoint duality relation over long time horizons and, as a result, forward-adjoint loops are not guaranteed to provide accurate or convergent gradients of the cost function.  
As a result, previous efforts have often been restricted to short optimization horizons \citep[e.g.][]{Xiao2017}.

In order to avoid some of the limitations of adjoint-variational methods, we develop an ensemble-variational (EnVar) algorithm to evaluate the nonlinearly most dangerous inflow disturbance.  
This technique does not require an adjoint solver and is therefore applicable to any forward model, e.g.\,direct and large-eddy simulations, nonlinear PSE,...etc. 
In addition, absent the requirement of an adjoint, the difficulties associated with storage requirements and the stability of the adjoint integration are no longer relevant.  
Applications of EnVar in fluid mechanics are recent, and are generally in the field of data assimilation \citep{Colburn2011,Suzuki2012,Kato2015,Yang2015,Mons2016,Gao2017}.
The current effort is the first to develop EnVar methods for computing the nonlinearly most dangerous disturbance, and the first to examine these disturbances in hypersonic boundary layers.

In \S\ref{sec:Theory}, the governing equations and a detailed description of the EnVar algorithm are presented.
Section \ref{sec:DNS} provides a summary of the parameters of the numerical simulations at the two levels of inflow energy.
The nonlinearly most dangerous routes to turbulence are discussed in detail in \S\ref{sec:Results}, and conclusions are provided in \S\ref{sec:Conclusions}.

\section{Theoretical formulation}
\label{sec:Theory}


The high-speed flow of a compressible fluid satisfies the Navier-Stokes equations, 
\begin{subequations}
	\begin{equation}
		\frac{\partial \rho}{\partial t} + \nabla . \left( \rho \textbf{u} \right) = 0 ,
	\label{Eq:Continuity}
	\end{equation}
	\begin{equation}
		\frac{\partial \rho \textbf{u} }{\partial t} + \nabla . \left( \rho \textbf{u} \textbf{u} + p \textbf{I} - \boldsymbol{\tau} \right) = 0 ,
	\label{Eq:Momentum}
	\end{equation}
	and
	\begin{equation}
		\frac{\partial E}{\partial t} + \nabla . \left( \textbf{u} \left[ E + p \right] + \boldsymbol{\theta} - \textbf{u} . \boldsymbol{\tau} \right) = 0 ,
	\label{Eq:Energy}
	\end{equation}
	where $\rho$ is the density, $\textbf{u}$ is the velocity vector, $p$ is the pressure, $\textbf{I}$ is the unit tensor, $E = \rho e + 0.5 \rho \textbf{u} \cdot \textbf{u}$ is the total energy, 
	$e$ is the specific internal energy, $\boldsymbol{\tau}$ is the viscous stress tensor, and $\boldsymbol{\theta}$ is the heat flux vector.
\label{Eq:Navier_Stokes}
\end{subequations}
To close the system of equations, the calorically perfect gas relations are used,
\begin{subequations}
	\begin{equation}
		p = \left( \gamma - 1 \right) \rho e ,
	\end{equation}
	and
	\begin{equation}
		T = \frac{\gamma - 1}{ R} e ,
	\end{equation}
	where $T$ is the temperature, $\gamma$ is the ratio of specific heats and $R$ is the gas constant.
\label{Eq:State_Equation}
\end{subequations}
Furthermore, the viscous stress tensor and the heat flux are defined as
\begin{equation}
	\boldsymbol{\tau} = \mu_s \left( \nabla \textbf{u} + ( \nabla \textbf{u})^{tr} \right) + \left( \mu_b - \frac{2}{3} \mu_s \right) ( \nabla . \textbf{u} ) \textbf{I} ,
\end{equation}
and
\begin{equation}
	\boldsymbol{\theta} = - \kappa \nabla T ,
\end{equation}
respectively, where $\mu_s$ is the dynamic shear viscosity, which is computed from the local temperature via Sutherland's law \citep{Sutherland1893}, $\mu_b$ is the bulk viscosity, and $\kappa$ is the thermal conductivity.
The viscosity variables are related via Stokes' hypothesis and the thermal conductivity is computed by assuming a constant Prandtl number and specific heat.
We will refer to our Navier-Stokes solution algorithm, which is discussed later in section \ref{sec:DNS}, in operator form as $\mathcal{N}$.

\begin{figure}
	\centerline{%
	\includegraphics[trim=0 0 0 0, clip,width=1.0\textwidth] {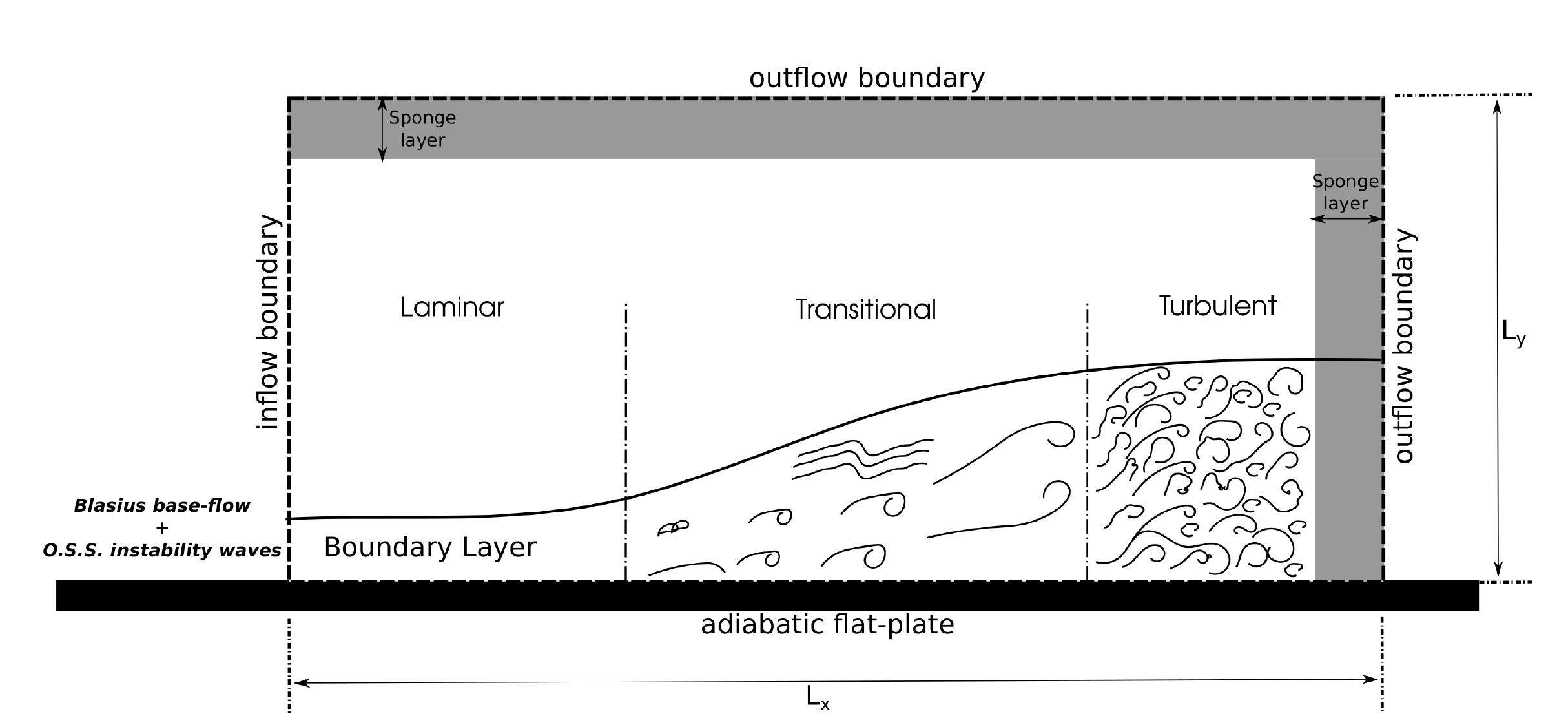}%
	}%
\caption{Schematic of the computational domain for a zero-pressure-gradient transitional boundary layer over a flat plate.}
\label{FIG:BL_Schematic}
\end{figure}

The flow configuration is a zero-pressure-gradient boundary layer over a flat plate, and is shown schematically in figure \ref{FIG:BL_Schematic}.
The computational domain starts downstream of the leading edge, and the inflow condition is a superposition of the compressible Blasius solution and perturbations,
\begin{equation}
	\boldsymbol{\Phi} = \boldsymbol{\Phi}_{B} + \boldsymbol{\varphi}^{\prime}
\label{Eq:Inflow_Boundary}
\end{equation}
where $\boldsymbol{\Phi} = \begin{bmatrix} \rho & u & v & w & T \end{bmatrix}^{tr}$ is the inflow variables, the subscript $B$ denotes Blasius base-state, $\boldsymbol{\varphi}^{\prime} = \begin{bmatrix} \rho^{\prime} & u^{\prime} & v^{\prime} & w^{\prime} & T^{\prime} \end{bmatrix}^{tr} $ is the perturbations with respect to this base-state, and the superscript $tr$ denotes transpose.
In the present work, the unknown is the most unstable disturbance vector, $\boldsymbol{\varphi}^{\prime}$, for a given level of initial energy.

\subsection{Problem definition}
\label{sec:problem_setup}

The non-linearly most dangerous disturbance is defined as the inflow perturbations (i) with a prescribed amount of total energy; (ii) that satisfies the Navies-Stokes equations; (iii) and undergoes the fastest breakdown to turbulence.
The first two elements of the definition are the constraints while the third is the objective.
This problem can be cast as a constrained optimization: find the control vector $\boldsymbol{\varphi}^{\prime}$ that optimizes the cost function
\begin{subequations}
	\begin{equation}
		\mathcal{J} \left( \boldsymbol{\varphi}^{\prime} \right) = \frac{1}{2} \Vert \mathcal{G} \left( \textbf{q} \right) \Vert^2 ,
	\label{Eq:OPT_1_1}
	\end{equation}
	while satisfying the constraints
	\begin{equation}
		\textbf{q} = \mathcal{N} \left( \boldsymbol{\varphi}^{\prime} \right) ,
	\label{Eq:OPT_1_2}
	\end{equation}
	and
	\begin{equation}
		\mathcal{E} \left( \boldsymbol{\varphi}^{\prime} \right) = E_0 ,
	\label{Eq:OPT_1_3}
	\end{equation}
	where $\textbf{q}$ is the state vector, $\mathcal{G} \left( \textbf{q} \right)$ is the observation vector that quantifies the breakdown to turbulence, $\mathcal{N}$ represents the Navier-Stokes solver, $\mathcal{E}$ is the energy operator, and $E_0$ is the constrained energy.
\label{Eq:Optimization_Problem_1}
\end{subequations}
Choices of $\mathcal{G} \left( \textbf{q} \right)$ can be formulated based on the skin friction, wall temperature, heat flux, or perturbation energy.
In the present work, the observation vector is proportional to the skin-friction coefficient, $ \mathcal{G} \left( \textbf{q} \right) = C_f \left({ dx / L_x }\right)^{1 \over 2} $.

The control vector is the inflow disturbance, which is expressed as a superposition of instability waves,
\begin{equation}
	\boldsymbol{\varphi}^{\prime} = \sum_{n,m} \text{Real} \left\lbrace c_{n,m} \hat{\boldsymbol{\varphi}}_{n,m} (x_0,y) e^{i(\beta_m z - \omega_n t - \theta_{n,m})} \right\rbrace ,
	\label{Eq:Inflow_Modes}
\end{equation}
where $\omega_n$ and $\beta_m$ are the frequency and spanwise wave number, and $\hat{\boldsymbol{\varphi}}_{n,m}$ is the complex mode shape which is obtained from solution of the Orr-Sommerfeld-Squire eigenvalue problem.  
At each $\left(n,m\right)$ pair, only the most unstable mode will be included at the inflow, which in the present study will be the slow mode \citep{Fedorov2011}.
The eigenmode of each instability wave is normalized to unit energy, 
\begin{align}
\begin{split}
	\frac{1}{2}
	\int_{0}^{L_y} \left( \overline{\rho} \left[ \hat{u}^* \hat{u} + \hat{v}^* \hat{v} + \hat{w}^* \hat{w} \right] + \frac{R}{\gamma-1} \left[ \frac{\overline{T}}{\overline{\rho}} \hat{\rho}^* \hat{\rho} + \frac{\overline{\rho}}{\overline{T}} \hat{T}^* \hat{T} \right] \right)_{n,m} dy = 1, 
\label{Eq:Normalization}
\end{split}
\end{align}
where star denotes the complex conjugate.
Assuming knowledge of the relevant range of frequencies and spanwise wave numbers, the only unknowns are the positive real-valued modal amplitudes $c_{n,m}$ and phases $ \theta_{n,m}$, which lead to the earliest transition location downstream.
The control vector can then be redefined in terms of those two parameters, 
\begin{equation}
\textbf{c} = \begin{bmatrix}
    c_{n,m} & \dots & \theta_{n,m} & \dots 
\end{bmatrix}^{tr} ,
\end{equation}
and its size is $2 M$, or twice the total number of instability modes.
In addition, the total energy of the perturbations is defined as
\begin{equation}
	\mathcal{E} \left( \boldsymbol{\varphi}^{\prime} \right) = \frac{1}{2} \int_{0}^{L_y} \left( \overline{\rho} \left[ \overline{u^{\prime 2}} + \overline{v^{\prime 2}} + \overline{w^{\prime 2}} \right] + \frac{R}{\gamma-1} \left[ \frac{\overline{T}}{\overline{\rho}} \overline{\rho^{\prime 2}} + \frac{\overline{\rho}}{\overline{T}} \overline{T^{\prime 2}} \right] \right) dy ,
\label{Eq:Total_Energy}
\end{equation}
where overbar denotes averaging in time and the homogeneous spanwise direction.
In terms of the new control vector, the perturbation energy (\ref{Eq:Total_Energy}) can be expressed as, 
\begin{equation}
	\mathcal{E} \left( \boldsymbol{\varphi}^{\prime} \right) = \frac{1}{2} \textbf{c}^{tr} \textbf{A}_1 \textbf{c}
\end{equation}
where
\begin{equation}
	\textbf{A}_1 = \begin{bmatrix}
			    \textbf{I} & \textbf{O} \\
			    \textbf{O} & \textbf{O}
			\end{bmatrix}_{2 M \times 2 M} ,
\label{Eq:dummy_Matrix_1}
\end{equation}
and $\textbf{O} $ and $\textbf{I}$ are zero and identity matrices of size $M \times M$, respectively.
Now the constrained optimization problem (\ref{Eq:Optimization_Problem_1}) can be rewritten as identifying $\textbf{c}$ that optimizes the cost function
\begin{subequations}
\begin{equation}
	\mathcal{J} \left( \textbf{c} \right) = \frac{1}{2} \Vert \mathcal{G} \left( \textbf{q} \right) \Vert^2 ,
\label{Eq:OPT_2_1}
\end{equation}
while satisfying the constraints
\begin{equation}
	\textbf{q} = \mathcal{N} \left( \textbf{c} \right) ,
\label{Eq:OPT_2_2}
\end{equation}
\begin{equation}
	\frac{1}{2} \textbf{c}^{tr} \textbf{A}_1 \textbf{c} = E_0 ,
\label{Eq:OPT_2_3}
\end{equation}
and
\begin{equation}
	\textbf{A}_1 \textbf{c} \geq 0.
\label{Eq:OPT_2_4}
\end{equation}
The constraint (\ref{Eq:OPT_2_4}) is introduced because $c_{n,m}$ in equation (\ref{Eq:Inflow_Modes}) are real positive values.
\label{Eq:Optimization_Problem_2}
\end{subequations}
An ensemble-variational (EnVar) approach is adopted to solve the optimization problem (\ref{Eq:Optimization_Problem_2}).

\subsection{Ensemble-variational optimization}
\label{sec:4DVarEns}

The optimization problem (\ref{Eq:Optimization_Problem_2}) starts with an estimate, or guess, of the solution $\textbf{c}^{(e)}$. 
An ensemble of $N_{en}$ variants, $\textbf{c}^{(r)}$ where $r = 1,...,N_{en}$, is also constructed whose mean is equal to $\textbf{c}^{(e)}$.  
The optimal control vector is then expressed as the weighted sum, 
\begin{equation}
	\textbf{c} = \textbf{c}^{(e)} + \textbf{E}^{\prime} \textbf{a}. 
\label{Eq:Ensemble}
\end{equation}
where $\textbf{E}^{\prime}$ is the deviation of the ensemble members from their mean,
\begin{equation}
	\textbf{E}^{\prime} = \begin{bmatrix}
		 \textbf{c}^{(1)} - \textbf{c}^{(e)} & \dots & c^{(r)} - \textbf{c}^{(e)} &\dots & \textbf{c}^{(N_{en})} - \textbf{c}^{(e)}
	\end{bmatrix}_{2M \times N_{en}} ,
\label{Eq:Deviation_Matrix}
\end{equation}
and $\textbf{a}$ is the weight vector,
\begin{equation}
	\textbf{a} = \begin{bmatrix}
	    a^{(1)} & \dots & a^{(r)} & \dots & a^{(N_{en})}
	\end{bmatrix}^{tr}.  
\end{equation}
Since the mean $\textbf{c}^{(e)}$ and members $\textbf{c}^{(r)}$ of the ensemble are known, the control vector becomes the optimal weights $\textbf{a}$.

The optimization problem can be re-written in terms of the new control vector.  
Using equation \ref{Eq:Ensemble}, the energy constraint (\ref{Eq:OPT_2_3}) becomes,
\begin{equation}
	\frac{1}{2} \textbf{c}^{(e) tr} \textbf{A}_1 \textbf{c}^{(e)} + \textbf{c}^{(e) tr} \textbf{A}_1 \textbf{E}^{\prime} \textbf{a} + \frac{1}{2} \textbf{a}^{tr} \textbf{E}^{\prime tr} \textbf{A}_1 \textbf{E}^{\prime} \textbf{a} = E_0. 
	\label{Eq:Energy_Constraint}
\end{equation}
Similarly, the inequality constraint (\ref{Eq:OPT_2_4}) is recast as,
\begin{equation}
	\textbf{A}_1 \textbf{c}^{(e)} + \textbf{A}_1 \textbf{E}^{\prime} \textbf{a} \geq 0 .
\end{equation}

The details of constructing the ensemble members $\textbf{c}^{(r)}$ are provided in Appendix \ref{Appendix_A}.
The procedure ensures the following three properties: 
(i) the mean of the ensemble members is $\textbf{c}^{(e)}$;  
(ii) each member and the mean satisfy the energy constraint, $\frac{1}{2} \textbf{c}^{(r) tr} \textbf{A}_1 \textbf{c}^{(r)} = E_0$ for $r = 1, \dots N_{en}$ and $\frac{1}{2} \textbf{c}^{(e) tr} \textbf{A}_1 \textbf{c}^{(e)} = E_0$;
(iii) the deviations of the members from the mean are controlled using a covariance matrix. 
It is important to note that the arithmetic mean can not satisfy (i) and (ii) simultaneously.  
Instead, the modal amplitudes and phases of the ensemble members and the mean control vector are related by, respectively,  
\begin{subequations}
\begin{equation}
	\textbf{A}_1 \textbf{c}^{(e)} \circ \textbf{A}_1 \textbf{c}^{(e)} = \frac{1}{N_{en}} \sum_{r=1}^{N_{en}} \textbf{A}_1 \textbf{c}^{(r)} \circ \textbf{A}_1 \textbf{c}^{(r)} ,
\end{equation}
and
\begin{equation}
	\textbf{A}_2 \textbf{c}^{(e)} = \frac{1}{N_{en}} \sum_{r=1}^{N_{en}} \textbf{A}_2 \textbf{c}^{(r)},
\end{equation}
where $\circ$ is element-wise product of two vectors and 
\label{Eq:Ensemble_Mean}
\end{subequations}
\begin{equation}
\textbf{A}_2 = \begin{bmatrix}
    \textbf{O} & \textbf{O} \\
    \textbf{O} & \textbf{I}
\end{bmatrix}_{2 M \times 2 M}.
\label{Eq:dummy_Matrix_3}
\end{equation}

The optimization procedure requires computation of the gradient of the cost function with respect to the control variable, and therefore $\mathcal{J}$ must be expressed as a differentiable function of $\textbf{a}$.
First, the governing Navier-Stokes equations \ref{Eq:OPT_2_2} are written in term of the new control vector,
\begin{equation}
	\textbf{q} = \mathcal{N}\left( \textbf{c}^{(e)} + \textbf{E}^{\prime} \textbf{a} \right).
\end{equation}
The equation is expanded around the mean vector using Taylor series,
\begin{equation}
	\textbf{q} = \textbf{q}^{(e)} + \left. \frac{\partial \mathcal{N}}{\partial \textbf{c}} \right\vert_{\textbf{c}^{(e)}} \textbf{E}^{\prime} \textbf{a} + \frac{1}{2 !} \left. \frac{\partial^2 \mathcal{N}}{\partial \textbf{c}^2 } \right\vert_{\textbf{c}^{(e)}} \left( \textbf{E}^{\prime} \textbf{a} \circ \textbf{E}^{\prime} \textbf{a} \right) + \dots ,
\label{Eq:Taylor}
\end{equation}
where $\textbf{q}^{(e)} = \mathcal{N} \left( \textbf{c}^{(e)} \right)$.
By assuming the members of the ensemble are close to the mean, or $\textbf{E}^{\prime} \textbf{a}$ is small, the high-order terms in equation (\ref{Eq:Taylor}) can be ignored, 
\begin{equation}
	\textbf{q} \approx \textbf{q}^{(e)} + \left. \frac{\partial \mathcal{N}}{\partial \textbf{c}} \right\vert_{\textbf{c}^{(e)}} \textbf{E}^{\prime} \textbf{a} .
\label{Eq:Linearized_StatVector}
\end{equation}
Substituting equation (\ref{Eq:Linearized_StatVector}) into (\ref{Eq:OPT_2_1}) and using Taylor series expansion, the cost function is approximated as, 
\begin{equation}
	\mathcal{J} \left( \textbf{a} \right) \approx \frac{1}{2} \left\Vert \mathcal{G} \left( \textbf{q}^{(e)} + \left. \frac{\partial \mathcal{N}}{\partial \textbf{c}} \right\vert_{\textbf{c}^{(e)}} \textbf{E}^{\prime} \textbf{a} \right) \right\Vert^2 \approx \frac{1}{2} \left\Vert \mathcal{G} \left(  \textbf{q}^{(e)} \right) + \left. \frac{\partial \mathcal{G}}{\partial \textbf{q}} \right\vert_{\textbf{q}^{(e)}} \left. \frac{\partial \mathcal{N}}{\partial \textbf{c}} \right\vert_{\textbf{c}^{(e)}} \textbf{E}^{\prime} \textbf{a} \right\Vert^2 .
\label{Eq:Cost_Function_approx}
\end{equation}
As long as the linear approximations in (\ref{Eq:Linearized_StatVector}) and (\ref{Eq:Cost_Function_approx}) are valid, constraint (\ref{Eq:OPT_2_2}) is satisfied and the observation matrix 
$ \textbf{H} \equiv \left. \frac{\partial \mathcal{G}}{\partial \textbf{q}} \right\vert_{\textbf{q}^{(e)}} \left. \frac{\partial \mathcal{N}}{\partial \textbf{c}} \right\vert_{\textbf{c}^{(e)}} \textbf{E}^{\prime}$ can be evaluated as,  
\begin{equation}
	\textbf{H} \approx \begin{bmatrix}
\mathcal{G} \left( \textbf{q}^{(1)} \right) - \mathcal{G} \left( \textbf{q}^{(e)} \right) & \dots & \mathcal{G} \left( \textbf{q}^{(N_{en})} \right) - \mathcal{G} \left( \textbf{q}^{(e)} \right)
\end{bmatrix}_{N_{\mathcal{G}}\times N_{en}} ,
\label{Eq:Observation_Matrix}
\end{equation}
where $N_{\mathcal{G}}$ is the size of observation vector.
Note that the validity of (\ref{Eq:Linearized_StatVector}) to (\ref{Eq:Observation_Matrix}) is predicated on $\textbf{E}^{\prime} \textbf{a}$ being small --- a condition that is ensured during the generation of the ensemble members (Appendix \ref{Appendix_A}).
Assembling matrix $\textbf{H}$ is the most expensive part of the algorithm since it requires $N_{en} + 1$ solutions of the governing equations.
While in the current study direct numerical simulations (DNS) are used to evaluate $\textbf{H}$, the derivation of (\ref{Eq:Observation_Matrix}) did not invoke any assumptions regarding the choice of the numerical approach to computing $\mathcal{G} \left( \textbf{q}^{(r)} \right)$.
As a result, other approaches such as large-eddy simulations and the nonlinear PSE can potentially be adopted.

In summary, the constrained optimization (\ref{Eq:Optimization_Problem_2}) can be recast as seeking the weight vector $\textbf{a}$ that optimizes the cost function,
\begin{subequations}
\begin{equation}
	\mathcal{J} \left( \textbf{a} \right) \approx \frac{1}{2} \left\Vert \mathcal{G} \left(  \textbf{q}^{(e)} \right) + \textbf{H} \textbf{a} \right\Vert^2 ,
\label{Eq:OPT_F_1}
\end{equation}
while satisfying the constraints
\begin{equation}
	\frac{1}{2} \textbf{c}^{(e) tr} \textbf{A}_1 \textbf{c}^{(e)} + \textbf{c}^{(e) tr} \textbf{A}_1 \textbf{E}^{\prime} \textbf{a} + \frac{1}{2} \textbf{a}^{tr} \textbf{E}^{\prime tr} \textbf{A}_1 \textbf{E}^{\prime} \textbf{a} - E_0 = 0 ,
\label{Eq:OPT_F_2}
\end{equation}
and
\begin{equation}
	- \textbf{A}_2 \textbf{E}^{\prime} \textbf{a} \leq \textbf{A}_2 \textbf{c}^{(e)} .
\label{Eq:OPT_F_3}
\end{equation}
Interior-point method \citep{Nocedal2006,Wachter2006} is used to solve (\ref{Eq:Optimization_Problem_Final}) by generating iterates that satisfy the inequality bounds, strictly. 
\label{Eq:Optimization_Problem_Final}
\end{subequations}
The full procedure to solve the optimization problem, and hence identify the nonlinearly most dangerous disturbance, is shown in algorithm \ref{Algorithm:Optimization}.

\begin{algorithm}
\vspace*{0.2in}
\SetAlgoLined
 \textbullet~{\it Iteration} = 0 \; 
   \textbullet~Establish a first guess for the control vector, $\textbf{c}^{(e)}$, that satisfies the constraints\;
   \textbullet~Simulate the evolution of the mean control vector, $\textbf{q}^{(e)} = \mathcal{N} \left( \textbf{c}^{(e)} \right) $\;
   \textbullet~Compute the mean observation vector $\mathcal{G}\left( \textbf{q}^{(e)} \right)$\;
    \While{The convergence condition is not satisfied}{
  \textbullet~Construct an ensemble of $\textbf{c}^{(r)}$ that satisfy the constraints (see Appendix \ref{Appendix_A})\;
  \textbullet~Compute the deviation matrix $\textbf{E}^{\prime}$ (equation \ref{Eq:Deviation_Matrix})\;
  \textbullet~Simulate the evolution of each ensemble member, $\textbf{q}^{(r)} = \mathcal{N} \left( \textbf{c}^{(r)} \right) $\;
  \textbullet~Compute the observation vectors of the members of the ensemble $\mathcal{G}\left( \textbf{q}^{(r)} \right)$\;
  \textbullet~Compute the observation matrix $\textbf{H}$ (equation \ref{Eq:Observation_Matrix})\;
  \textbullet~Solve the constrained optimization (\ref{Eq:Optimization_Problem_Final}) using interior-point method\;
  \textbullet~Update the mean control vector as $\textbf{c}^{(e)} \rightarrow \textbf{c}^{(e)} + \textbf{E}^{\prime} \textbf{a} $\;
 \textbullet~Simulate the evolution of the mean control vector, $\textbf{q}^{(e)} = \mathcal{N} \left( \textbf{c}^{(e)} \right) $\;
 \textbullet~Compute the mean observation vector $\mathcal{G}\left( \textbf{q}^{(e)} \right)$\;
  \textbullet~Check for convergence\;
  \textbullet~{\it Iteration} = {\it Iteration} + 1 \;
  }
 \caption{Finding the nonlinearly most dangerous perturbation using an ensemble-variational optimization technique.}
 \label{Algorithm:Optimization}
\end{algorithm}

\section{Computational aspects}
\label{sec:DNS}

During the search for the nonlinearly most dangerous inflow disturbance, the optimization algorithm involves the solution of the Navier-Stokes equations for every ensemble member and observation of the wall friction.  
The simulations are performed using code {\it Hybrid} \citep{Johnsen2010}, which solves the compressible Navier-Stokes equations (\ref{Eq:Navier_Stokes}) on a structured grid and is designed for accurate simulations of high-speed transitional and fully turbulent flows.
It is nominally free of numerical dissipation, which is important for accurate prediction of the evolution of instability waves and transition onset.
The time advancement is $4^{\textrm{th}}$ order accurate using the Runge$-$Kutta scheme, and the spatial discretization in absence of shocks is $6^{\textrm{th}}$ order central with a split form which improves nonlinear stability, and a $5^{\textrm{th}}$ order WENO scheme with Roe flux-splitting is used near shocks.
A more detailed description of the numerical methods used in {\it Hybrid} is provided in \cite{Johnsen2010}.

A schematic of the simulated domain is shown in figure \ref{FIG:BL_Schematic}.
The boundary conditions are periodic in the spanwise $(z)$ direction, while characteristic conditions are adopted on the remaining four boundaries in the streamwise ($x$) and wall-normal ($y$) directions.
As indicated in the figure, sponge regions are used along the outflow boundaries to minimize reflections of disturbances back into the computational domain.
Along these layers, a relaxation term is added to the governing equations to force the solution towards the spanwise average at each time step.

Two cases are examined to highlight the effect of the energy level on the nonlinearly most dangerous inflow perturbations.
Table \ref{TABLE:Simulation_Parameters} summarizes the computational and physical parameters of each case, which are identical except for the inflow disturbance energy.
The free-stream values are adopted as reference scales for normalization of velocity, temperature, density, viscosity, specific heat and conductivity.
The reference length is the Blasius length-scale, $( \tilde{\mu}_{\infty} \tilde{x}_0 / \tilde{\rho}_{\infty} \tilde{U}_{\infty} )^{1/2}$, at the inlet location, where the $\tilde{.}$ represents dimensional quantities and $\tilde{x}_0$ is the start location of the simulation relative to the virtual boundary-layer origin.

The free-stream Mach number in all simulations is $\textrm{Ma}_\infty = 4.5$.  
The Reynolds number based on inlet is $\textrm{Re}_{x_0}$ and the streamwise position of the inflow plane is $x_0 = \sqrt{ \textrm{Re}_{x_0} }=1800$, which was selected based on the transition Reynolds numbers in high-altitude flight tests being $\sqrt{\textrm{Re}_{x,tr}} > 2000$ for $\textrm{Ma}_{\infty} > 4$ \citep{Harvey1978,Schneider1999}.
Starting the simulations from smaller Reynolds numbers requires a much longer domain size, which poses prohibitively high computational requirements.
The lengths of computational domain in $x$, $y$, and $z-$directions are $L_x$, $L_y$ and $L_z$, with grid points $N_x$, $N_y$ and $N_z$.
The computational grid is uniform in $x$ and $z-$directions, and a hyperbolic tangent stretching is used in $y-$direction.
The sizes of the sponge layers in $x$ and $y$ are denoted $L_{s,x}$ and $L_{s,y}$.

\begin{table}
\begin{center}
\def~{\hphantom{0}}
\begin{tabular}{lccccccccccccccc}
Case & $\textrm{Ma}_{\infty}$ & $ \sqrt{\textrm{Re}_{x_0}} $ & $L_x$ & $L_y$ & $L_z$ & $N_x$ & $N_y$ & $N_z$ & $L_{s,x}$ & $L_{s,y}$ & $E_0 \times 10^{5} $  & $F_{l}$ & $F_{u}$  & $k_{z,l}$ & $k_{z,u}$ \\ [3pt]
\hline
\\
E1 & $4.5$ & $1800$ & $2983.5$ & $ 200 $ & $150$ & $1989$ & $130$ & $108$ & $250$ & $30$ & $2 $ & $10$ & $250$ & $0$ & $15$ \\
E2 & $4.5$ & $1800$ & $2983.5$ & $ 200 $ & $150$ & $1989$ & $130$ & $108$ & $250$ & $30$ & $100$ & $10$ & $250$ & $0$ & $15$ \\
\hline
\end{tabular}
\caption{Physical and computational parameters for the cases of the present study.}
\label{TABLE:Simulation_Parameters}
\end{center}
\end{table}

The two cases, E1 and E2, are distinguished by the total disturbance energy at the inflow plane, $E_0$, which is prescribed as a constraint in the optimization problem (constraint \ref{Eq:OPT_F_2}).
Comparison of E1 and E2 highlights the effect of the energy constrain on the spectral makeup of the nonlinearly most unstable inflow perturbation, and the associated transition mechanism. 
The amount of energy for case E1 is chosen to be of the same order as the turbulence kinetic energy in stratospheric layers, measured during a recent experimental campaign by \cite{Haack2014} with the balloon-borne instrument Leibniz Institute Turbulence Observations in the Stratosphere (LITOS).
The normalized frequencies of the disturbance (\ref{Eq:Inflow_Modes}) are assumed to range from $F_l$ to $F_u$ with an increment $\Delta F = F_l = 10$, where
\begin{equation}
	F = \frac{\omega_n}{\sqrt{\textrm{Re}_{x_0}}} \times 10^{6} = n F_l \;\;\;\;\;\; \textrm{for} \;\;\;\;\;\; n = 1,2,\dots,\frac{F_u}{F_l} .
\label{Eq:Normalized_Freq}
\end{equation}
Similarly, $k_{z,l}$ and $k_{z,u}$ in table \ref{TABLE:Simulation_Parameters} are the lower and upper bounds of the integer spanwise wavenumbers of the inflow disturbance, where
\begin{equation}
	k_{z} = \frac{\beta_{m} L_z}{2 \pi} = m \;\;\;\;\;\; \textrm{for} \;\;\;\;\;\; m = k_{z,l},\dots,k_{z,u} .
\label{Eq:spanwise_wavenum}
\end{equation}
For the three-dimensional modes at the inlet, $k_z > 0$, the modal energy is divided equally between the positive and negative spanwise wave-numbers, $\pm \beta_m$, with the same phase $\theta_{n,m}$ in equation (\ref{Eq:Inflow_Modes}).
Therefore, hereafter only $k_{z} > 0$ will be reported since they are representative of the pairs of oblique modes.
The operating gas is air and the free-stream temperature used in Sutherland's law is $65.15$ Kelvin.

\begin{figure}
\centerline{%
\includegraphics[trim=0 0 0 0, clip,width=0.8\textwidth] {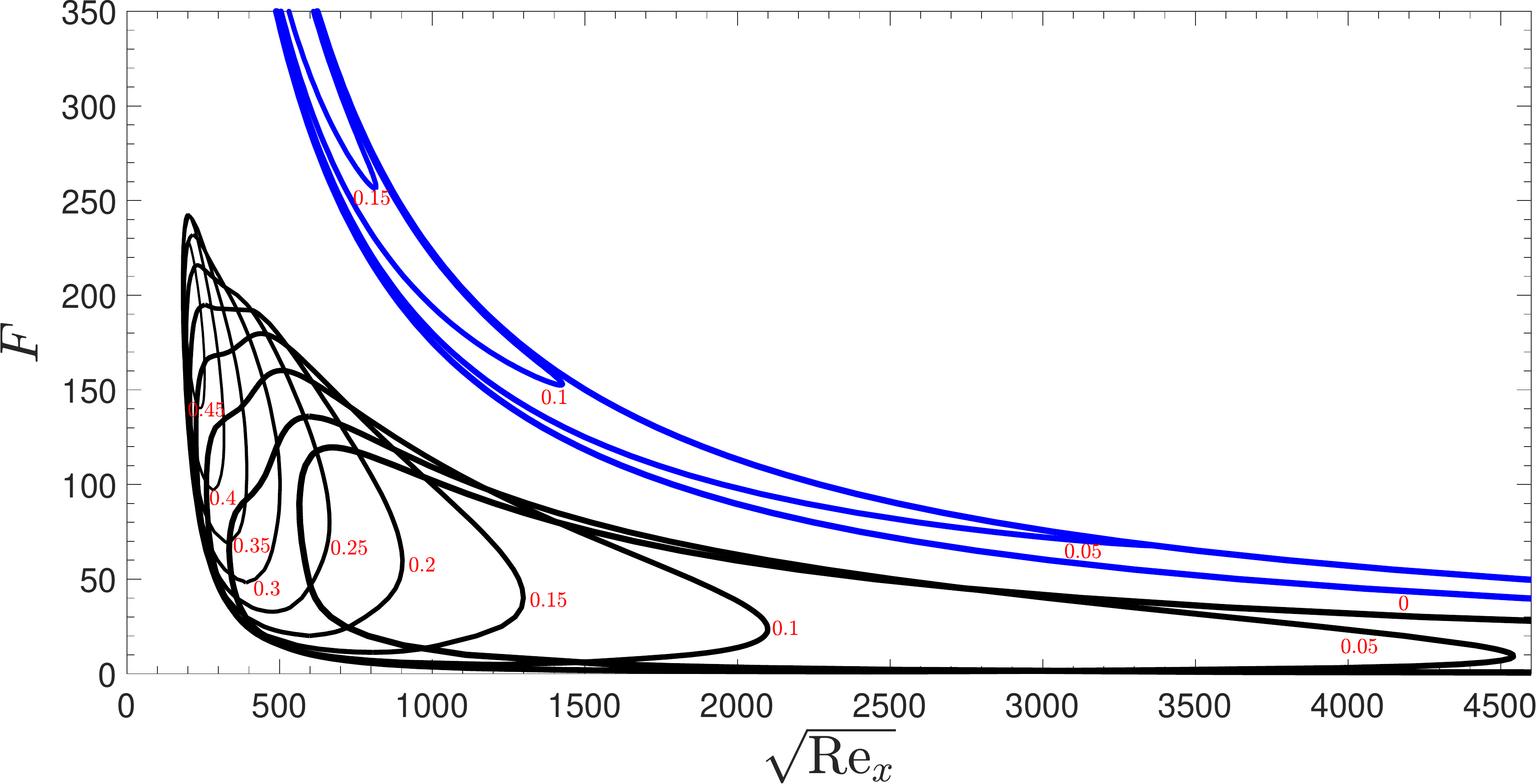}%
}%
\caption{Neutral Curves of a zero-pressure-gradient boundary layer at $\textrm{Ma} = 4.5$. The values marked on the curves correspond to the normalized spanwise wave number, $b = \frac{\beta}{\sqrt{\textrm{Re}_{x}}} \times 10^3$.}
\label{FIG:Neutral_Curve}
\end{figure}

Figure \ref{FIG:Neutral_Curve} shows the neutral curves from linear spatial stability analysis of a zero-pressure-gradient boundary layer at $\textrm{Ma}_{\infty} = 4.5$.
The lower unstable regions in the figure are commonly referred to as (Mack's) first-mode instability regions, whereas the upper unstable regions are called (Mack's) second-mode disturbances \citep{Mack1984}.
The ranges of $F$ and $k_z$ in table \ref{TABLE:Simulation_Parameters} are appropriately selected such that they include all the relevant frequencies and spanwise wave numbers at the inlet location.
For every pair of frequency and spanwise wavenumber, the most unstable discrete mode in the Orr-Sommerfeld and Squire spectra is included; these instabilities correspond to slow modes \citep{Fedorov2011}.

\section{Results and discussions}
\label{sec:Results}

\subsection{Nonlinearly most unstable inflow spectra}
\label{sec:All_Cases}

The optimization algorithm \ref{Algorithm:Optimization} seeks the nonlinearly most unstable disturbance by maximizing the cost-function $\mathcal{J} = \frac{1}{2} \Vert \mathcal{G} ( \textbf{q} ) \Vert^2 = \frac{1}{2 L_x} \int_{x_0}^{x_f} C_f^2 dx$ where, 
\begin{equation}
	C_f = \frac{\tau_{wall}}{\frac{1}{2} \rho_{\infty} U_{\infty}^2 }. 
\end{equation}
The initial guess of the disturbance spectra was equipartition of the energy among all instability waves and randomly assigned phases. 
In addition to the mean, twenty ensemble members were used in each iteration. 
The stopping criterion of the optimization was $(\mathcal{J}_{i} - \mathcal{J}_{i-1}) / \mathcal{J}_{i} < 10^{-3}$, where $i$ is the iteration number.

\begin{figure}
\centerline{%
\includegraphics[trim=0 0 0 0, clip,width=0.45\textwidth] {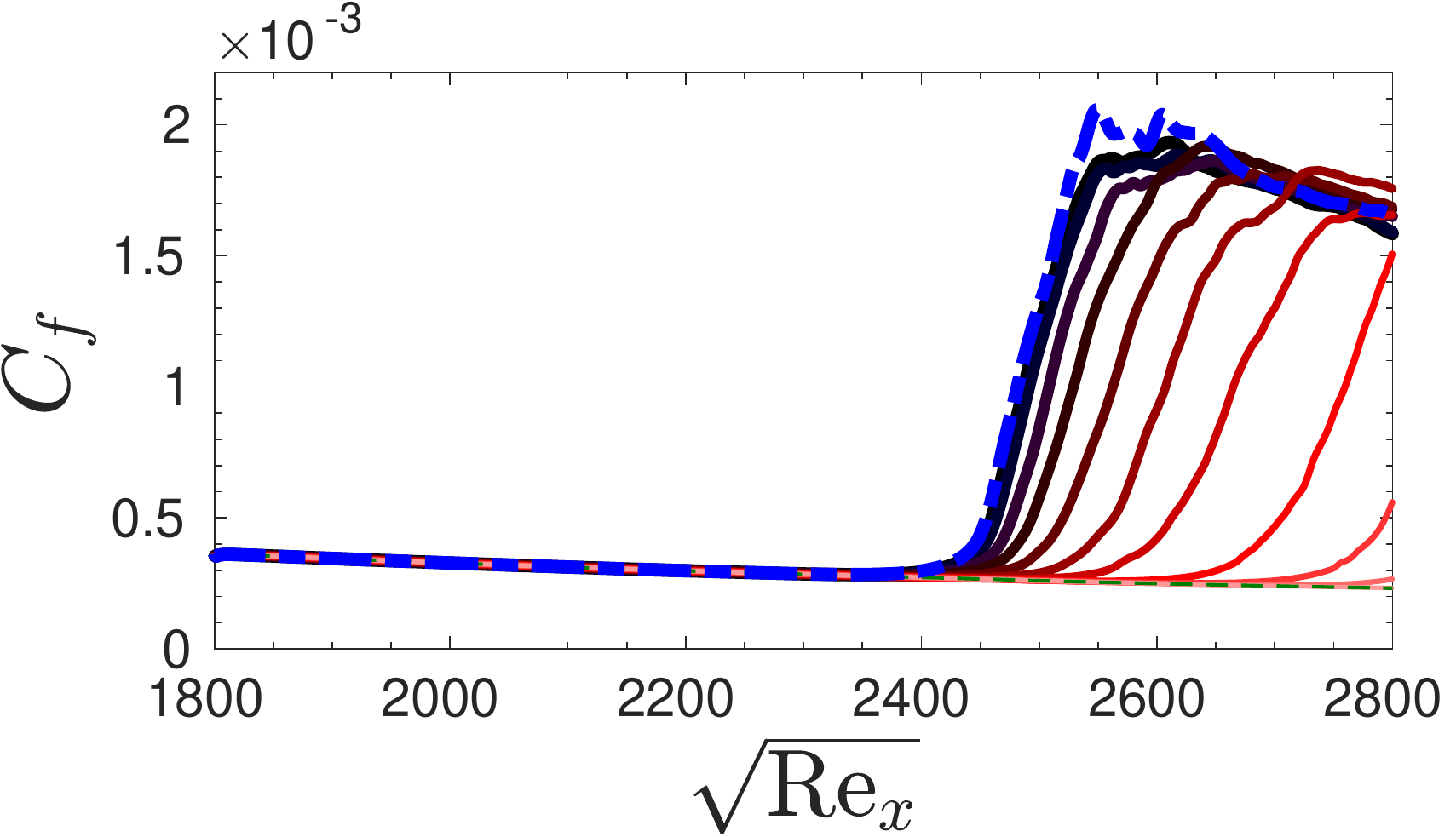}%
\hspace{1.00mm}
\includegraphics[trim=0 0 0 0, clip,width=0.45\textwidth] {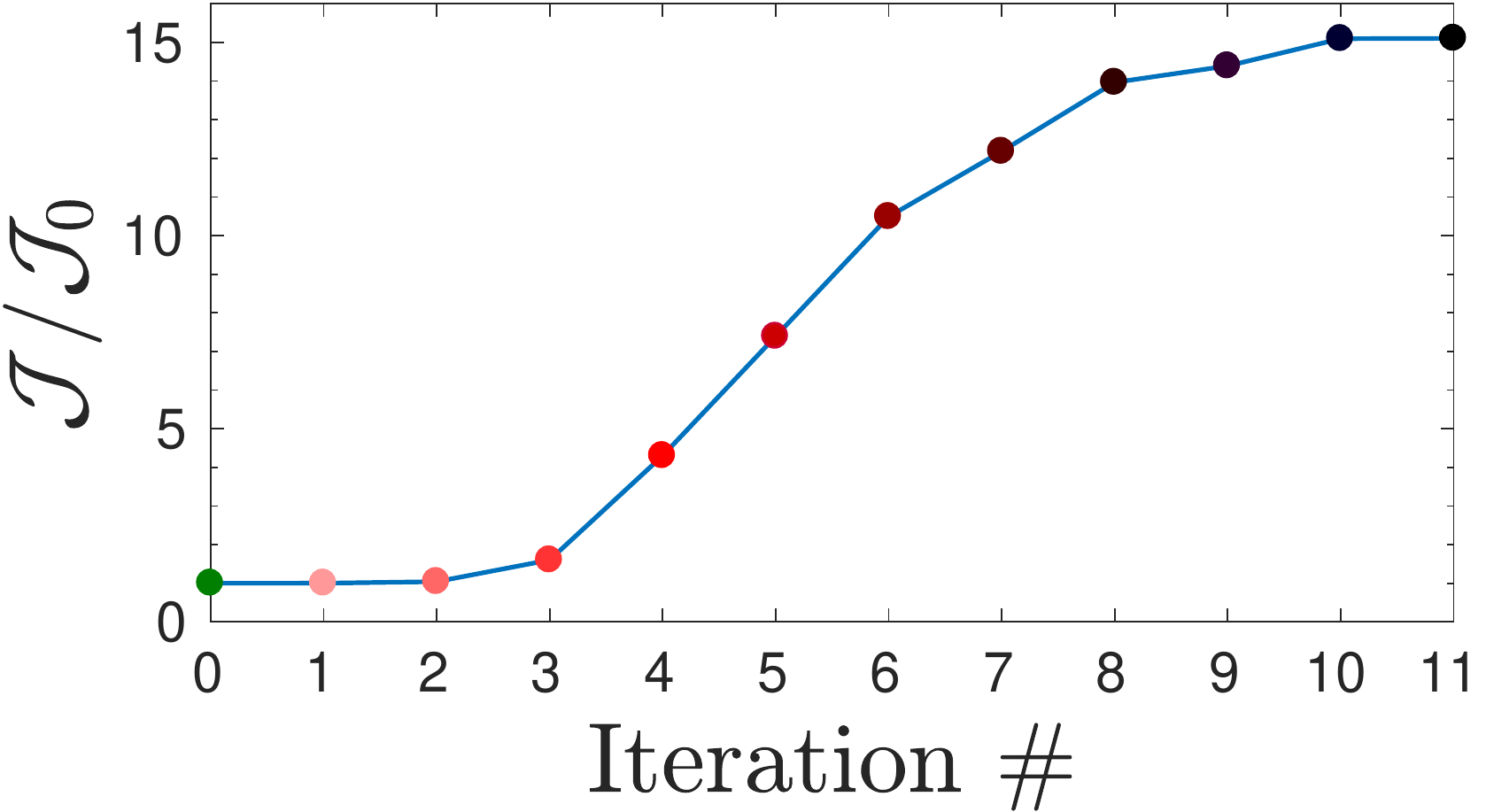}%
\put(-360.0,90.0){$(a)$}
}%
\centerline{%
\includegraphics[trim=0 0 0 0, clip,width=0.45\textwidth] {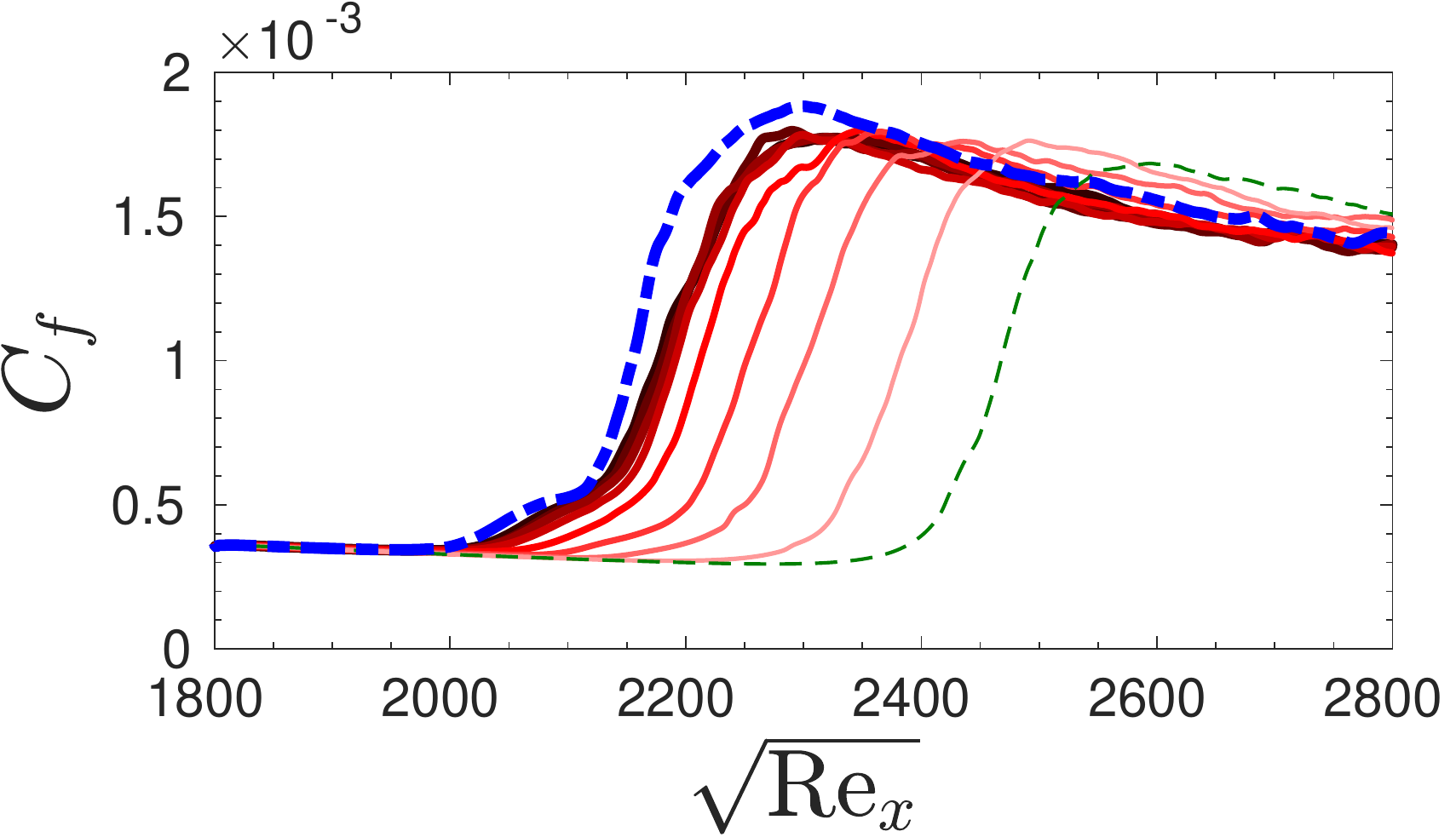}%
\hspace{1.00mm}
\includegraphics[trim=0 0 0 0, clip,width=0.45\textwidth] {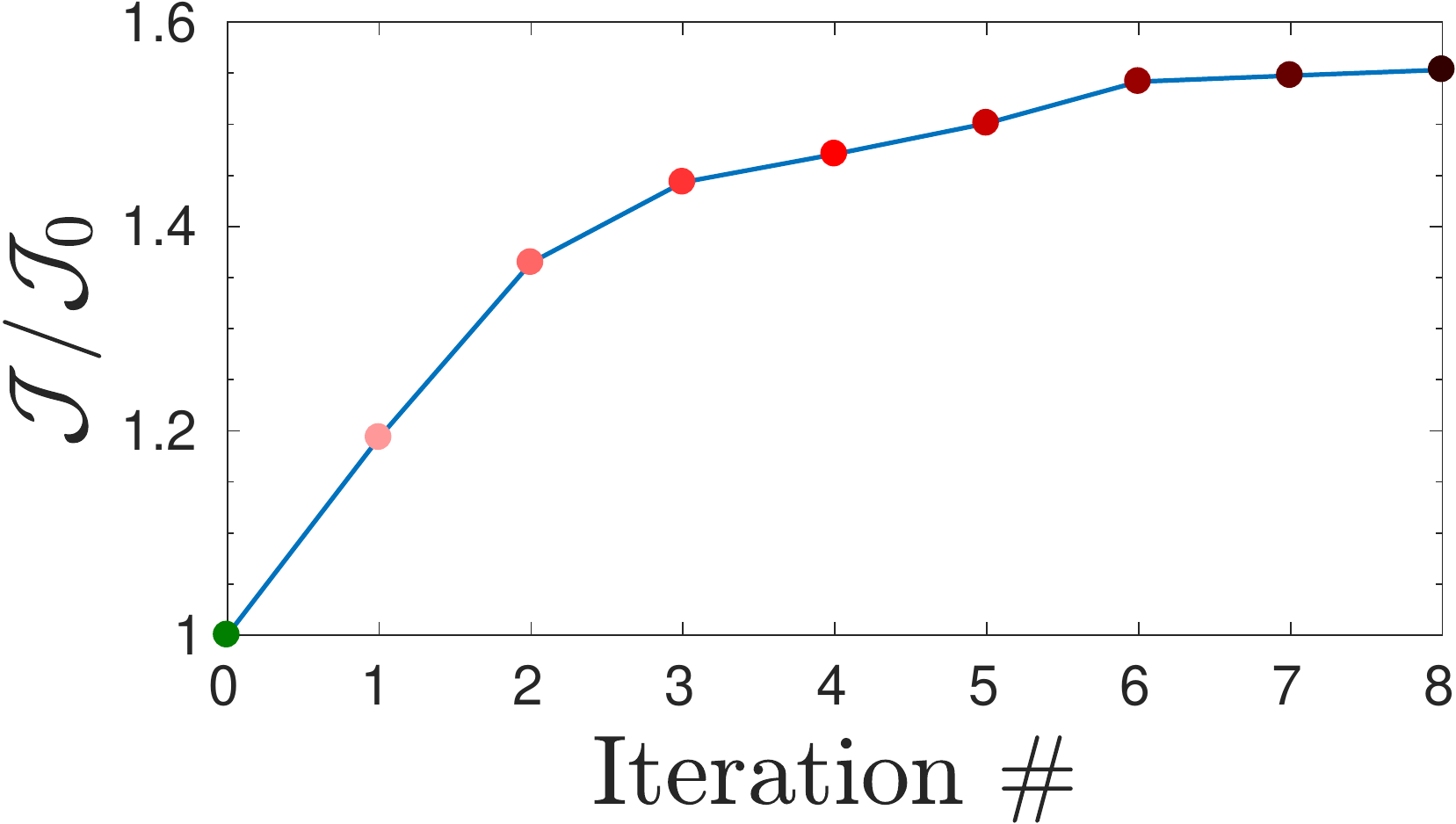}%
\put(-360.0,90.0){$(b)$}
}%
\caption{$C_f$ (left) and normalized cost function (right) for cases $(a)$ E1 and $(b)$ E2. In the $C_f$ plots, the dashed green profiles correspond to the initial guess, and thin to thick (also light to dark coloured) solid lines represent the result for the mean control vector after consecutive iterations.  The dashed-dotted blue curves correspond to the nonlinearly most unstable inflow spectra for each case (see table \ref{TABLE:NonLinearly_Most_Unstable}).}
\label{FIG:Cf_Curves_Cost_Function}
\end{figure}

For each iteration, both $C_f$ and $\mathcal{J}/\mathcal{J}_0$ associated with the mean control vector $\textbf{c}^{(e)}$ are plotted in figure  \ref{FIG:Cf_Curves_Cost_Function}. 
For case E1, the flow remains laminar throughout the computational domain for the initial guess and first iteration.
As a result, the $C_f$ curves of these two inflow perturbations nearly coincide and, while $(\mathcal{J}_{1} - \mathcal{J}_{0}) / \mathcal{J}_{1} \approx 9.99 \times 10^{-4} < 10^{-3}$, further iterations are performed. 
The convergence rate of the optimization algorithm is highly dependent on the observation matrix, defined in equation (\ref{Eq:Observation_Matrix}).
Figure \ref{FIG:Cf_Curves_Cost_Function}$a$ shows that the convergence rate of the algorithm for case E1 is slow during the first two iterations.
This is due to the fact that the observation matrix formed by the skin friction profiles is close to a singular matrix for these two iterations.
Generally, transition to turbulence advances upstream with successive iterations, with convergence observed for both E1 and E2.  
For E1, transition location is nearly unchanged from the tenth to the eleventh iteration, and for case E2 a similar behaviour is observed from the seventh to eighth iteration.

\begin{figure}
\centerline{%
\includegraphics[trim=0 0 0 0, clip,width=0.45\textwidth] {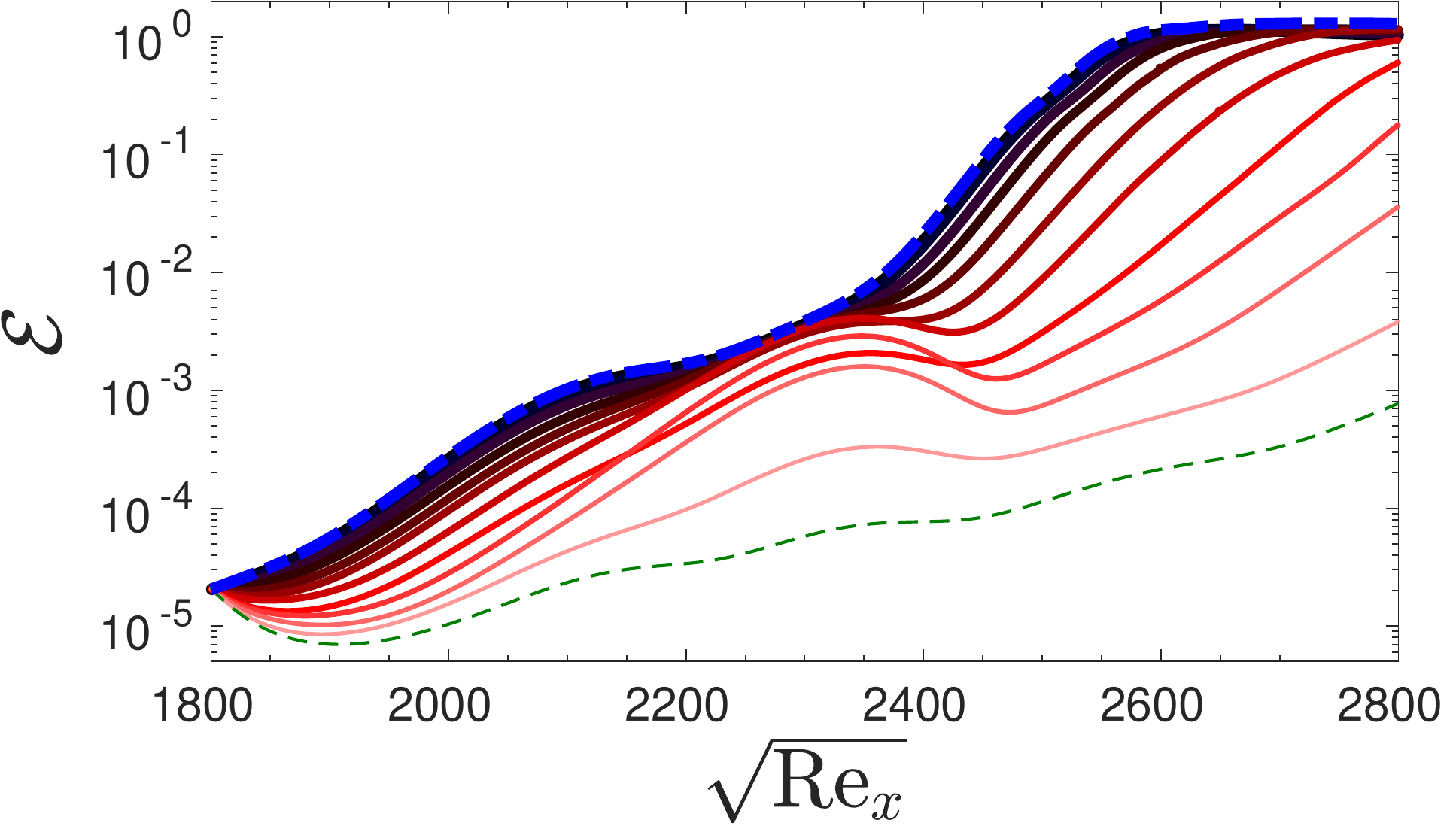}%
\hspace{1.00mm}
\includegraphics[trim=0 0 0 0, clip,width=0.45\textwidth] {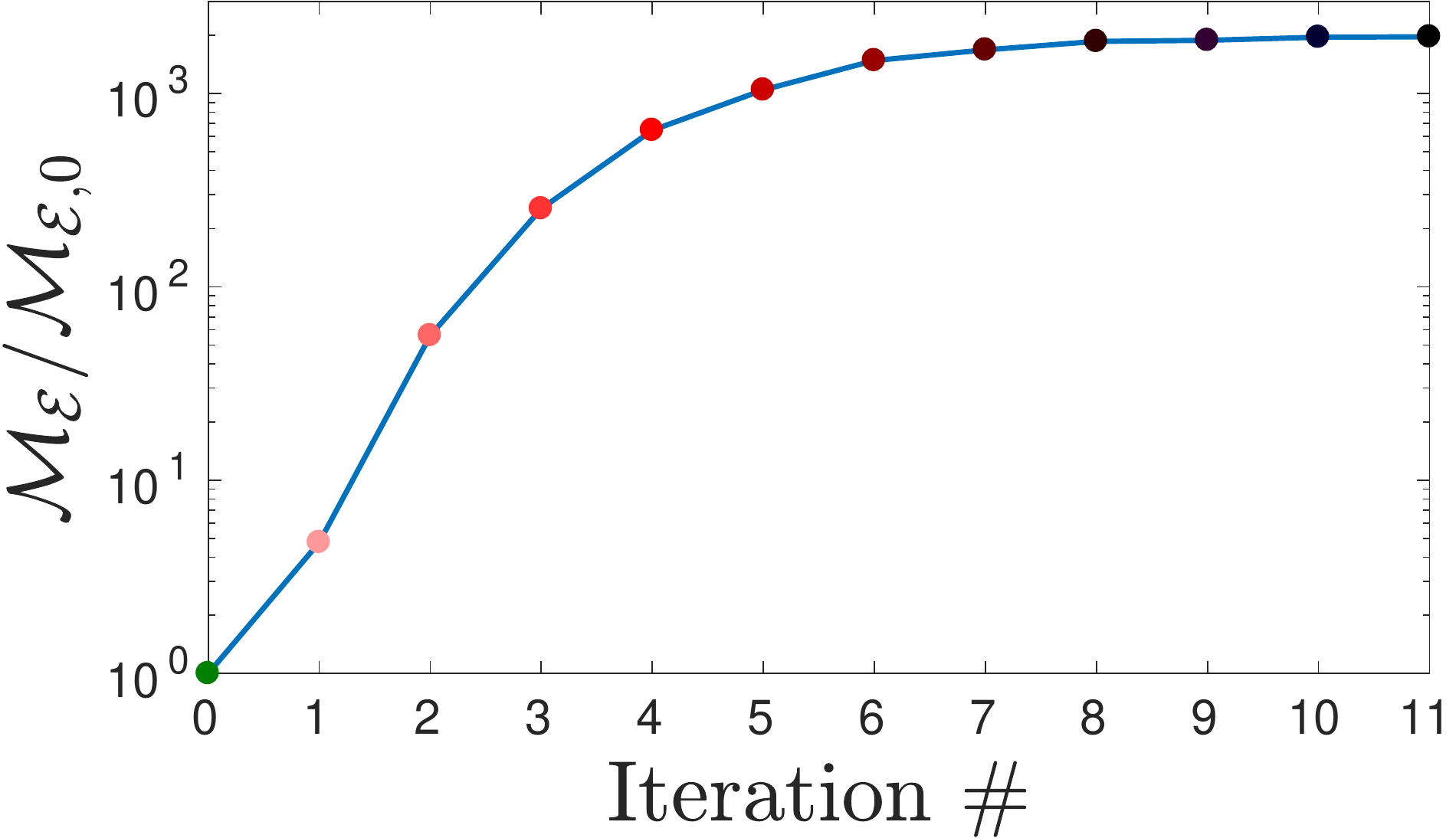}%
\put(-360.0,90.0){$(a)$}
}%
\centerline{%
\includegraphics[trim=0 0 0 0, clip,width=0.45\textwidth] {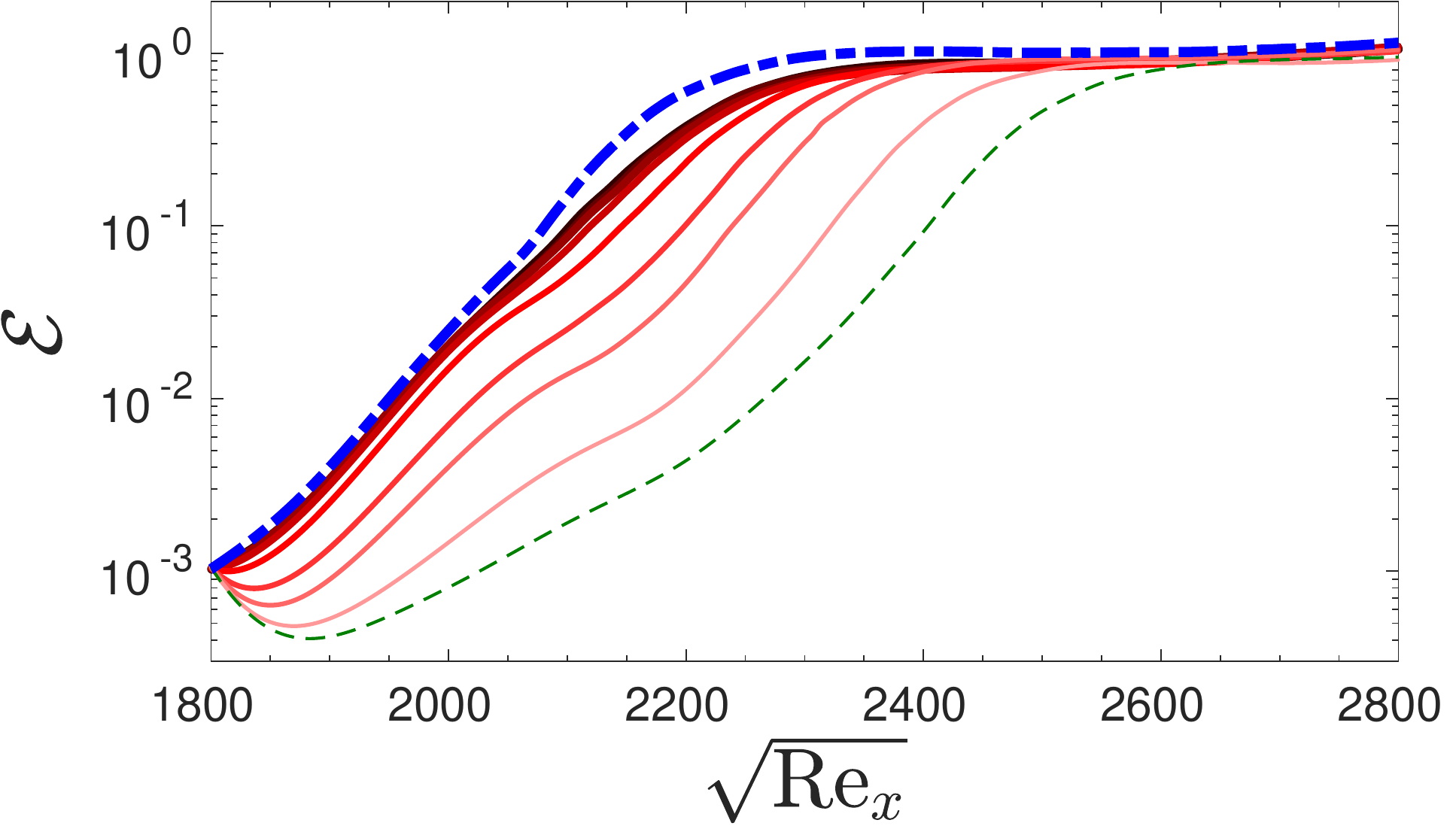}%
\hspace{1.00mm}
\includegraphics[trim=0 0 0 0, clip,width=0.45\textwidth] {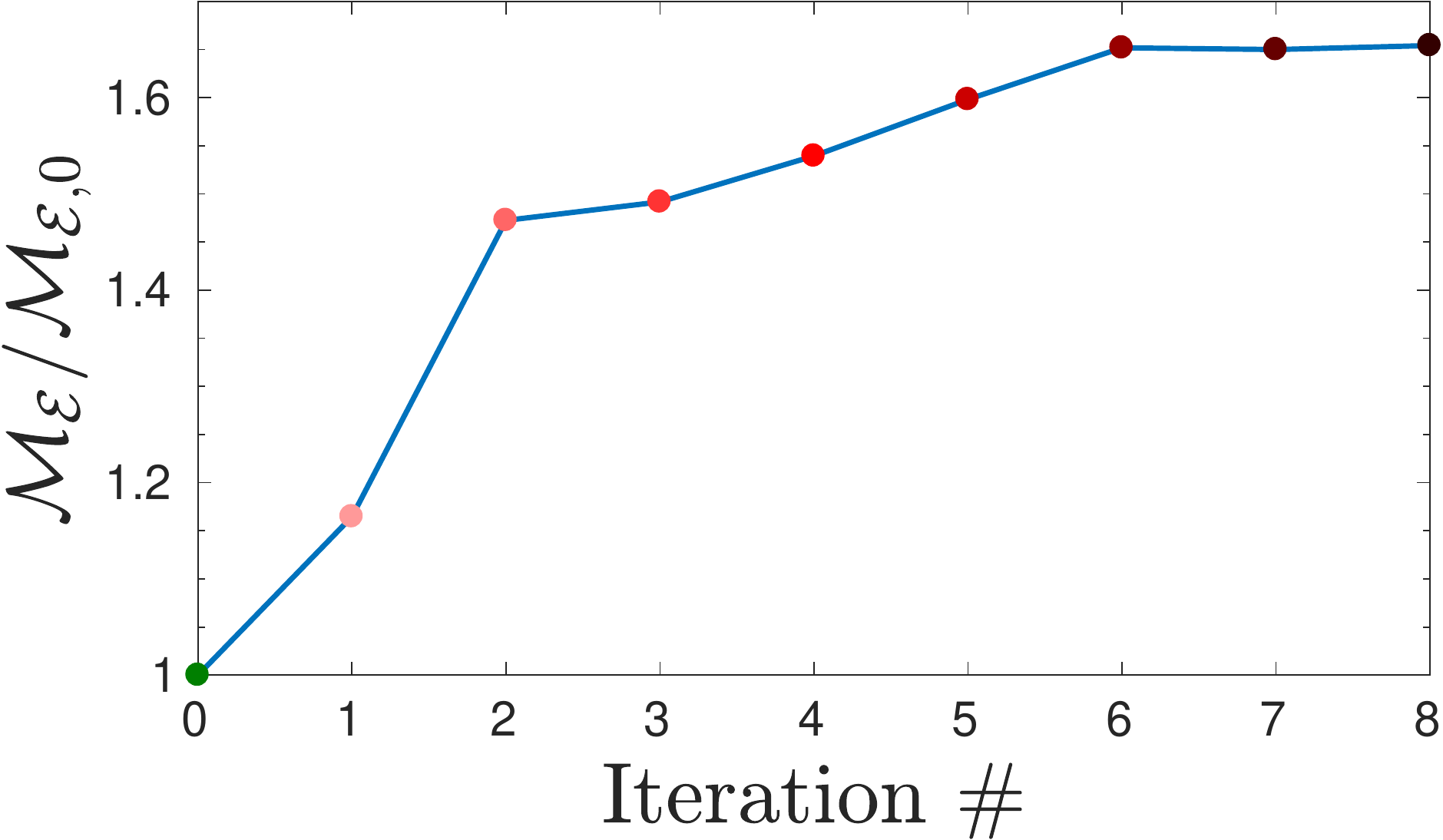}%
\put(-360.0,90.0){$(b)$}
}%
\caption{$\mathcal{E}$ curves (left plots) and the value of the norm function (right plots), computed as $ \mathcal{M}_{\mathcal{E}} = \frac{1}{L_x} \int_{x_0}^{x_f} \mathcal{E} dx $, corresponding to cases $(a)$ E1 and $(b)$ E2. In the $\mathcal{E}$ plots, the dashed green profiles correspond to the initial guess (iteration \# 0 in algorithm \ref{Algorithm:Optimization}), thin to thick (also light to dark coloured) solid lines represent the optimal solutions after consecutive iterations, and the dashed-dotted blue profiles correspond to the nonlinearly most unstable inflow spectra for each case (the spectra in table \ref{TABLE:NonLinearly_Most_Unstable}).}
\label{FIG:Et_Curves_Cost_Function}
\end{figure}

\begin{figure}
\centerline{%
\includegraphics[trim=0 0 0 0, clip,width=0.45\textwidth] {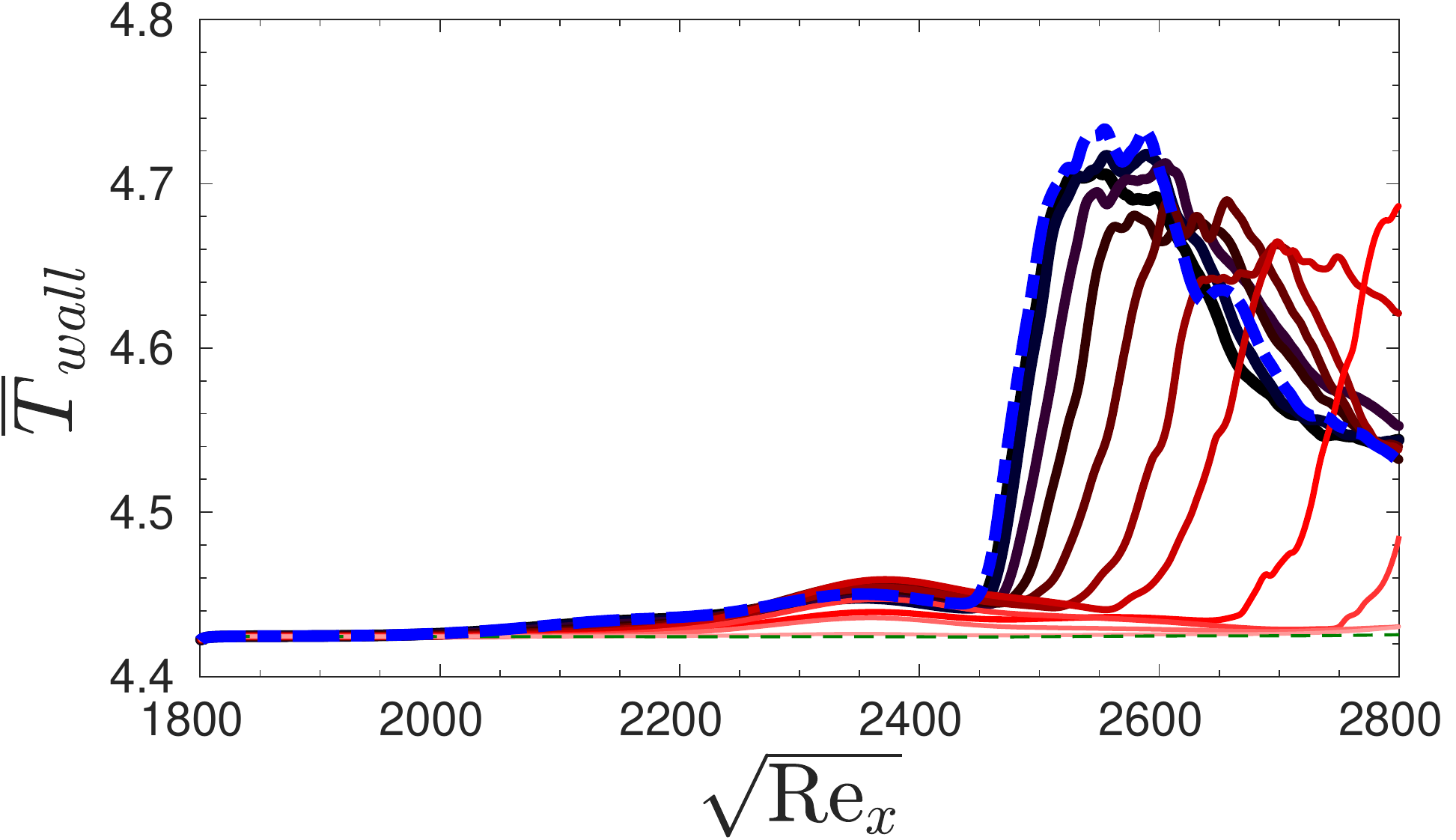}%
\hspace{1.00mm}
\includegraphics[trim=0 0 0 0, clip,width=0.45\textwidth] {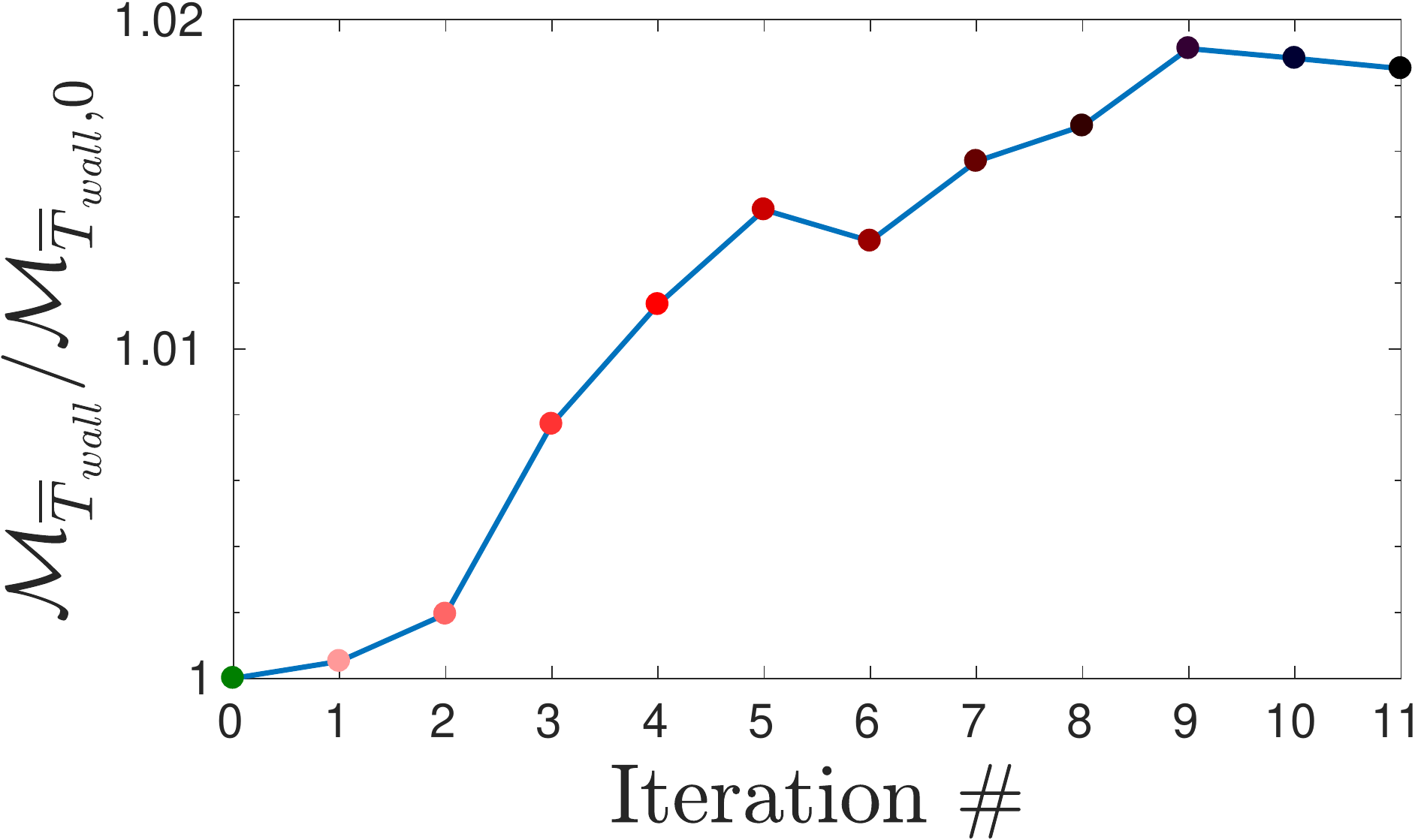}%
\put(-360.0,90.0){$(a)$}
}%
\centerline{%
\includegraphics[trim=0 0 0 0, clip,width=0.45\textwidth] {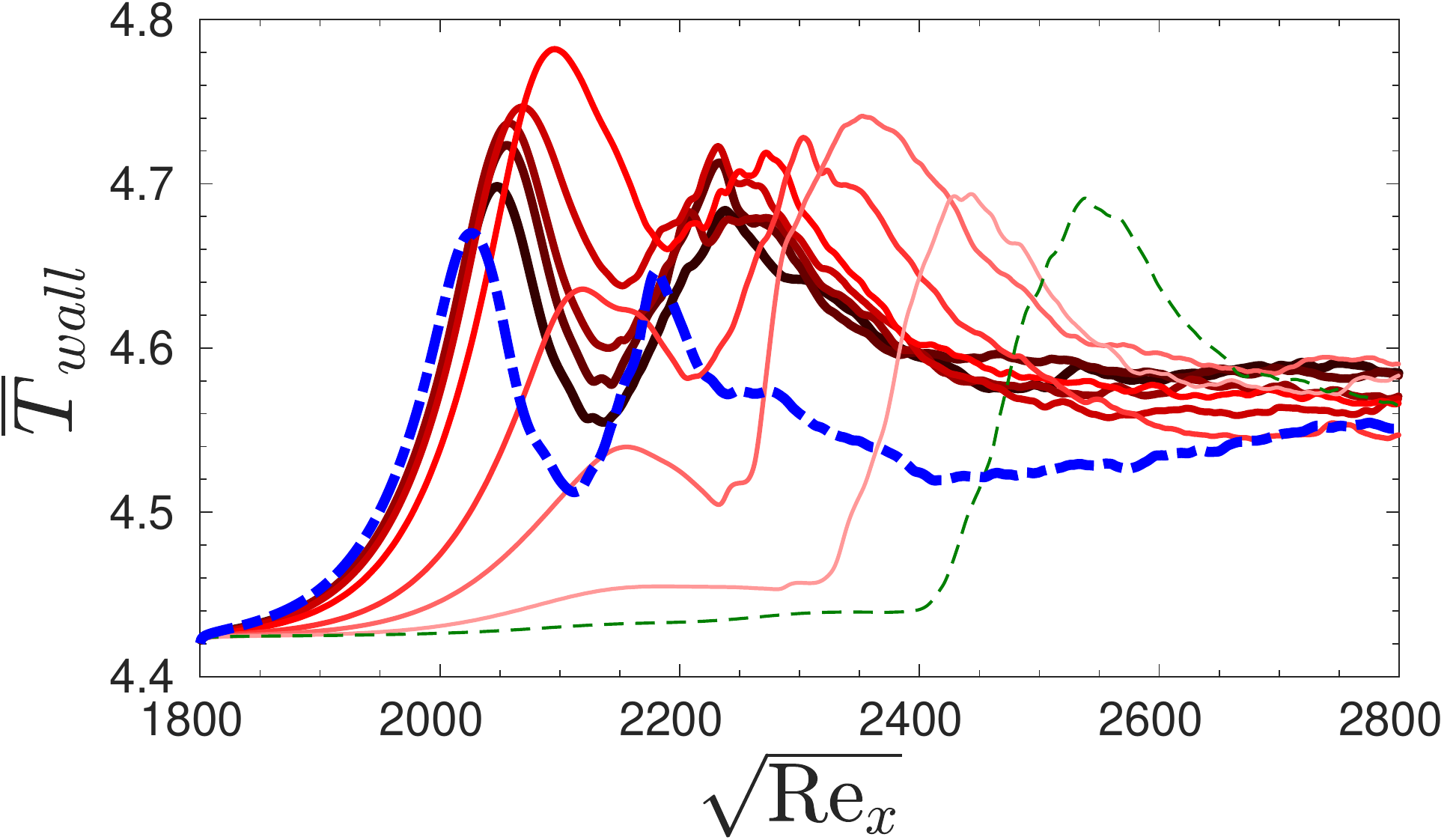}%
\hspace{1.00mm}
\includegraphics[trim=0 0 0 0, clip,width=0.45\textwidth] {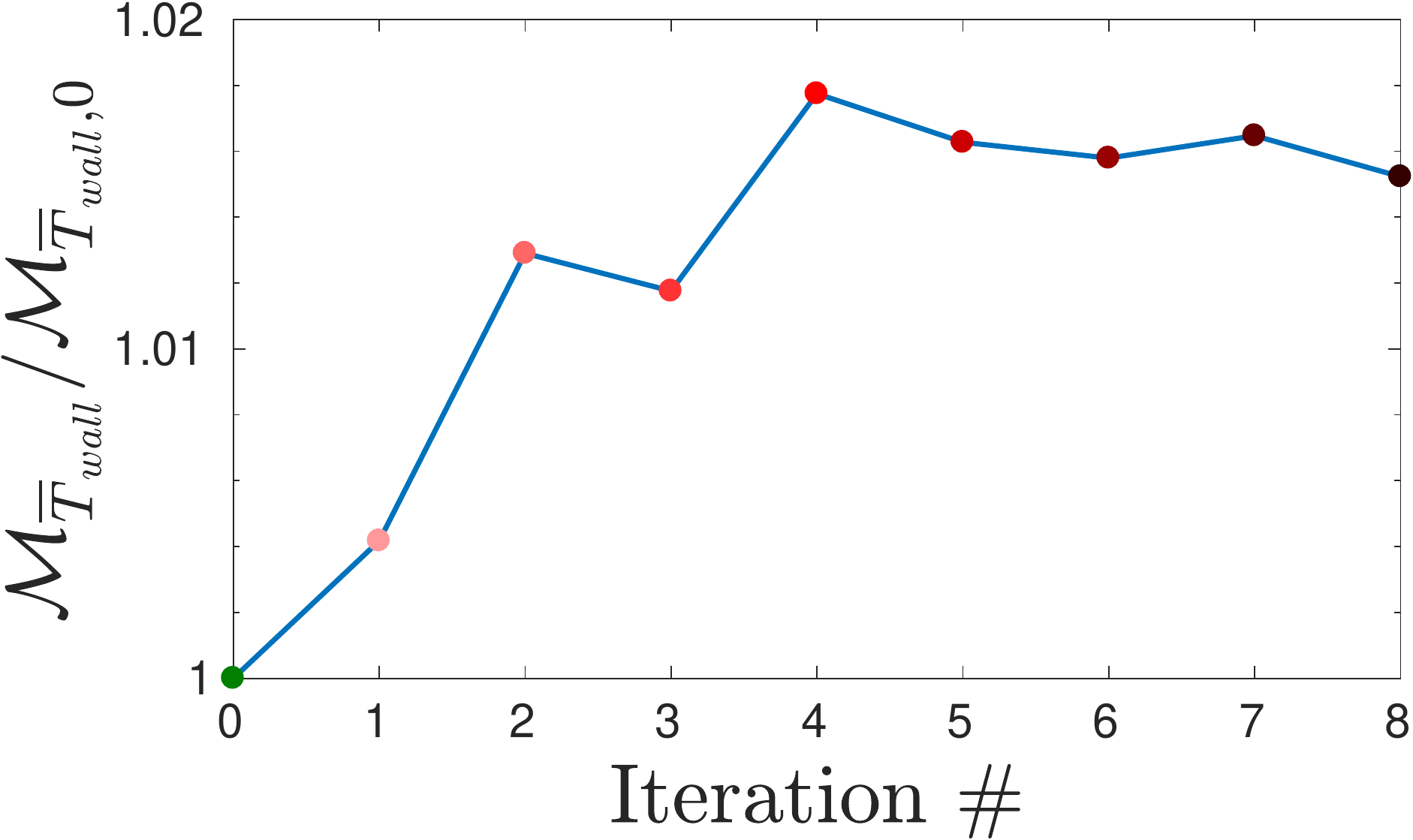}%
\put(-360.0,90.0){$(b)$}
}%
\caption{$\overline{T}_{wall}$ curves (left plots) and the value of the norm function (right plots), computed as $\mathcal{M}_{\overline{T}_{wall}} = \frac{1}{L_x} \int_{x_0}^{x_f} \overline{T}_{wall} dx$, corresponding to cases $(a)$ E1 and $(b)$ E2. In the $\overline{T}_{wall}$ plots, the dashed green profiles correspond to the initial guess (iteration \# 0 in algorithm \ref{Algorithm:Optimization}), thin to thick (also light to dark coloured) solid lines represent the optimal solutions after consecutive iterations, and the dashed-dotted blue profiles correspond to the nonlinearly most unstable inflow spectra for each case (the spectra in table \ref{TABLE:NonLinearly_Most_Unstable}).}
\label{FIG:Twall_Curves_Cost_Function}
\end{figure}

The choice of cost function in algorithm \ref{Algorithm:Optimization} is not unique.
Depending on the application of interest, measures based on the total energy of the perturbations within the domain or the wall temperature could have been adopted.
Every new cost function, however, requires repeating the entire optimization procedure, which is not performed here.  
Instead, for each iterate of the optimization where the cost function was proportional to skin friction, the average integrated energy and the average wall temperature were recorded and are plotted in figures \ref{FIG:Et_Curves_Cost_Function} and \ref{FIG:Twall_Curves_Cost_Function}.  
Note that in figure \ref{FIG:Et_Curves_Cost_Function}, the energy $\mathcal{E}$ is computed from equation (\ref{Eq:Total_Energy}) with fluctuations evaluated relative to a laminar solution over the entire plate. 
As a result, a base-state distortion term with frequency and spanwise wavenumber $\left< 0 , 0 \right>$ contributes to the perturbation energy.

The streamwise integrals of the energy and wall temperature are also plotted in figures \ref{FIG:Et_Curves_Cost_Function} and \ref{FIG:Twall_Curves_Cost_Function}, normalized by their values from the initial guess. 
Note, however, that these norms were not used in an optimization procedure.
The normalized integral of the energy, $\mathcal{M}_{\mathcal{E}}/\mathcal{M}_{\mathcal{E},0}$, increases with every iteration for both inflow perturbation levels. 

Comparison of figures \ref{FIG:Et_Curves_Cost_Function} and \ref{FIG:Cf_Curves_Cost_Function} highlights the rationale for choosing $C_f$ for the definition of the cost function. 
The increase in the perturbation energy takes place throughout the domain, initially through the amplification of primary instabilities, subsequent secondary instabilities and ultimately breakdown to turbulence if it takes place within the domain.  
For case E1, the initial guess yields an increasing perturbation energy even though the flow is laminar throughout the computational domain; the second iterate too yields a laminar solution, even though its energy norm is higher.   
Therefore, using $\mathcal{E}$ as the observation vector may not in theory guarantee the fastest breakdown to turbulence, although in practice it is effective \citep{Pringle2010}.
In contrast, the skin-friction curve reproduces the laminar behaviour throughout the pre-transitional region and only increases towards the turbulent correlation when intermittency is finite, 
where intermittency is the fraction of time that the flow is turbulent.  
As a result, a cost function based on $C_f$ only shows appreciable increase in case E1 when the flow undergoes transition to turbulence.  
In addition, increasing the cost function based on $C_f$ is directly tied to generation of turbulence early upstream, unlike a cost function based on the perturbation energy that includes contributions from laminar disturbances.

The average wall temperature is plotted versus downstream Reynolds number in figure \ref{FIG:Twall_Curves_Cost_Function}, along with the associated normalized norm at each iterate and for both levels of the inflow perturbation energy.  
The figure shows that, while a cost function based on skin friction promotes early breakdown to turbulence, it does not guarantee that the norm of the wall temperature is highest. 
Based on actual temperature profiles, while a cost function based on skin friction promotes early breakdown to turbulence, it does not guarantee that the peak temperature in the transition zone is largest.  
In addition, it does not guarantee that the temperature norm is highest at convergence either.  
These result optimizations for minimizing losses and for robust thermal design in hypersonic applications might not necessarily lead to the same outcome;  an appropriate cost function must be adopted for each choice and the results compared.

\begin{figure}
\centerline{%
\includegraphics[trim=0 0 0 0, clip,width=0.5\textwidth] {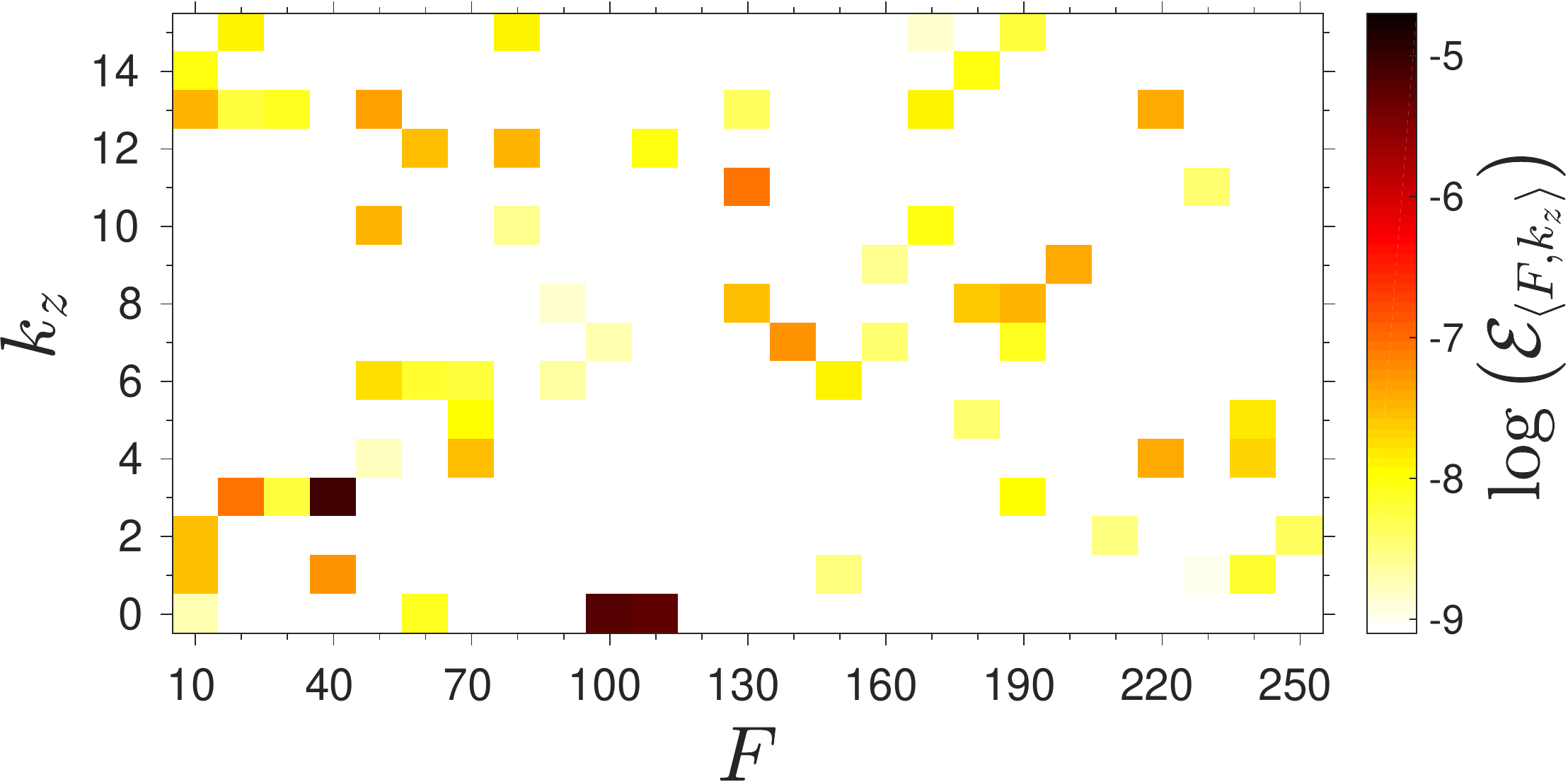}%
\hspace{1.00mm}
\includegraphics[trim=0 0 0 0, clip,width=0.5\textwidth] {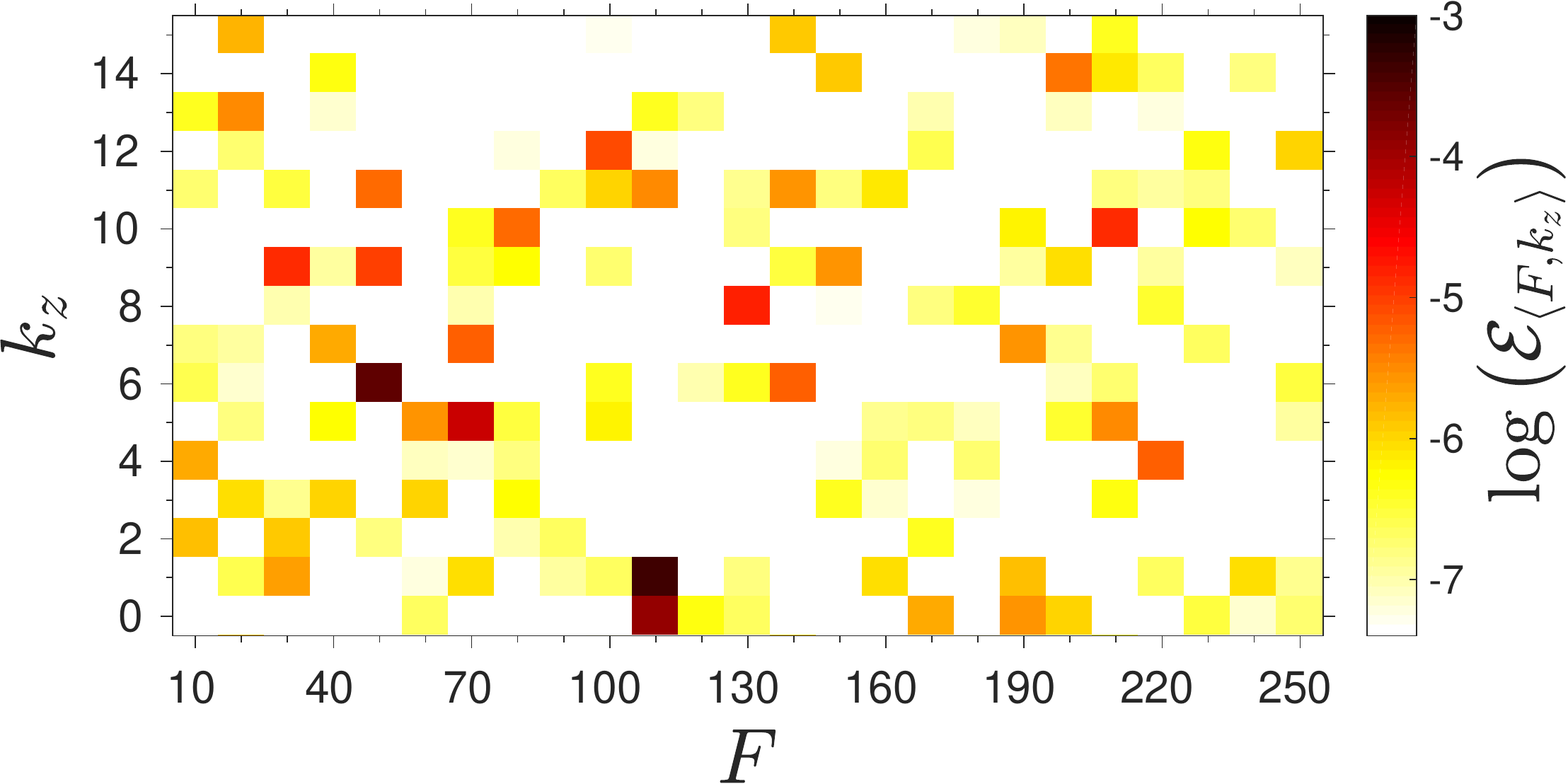}%
\put(-395.0,90.0){$(a)$}
\put(-195.0,90.0){$(b)$}
}%
\caption{The energy distribution amongst the instability modes at the inflow corresponding to the optimal solutions after the final iteration for cases $(a)$ E1 and $(b)$ E2.}
\label{FIG:Modes_Energy_Optimal_Solution}
\end{figure}

The focus is hereafter placed on the results from the optimization algorithm, where the cost function is defined using the wall friction.
Figure \ref{FIG:Modes_Energy_Optimal_Solution} shows the energy spectra of the optimal inflow disturbances, 
\begin{align}
\begin{split}
   \mathcal{E}_{\left< F , k_z \right>} = \frac{1}{2}
	\int_{0}^{L_y} \hat{\boldsymbol{\psi}}_{\left< F , k_z \right>}^*   \textbf{diag} \left( \boldsymbol{\xi} \right) \hat{\boldsymbol{\psi}}_{\left< F , k_z \right>} dy, 
\label{Eq:FourierTransform_Energy_Mode_FKz}
\end{split}
\end{align}
where $\hat{\boldsymbol{\psi}}_{\left< F , k_z \right>}$ are the two-dimensional Fourier coefficients of the perturbation vector $\boldsymbol{\psi}^{\prime} \equiv \begin{bmatrix} \rho^{\prime} & u^{\prime} & v^{\prime} & w^{\prime} & T^{\prime} \end{bmatrix}^{tr} $, the subscripts correspond to the frequency $F \equiv 10^6\, \omega / \sqrt{Re_{x_0}}$ and integer spanwise wavenumber $k_z \equiv \beta / \left(2 \pi / L_z \right)$, and $\hat{\boldsymbol{\psi}}^*$ is the complex-conjugate transpose. 
In (\ref{Eq:FourierTransform_Energy_Mode_FKz}), the term $\boldsymbol{\xi} = \begin{bmatrix} \frac{R}{\gamma-1}\frac{\overline{T}}{ \overline{\rho}} & \overline{\rho} & \overline{\rho} & \overline{\rho} & \frac{R}{\gamma-1} \frac{\overline{\rho}}{\overline{T}} \end{bmatrix}^{tr}$ ensures that the kinetic and internal energy are appropriately weighted; note that overline in $\boldsymbol{\xi}$ denotes the laminar state.

For case E1 (figure \ref{FIG:Modes_Energy_Optimal_Solution}$a$), most of the available energy is allocated to modes $\left< 40,3 \right>$, $\left< 100,0 \right>$ and $\left< 110,0 \right> $.
The energy ratio among these modes is $\left( 1.66:1.18:1 \right)$.
The downstream evolution of the spectra for all 400 instability waves that comprise the optimal perturbation was evaluated from the direct numerical simulations data (not shown), which confirmed that the three identified modes are indeed the most important instability waves at the inlet.  
Collectively they are responsible for the earliest transition to turbulence for case E1.
In order to verify this assertion, a few test cases were performed with variations to the inlet spectra that were motivated by figure \ref{FIG:Modes_Energy_Optimal_Solution}$a$ and also by analysis of the downstream development of instability waves.
All tests confirmed that a disturbance comprised of mode $\left< 40,3 \right>$  with 43\% of total energy, mode $\left< 100,0 \right>$  with 31\% of total energy, and mode $\left< 110,0 \right>$ with 26\% of total energy is the most potent inflow spectra for case E1.  
This reduced form of the disturbance is reported in table \ref{TABLE:NonLinearly_Most_Unstable}, and will be referred to as E1N, and is examined in detail in \S\ref{sec:Transition_R1E1_R1E2}.
The associated skin-friction coefficient is reported in figure \ref{FIG:Cf_Curves_Cost_Function}$a$ (dashed blue line).

\begin{table}
\begin{center}
\def~{\hphantom{0}}
\begin{tabular}{lcccc}
Case & $E_0 \times 10^{5}$ & Inlet's modes & $\mathcal{E}_{\left< F , k_z \right>} \times E_0^{-1}$ & $\theta_{\left< F , k_z \right>} \times \pi^{-1} $ \\
\hline
\\
E1N & $2 $ & \begin{tabular}{c} $\left< 40 , 3 \right>$ \\ $\left< 100 , 0 \right>$ \\ $\left< 110 , 0 \right>$ \end{tabular} & \begin{tabular}{c} $0.43$ \\ $0.31$ \\ $0.26$ \end{tabular} & \begin{tabular}{c} $0.09$ \\ $0.00$ \\ $0.12$ \end{tabular} \\
\\
\rowcolor{blue!10} E2N & $100$ & \begin{tabular}{c} $\left< 50 , 6 \right>$ \\ $\left< 110 , 1 \right>$ \end{tabular} & \begin{tabular}{c} $0.32$ \\ $0.68$ \end{tabular} & \begin{tabular}{c} $0.00$ \\ $0.02$ \end{tabular} \\
\\
\hline
\hline
\\
E1L1 & $2$ & \begin{tabular}{c} $\left< 100 , 0 \right>$ \\ broadband \end{tabular} & \begin{tabular}{c} $0.99$ \\ $0.01$ \end{tabular} & randomly assigned \\
\\
E1L2 & $2$ & \begin{tabular}{c} $\left< 90 , 0 \right>$ \\ broadband \end{tabular} & \begin{tabular}{c} $0.99$ \\ $0.01$ \end{tabular} & randomly assigned \\
\\
E1L3 & $2$ & \begin{tabular}{c} $\left< 20 , 2 \right>$ \\ broadband \end{tabular} & \begin{tabular}{c} $0.99$ \\ $0.01$ \end{tabular} & randomly assigned \\
\\
\rowcolor{blue!10} E2L & $100$ & \begin{tabular}{c} $\left< 110 , 0 \right>$ \\ broadband \end{tabular} & \begin{tabular}{c} $0.99$ \\ $0.01$ \end{tabular} & randomly assigned \\
\hline
\end{tabular}
\caption{E1N and E2N are the nonlinearly most unstable inflow spectra for the cases of the present study. E1L1, E1L2, E1L3 and E2L are the linearly most unstable inflow spectra for the cases examined here. The numbers in the brackets correspond to $\left< F , k_z \right>$.}
\label{TABLE:NonLinearly_Most_Unstable}
\end{center}
\end{table}

The spectra of case E2 at convergence of the optimization algorithm is reported in figure \ref{FIG:Modes_Energy_Optimal_Solution}$b$:  most of the available energy is in modes $\left< 50,6 \right>$ and $\left< 110,1 \right>$.
The ratio of their energy content is $\left( 1:2.17 \right)$.
Similar to E1, the downstream evolution of the spectra for E2 was computed from direct numerical simulation data, for all the 400 wavenumber pairs at the inflow and which can potentially be excited downstream (not shown).  
Careful assessment of the spectra indicated that modes $\left< 50,6 \right>$ and $\left< 110,1 \right>$ are the most important elements of the inflow spectra, and are responsible for the earliest transition to turbulence for case E2.
This conclusion was verified using several additional simulations where the inflow spectra was modified and transition location was compared to the optimal.
In one of the tests, the spanwise size of the domain was doubled and the energy of mode $\left< 110,1 \right>$ was reassigned to mode $\left< 110,\frac{1}{2} \right>$ in order to ensure that the width of the domain did not influence the outcome of the optimization procedure.
All tests confirmed that an inflow disturbance spectra composed of modes $\left< 50,6 \right>$ and $\left< 110,1 \right>$  with 32\% and 68\% of total energy, respectively, is the most potent inflow spectra for case E2.
The skin-friction coefficient associated with this disturbance is reported in figure \ref{FIG:Cf_Curves_Cost_Function}$b$ by the dashed blue line.  
It will be designated E2N in table \ref{TABLE:NonLinearly_Most_Unstable}, and a detailed description of its transition mechanism is discussed in \S\ref{sec:Transition_R1E1_R1E2}.

\begin{figure}
\centerline{%
\includegraphics[trim=2 40 2 0, clip,width=0.99\textwidth] {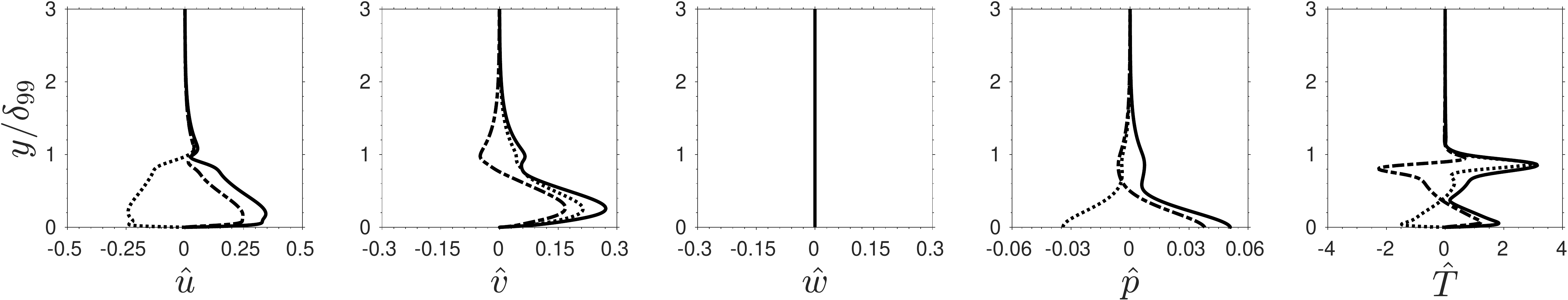}%
\put(-390.0,55.0){$(a)$}
}%
\centerline{%
\includegraphics[trim=2 40 2 0, clip,width=0.99\textwidth] {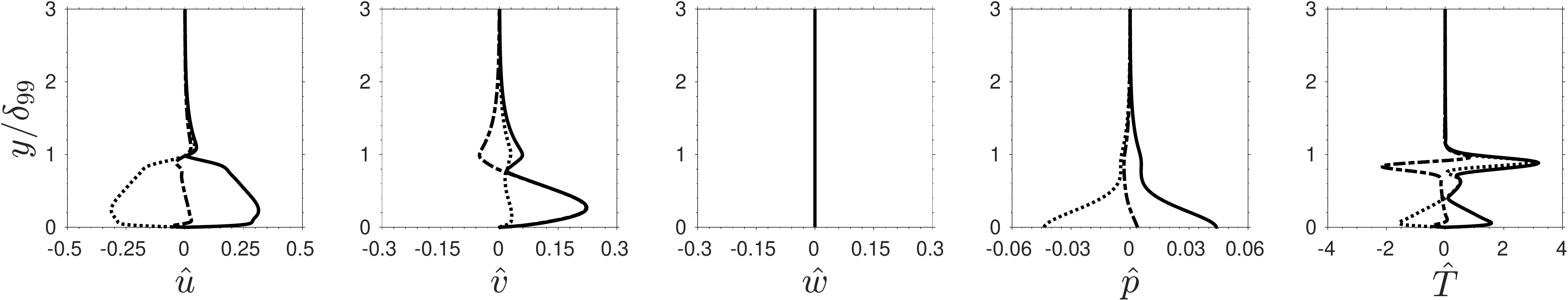}%
\put(-390.0,55.0){$(b)$}
}%
\centerline{%
\includegraphics[trim=2 0 2 0, clip,width=0.99\textwidth] {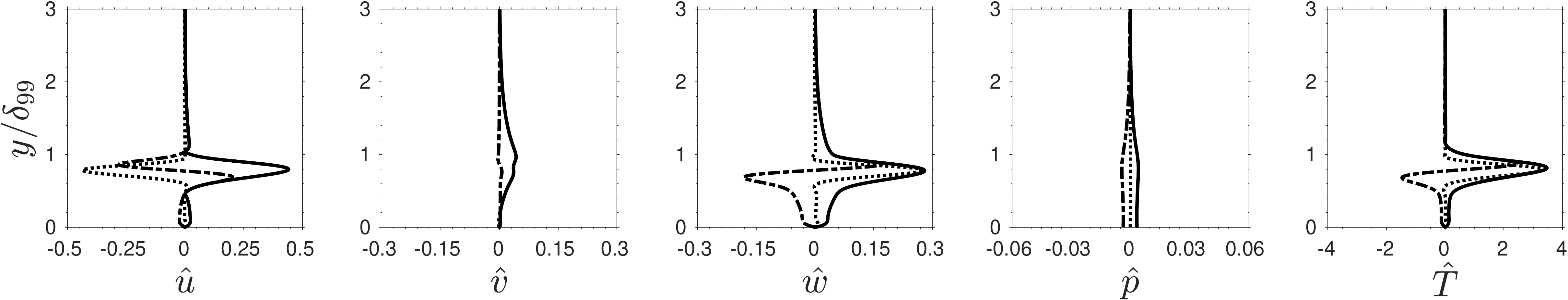}%
\put(-390.0,65.0){$(c)$}
}%
\caption{The mode shapes of the insatiability waves of case E1N at the inlet, $\sqrt{\textrm{Re}_x} = 1800$. $(a)$ mode $ \left< 110 , 0 \right>$, $(b)$ mode $ \left< 100 , 0 \right>$, and $(c)$ mode $ \left< 40 , 3 \right>$. The solid, dashed-dotted, dotted lines are the magnitude, real part, and imaginary part of the mode shapes respectively.}
\label{FIG:ModeShapes_R1E1}
\end{figure}

\begin{figure}
\centerline{%
\includegraphics[trim=2 40 2 0, clip,width=0.99\textwidth] {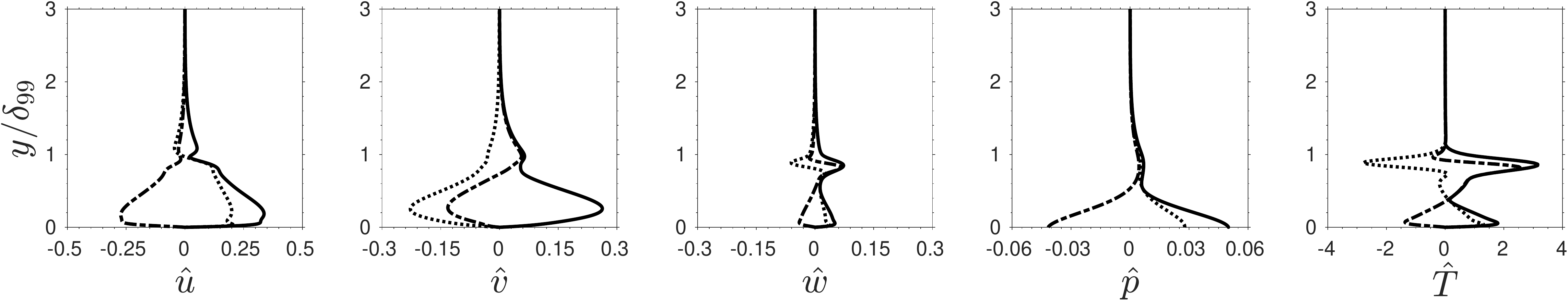}%
\put(-390.0,55.0){$(a)$}
}%
\centerline{%
\includegraphics[trim=2 0 2 0, clip,width=0.99\textwidth] {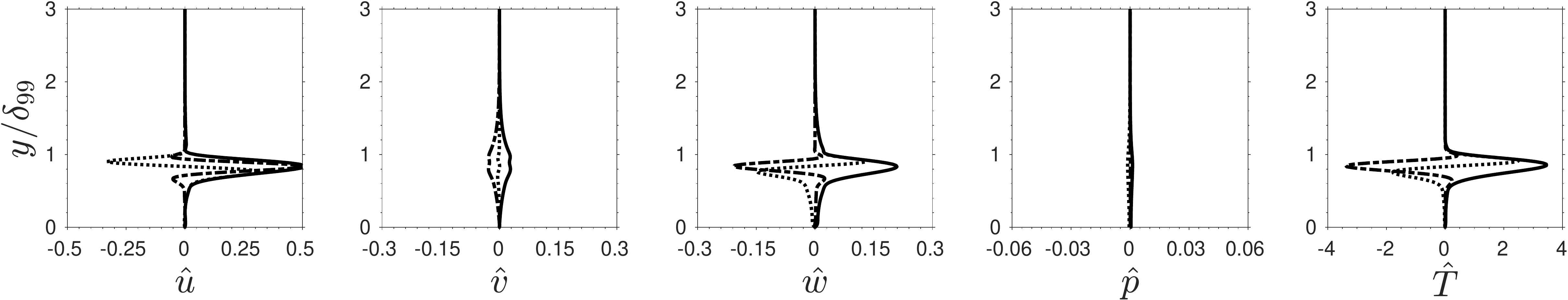}%
\put(-390.0,65.0){$(b)$}
}%
\caption{The mode shapes of the insatiability waves of case E2N at the inlet, $\sqrt{\textrm{Re}_x} = 1800$. $(a)$ mode $ \left< 110 , 1 \right>$, and $(b)$ mode $ \left< 50 , 6 \right>$. The solid, dashed-dotted, dotted lines are the magnitude, real part, and imaginary part of the mode shapes respectively.}
\label{FIG:ModeShapes_R1E2}
\end{figure}

The mode shapes that play an important role in cases E1N and E2N are plotted in figures \ref{FIG:ModeShapes_R1E1} and \ref{FIG:ModeShapes_R1E2}, respectively.
As the $\hat{p}$ profiles in figures \ref{FIG:ModeShapes_R1E1} and \ref{FIG:ModeShapes_R1E2} indicate, modes $\left< 110 , 0 \right>$, $\left< 100 , 0 \right>$ and $\left< 110 , 1 \right>$ with one zero crossing of $\textrm{real} (\hat{p})$ in the wall normal direction are the so-called second-mode instabilities \citep{Mack1984}.
These are generally sound (acoustic) waves that reflect back and forth between the wall and the sonic line of the relative flow.
Modes $\left< 40 , 3 \right>$ and $\left< 50 , 6 \right>$ are first-mode instabilities which for $\textrm{Ma}_{\infty} < 5$ are vorticity waves that can cause strong inflectional instability.

Depending on the wavelength of the modes and the local boundary-layer thickness, the growth rate of the acoustic waves can be much larger than the typical vorticity modes.
It is also evident in figures \ref{FIG:ModeShapes_R1E1} and \ref{FIG:ModeShapes_R1E2} that,
while the second-mode instabilities are large across the boundary layer, the first-mode instabilities have appreciable magnitude only close to its edge.
These observations have important implications on the mechanism of transition to turbulence in cases E1N and E2N.
For instance, a commonly observed feature in hypersonic and supersonic boundary-layer transition is the generation of elongated streaks from the nonlinear interaction of oblique instability waves \citep{Sivasubramanian2016,Franko2013,Mayer2011,Jiang2006,Thumm1990}.
Since streaks are the boundary-layer response to vortical forcing by three-dimensional modes, they are anticipated in case E2N in response to mode $\left< 110 , 1 \right>$. 
Also note that mode $\left< 110 , 1 \right>$ has strong wall-normal perturbation which is essential for the lift-up process that leads to the streak response \citep{ZakiDurbin2006,ZakiDurbin2005}.

\subsection{Transition mechanisms}
\label{sec:Transition_R1E1_R1E2}

In this section we discuss the two transition mechanisms due to the nonlinearly most unstable inflow spectra for cases E1N and E2N (table \ref{TABLE:NonLinearly_Most_Unstable}), respectively.
In the former case, the inflow is comprised of a pair of two-dimensional second-mode disturbances and an oblique first-mode instability wave.
We will demonstrate that the resulting breakdown to turbulence can not be categorized as fundamental and/or oblique, and is hence a new mechanism. 
Focus will therefore be placed on this case.  
At the higher energy level, transition follows a typical second-mode oblique breakdown which will be summarized.

\begin{figure}
\centerline{%
\includegraphics[trim=0 0 0 0, clip,width=0.999\textwidth] {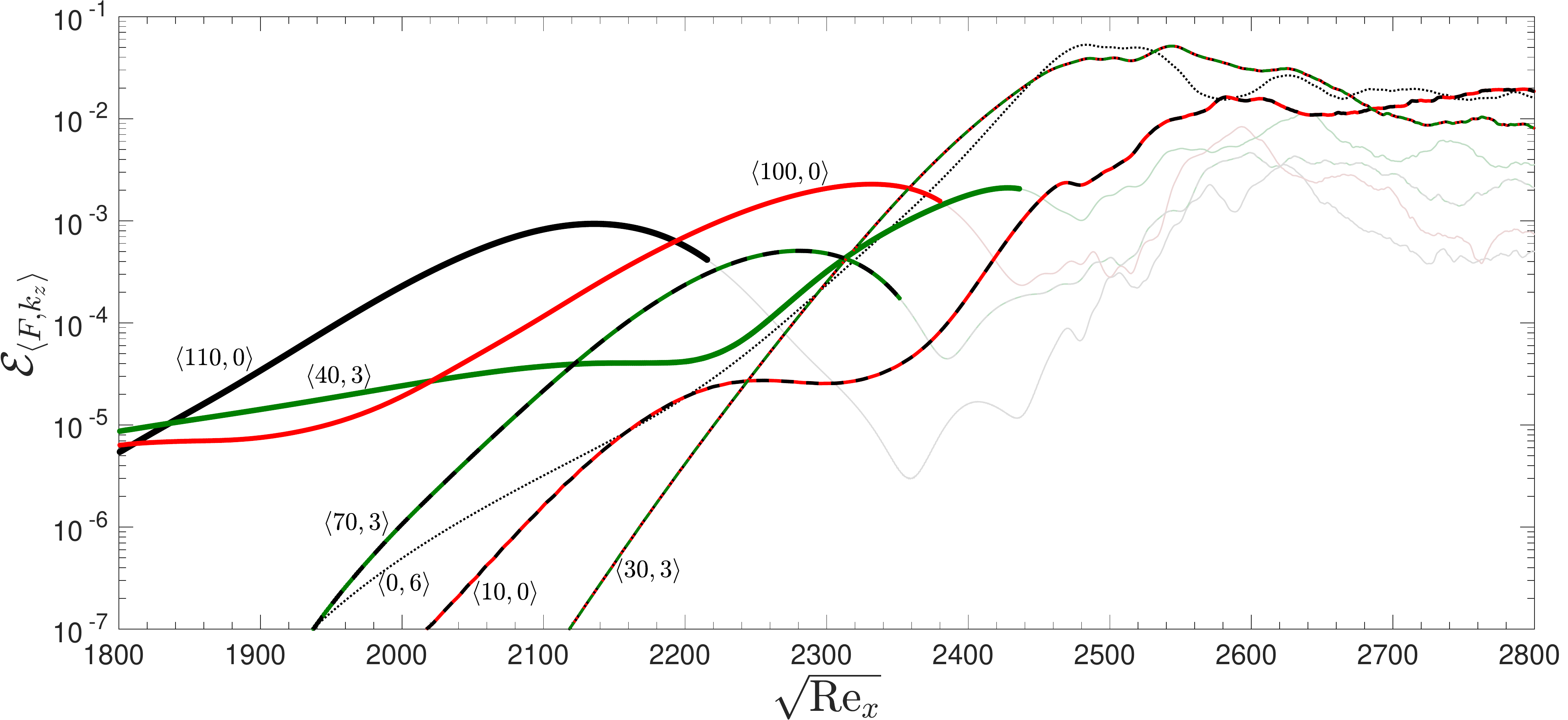}%
\put(-399.0,170.0){$(a)$}
}%
\centerline{%
\includegraphics[trim=45 200 20 210, clip,width=0.999\textwidth] {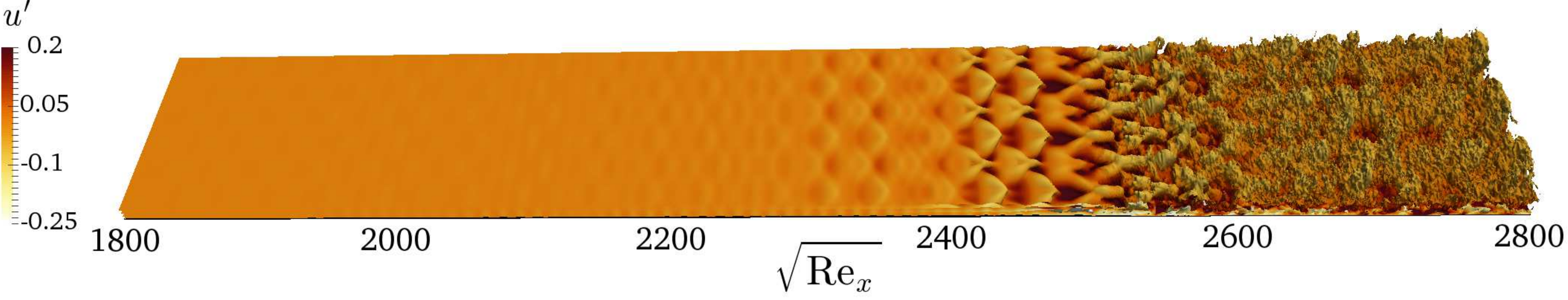}%
\put(-400.0,60.0){$(b)$}
}%
\centerline{%
\includegraphics[trim=45 200 20 210, clip,width=0.999\textwidth] {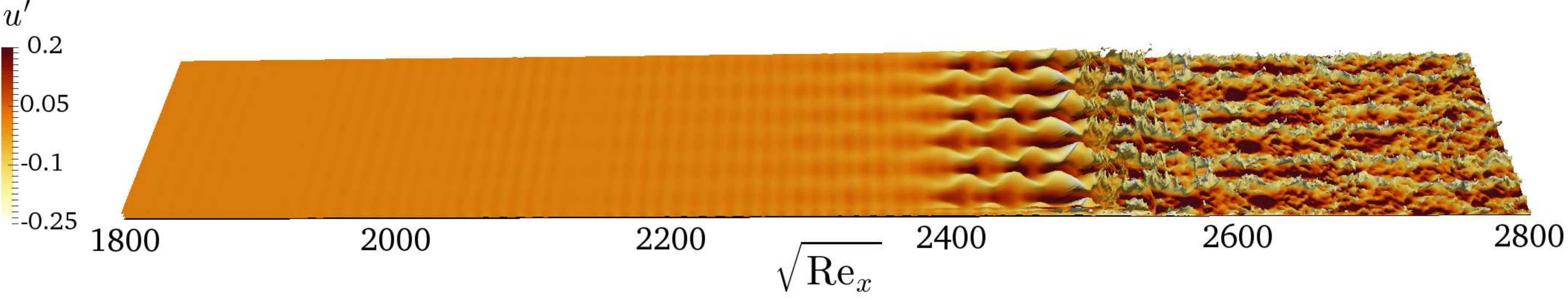}%
\put(-400.0,60.0){$(c)$}
}%
\centerline{%
\includegraphics[trim=45 200 20 210, clip,width=0.999\textwidth] {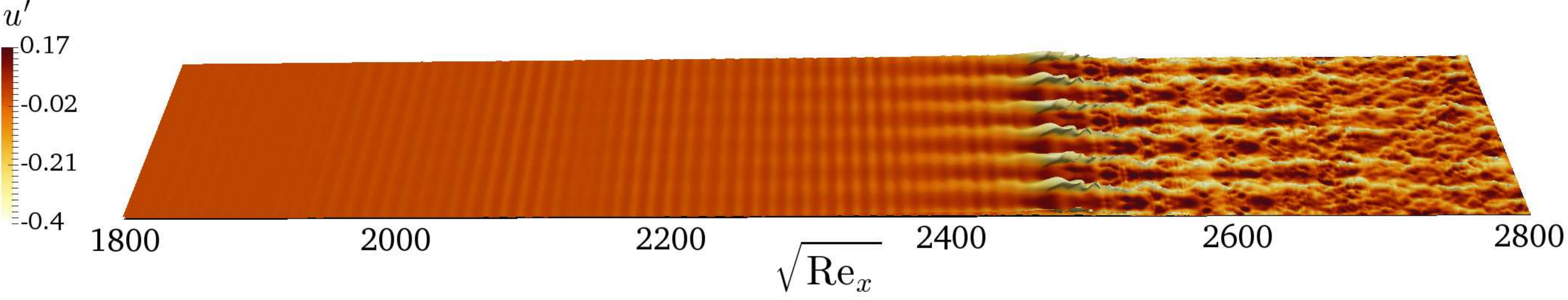}%
\put(-400.0,60.0){$(d)$}
}%
\centerline{%
\includegraphics[trim=0 0 0 0, clip,width=0.7\textwidth] {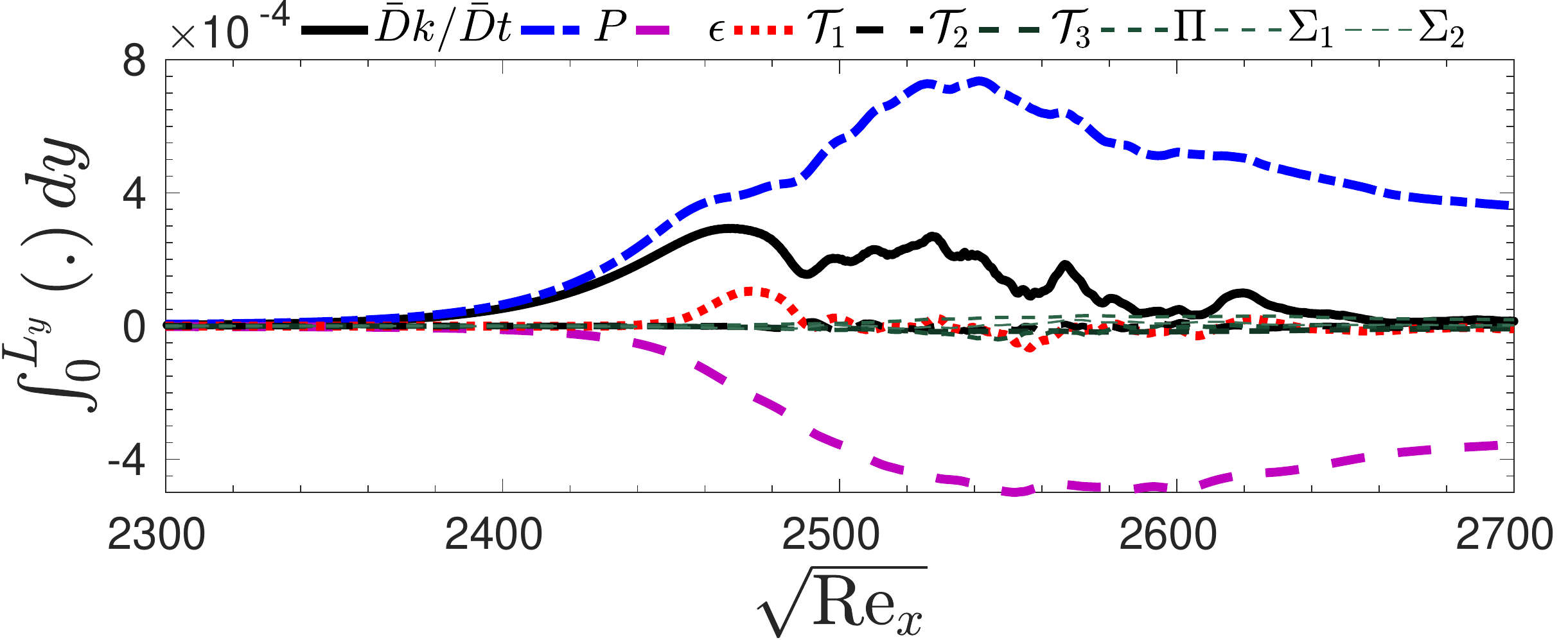}%
\put(-275.0,100.0){$(e)$}
}%
\caption{Results corresponding to case E1N: $(a)$ fluctuations' energy content for selected instability modes, $\mathcal{E}_{ \left< F , k_z \right>}$, versus the streamwise coordinate. $F$ is the normalized frequency and $k_z$ is the integer spanwise wave number, defined in Eqs. \ref{Eq:Normalized_Freq} and \ref{Eq:spanwise_wavenum}. $(b,c,d)$ Instantaneous iso-surfaces of streamwise velocity component, $(b)$ $u = 0.9$, $(c)$ $u = 0.6$ and $(d)$ $u = 0.3$, colorued by the streamwise velocity fluctuations $u^{\prime}$. $(e)$ The turbulent-kinetic-energy transport terms integrated in wall-normal direction along the transition zone.}
\label{FIG:Disturbances_Spectral_Evolution_R1E1}
\end{figure}

\begin{figure}
\centerline{%
\includegraphics[trim=0 0 0 0, clip,width=0.999\textwidth] {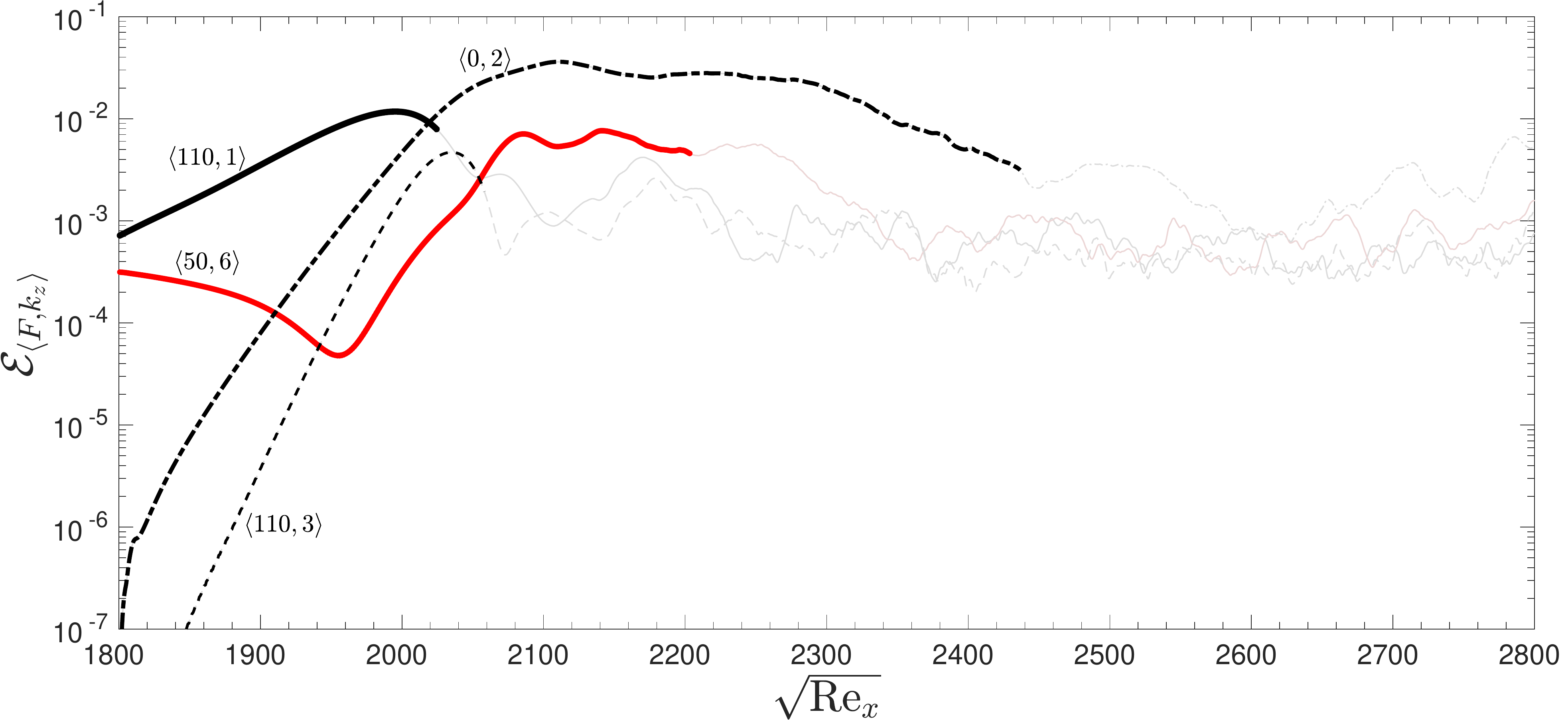}%
\put(-399.0,170.0){$(a)$}
}%
\centerline{%
\includegraphics[trim=45 200 20 210, clip,width=0.999\textwidth] {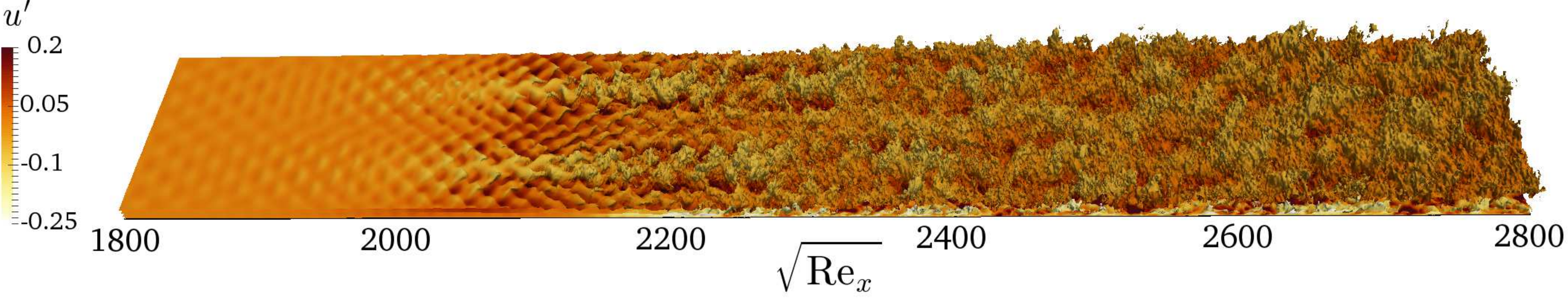}%
\put(-400.0,60.0){$(b)$}
}%
\centerline{%
\includegraphics[trim=45 200 20 210, clip,width=0.999\textwidth] {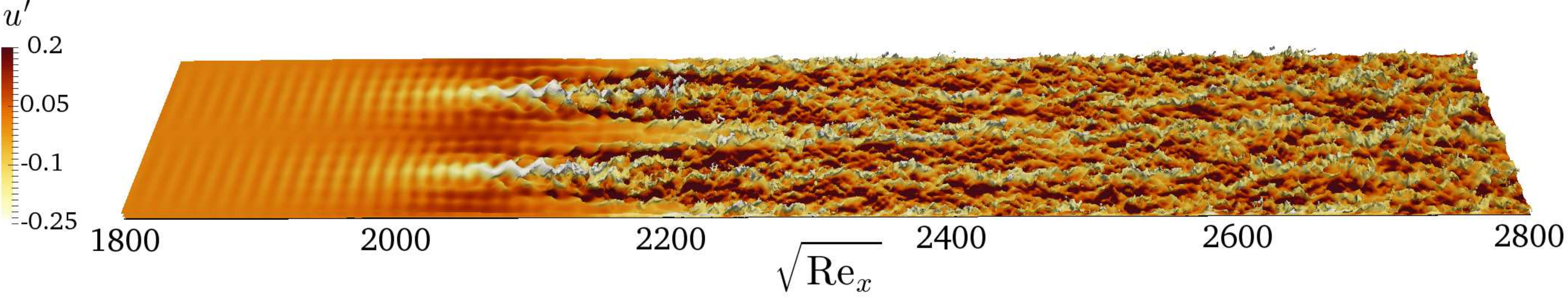}%
\put(-400.0,60.0){$(c)$}
}%
\centerline{%
\includegraphics[trim=45 200 20 210, clip,width=0.999\textwidth] {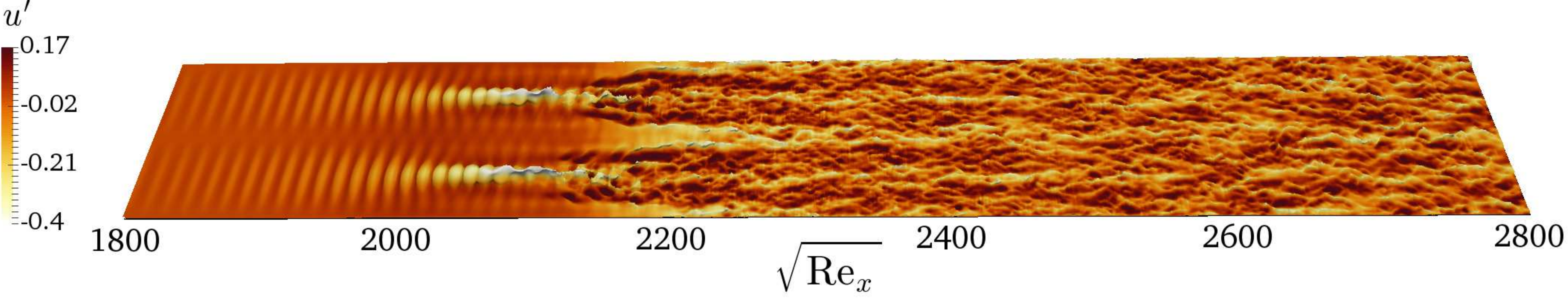}%
\put(-400.0,60.0){$(d)$}
}%
\centerline{%
\includegraphics[trim=0 0 0 0, clip,width=0.7\textwidth] {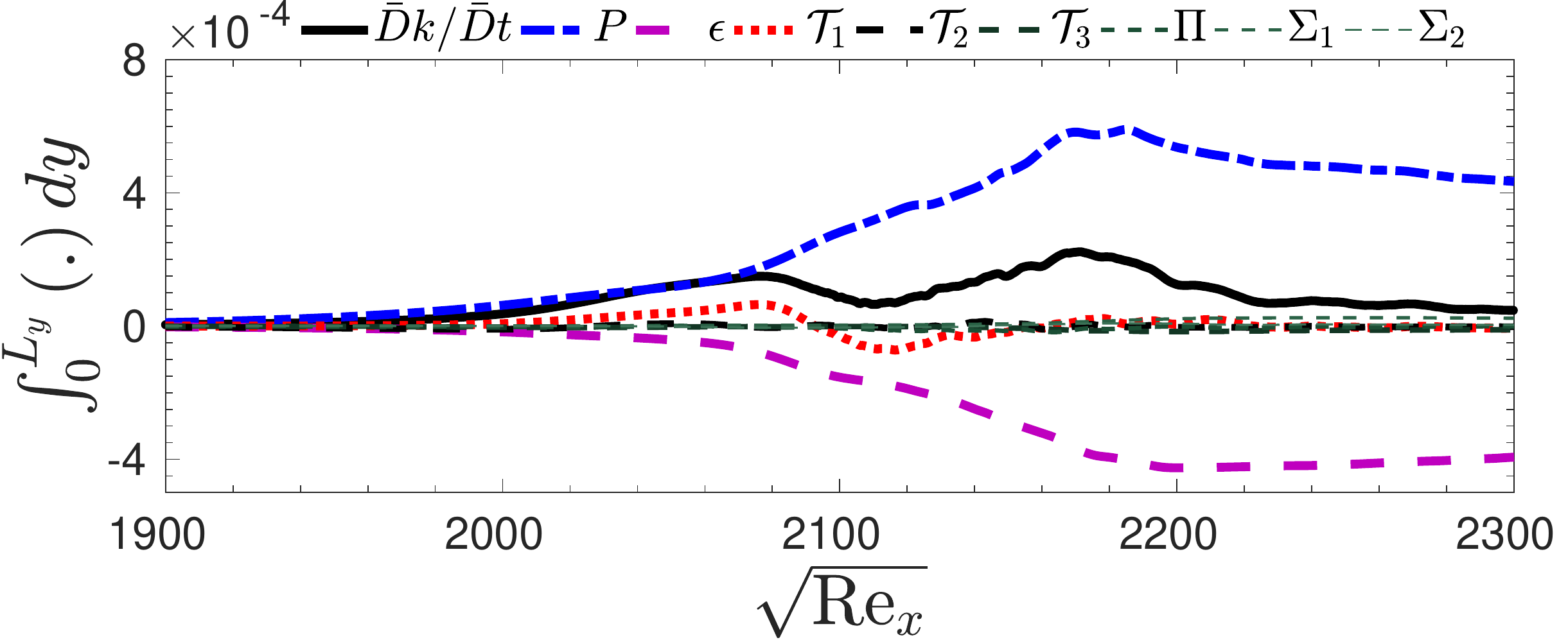}%
\put(-275.0,100.0){$(e)$}
}%
\caption{Results corresponding to case E2N: $(a)$ fluctuations' energy content for selected instability modes, $\mathcal{E}_{ \left< F , k_z \right>}$, versus the streamwise coordinate. $F$ is the normalized frequency and $k_z$ is the integer spanwise wave number, defined in Eqs. \ref{Eq:Normalized_Freq} and \ref{Eq:spanwise_wavenum}. $(b,c,d)$ Instantaneous iso-surfaces of streamwise velocity component, $(b)$ $u = 0.9$, $(c)$ $u = 0.6$ and $(d)$ $u = 0.3$, coloured by the streamwise velocity fluctuations $u^{\prime}$. $(e)$ The turbulent-kinetic-energy transport terms integrated in wall-normal direction along the transition zone.}
\label{FIG:Disturbances_Spectral_Evolution_R1E2}
\end{figure}

The downstream evolution of $\mathcal{E}_{\left< F , k_z \right>}$ of modes that play a principal role in transition was evaluated using (\ref{Eq:FourierTransform_Energy_Mode_FKz}) and is plotted in figure \ref{FIG:Disturbances_Spectral_Evolution_R1E1} and \ref{FIG:Disturbances_Spectral_Evolution_R1E2}.
For clarity, the curves become thin and faint coloured beyond the location where their interactions are discussed.  
Figures \ref{FIG:Disturbances_Spectral_Evolution_R1E1} and \ref{FIG:Disturbances_Spectral_Evolution_R1E2} also include instantaneous iso-surfaces of streamwise velocity $u$ coloured by its fluctuation $u^{\prime}$.
The bottom panels in both figures feature the downstream evolution of integrated terms in the turbulence kinetic energy (TKE) equation, 
\begin{eqnarray}
\nonumber
\underbrace{ \frac{1}{2} \frac{\partial \overline{\rho u_i^{\prime\prime} u_i^{\prime\prime}} }{\partial t} + \frac{1}{2} \frac{\partial \overline{ u_j }^f \overline{\rho u_i^{\prime\prime} u_i^{\prime\prime}} }{\partial x_j} }_{ \bar{D} k / \bar{D} t } &=& \underbrace{ - \overline{ \rho u_i^{\prime \prime} u_j^{\prime \prime} } \frac{\partial \overline{ u_i }^f}{\partial x_j} }_{P} \underbrace{ - \overline{ \tau_{ij}^{\prime} \frac{\partial u_i^{\prime\prime} }{\partial x_j} } }_{\epsilon} \underbrace{- \frac{1}{2} \frac{\partial \overline{ \rho u_i^{\prime\prime} u_i^{\prime\prime} u_j^{\prime\prime} } }{\partial x_j} }_{\mathcal{T}_1}  \\
 && \underbrace{- \frac{\partial \overline{ p^{\prime} u_i^{\prime} \delta_{ij} } }{\partial x_j}}_{\mathcal{T}_2} + \underbrace{ \frac{\partial \overline{ \tau_{ij}^{\prime} u_i^{\prime\prime} } }{\partial x_j} }_{\mathcal{T}_3} + \underbrace{ \overline{ p^{\prime} \frac{\partial u_i^{\prime\prime}}{\partial x_i} } }_{\Pi} + \underbrace{\overline{ u_{i}^{\prime\prime} } \frac{\partial \overline{ \tau_{ij} } }{\partial x_j}}_{\Sigma_1} \underbrace{- \overline{ u_{i}^{\prime\prime} } \frac{\partial \overline{ p } }{\partial x_i}}_{\Sigma_2}.  
\label{Eq:TKE_Transport}
\end{eqnarray}
For an arbitrary quantity $\mathcal{X}$, the Reynolds average is denoted by $\overline{\mathcal{X}}$, the Favre (density-weighted) average is identified by $\overline{\mathcal{X}}^f$, and fluctuations with respect to the Reynolds and Favre averages are marked as $\mathcal{X}^{\prime}$ and $\mathcal{X}^{\prime\prime}$, respectively.
The terms on the right-hand side of (\ref{Eq:TKE_Transport}) correspond to production rate $P$, dissipation rate $\epsilon$, transport by velocity fluctuations $\mathcal{T}_1$, transport by pressure fluctuation $\mathcal{T}_2$, transport by viscous diffusion $\mathcal{T}_3$, pressure dilatation (pressure-strain) $\Pi$, viscous mass flux coupling $\Sigma_1$, and pressure mass flux coupling $\Sigma_2$.

In case E1N, transition involves a series of instability waves that are activated at various downstream locations (figure \ref{FIG:Disturbances_Spectral_Evolution_R1E1}$a$): the inflow modes amplify first and, downstream, spur other instabilities which were not prominently featured in the inlet condition.  
Specifically, the inlet pair $\left< 100 , 0 \right> $ and$ \left< 110 , 0 \right>$ activate mode $\left< 10 , 0 \right>$\textemdash an interaction that we will denote $\mathcal{I}^{(1)}$.
The resulting mode $\left< 10 , 0 \right>$ is manifest as a low-frequency modulation of the near-wall waves in panel \ref{FIG:Disturbances_Spectral_Evolution_R1E1}$d$.  
In a subsequent interaction $\mathcal{I}^{(2)}$, the pair $\left< 40 , 3 \right>$ and $\left< 110 , 0 \right>$ give rise to mode $\left<70 , 3 \right>$, which is visible in panel \ref{FIG:Disturbances_Spectral_Evolution_R1E1}$b$ near $\sqrt{\textrm{Re}_x} = 2300$.
The next interactions $\mathcal{I}^{(3)}$ involves the newly formed three-dimensional mode and the inflow wave $\left< 100 , 0 \right>$.  
It spurs the instability $\left< 30 , 3 \right> $ which grows at the highest rate, is evident in panel \ref{FIG:Disturbances_Spectral_Evolution_R1E1}$b$ in the range $2400 < \sqrt{\textrm{Re}_x} < 2550$, and is the seat of breakdown to turbulence (also see movie \# 1 in the supplementary material).  
Note that a triad $\mathcal{I}^{(4)}$ is established between the first two nonlinearly generated waves, $\left< 10 , 0 \right> $ and $\left< 40 , 3 \right>$, and the fastest amplifying instability $\left< 30 , 3 \right>$.
Flow structures resembling an interweaving rope-shaped waves evolve far from the wall prior to breakdown to turbulence in movie \# 1, and bear resemblance to those observed experimentally in second-mode transition at Mach 5-8 \citep{Casper2016,Laurence2016}.

Streaks feature in the instantaneous field (figures \ref{FIG:Disturbances_Spectral_Evolution_R1E1}$c$-$d$). 
In the spectra, they correspond to mode $\left< 0 , 6 \right>$, which can be generated by the nonlinear interaction of modes $\left<\pm F,3\right>$; their spanwise size is approximately equal to the thickness of boundary layer at $\sqrt{\textrm{Re}_{x}} \approx 2550$.  
The energy of the streaks, or mode $\left< 0, 6 \right>$, shadows that of mode $\left<30,3\right>$ which was discussed in connection with the outer $\Lambda$-shaped structure and breakdown to turbulence.  
The breakdown of the streaks follows almost immediately after that of the outer $\Lambda$'s.

The interactions $\mathcal{I}$ described above can only be hypothesized based on the spectra.  
For example, whether the streaks $\left< 0 , 6 \right> $ are due to the self interactions of $\left< \pm30 , 3 \right>$, $\left< \pm40 , 3 \right>$ or $\left<\pm 70 , 3 \right>$ can not be conclusively determined; 
the similarity between  $\left< \pm30 , 3 \right>$ and  $\left< \pm 0 , 6 \right>$ suggests a connection but does not guarantee it.  
In order to quantify the nonlinear interactions, we compute the energy transfer terms among wavenumber triads.  
Starting from $\mathcal{T}_1$ in (\ref{Eq:TKE_Transport}), the following expression for $\mathcal{I}$ can be derived (see Appendix \ref{Appendix_B} for details), 
\begin{equation}
	\mathcal{I}_{\left< F , k_{z} \right>} = \int_{0}^{L_y} \left\vert \boldsymbol{\hat{\mathcal{A}}}^*_{\left< F_1 , k_{z,1} \right>} \boldsymbol{\hat{\mathcal{F}}}_{\left< F_2 , k_{z,2} \right>} \right\vert dy
\label{Eq:NonLinearEnergyTransfer}
\end{equation}
where $F = F_1 \pm F_2$ and $k_z = k_{z,1} \pm k_{z,2}$.  
In equation (\ref{Eq:NonLinearEnergyTransfer}), $\boldsymbol{\hat{\mathcal{A}}}$ and $\boldsymbol{\hat{\mathcal{F}}}$ are vector quantities containing the Fourier coefficients of 
$\boldsymbol{\mathcal{A}} = 
	\begin{bmatrix}
		\rho u^{\prime\prime} u^{\prime\prime} & \rho u^{\prime\prime} v^{\prime\prime} & \rho u^{\prime\prime} w^{\prime\prime}
	\end{bmatrix}^{tr} $
and of 
$\boldsymbol{\mathcal{F}} = 
	\begin{bmatrix}
		u^{\prime\prime} & v^{\prime\prime} & w^{\prime\prime}
	\end{bmatrix}^{tr}.$
The quantity $\mathcal{I}_{\left< F , k_z \right>}$ thus represents the nonlinear energy transfer among modes $\left< F , k_{z} \right>$, $\left< F_1 , k_{z,1} \right>$ and $\left< F_2 , k_{z,2} \right>$ \citep{Cheung2010}.
Since $\mathcal{I}_{\left< F , k_z \right>}$ is independent of the ordering of modes within the triad, it only measures the energy transfer among the three modes but not the direction of the transfer.

\begin{figure}
\centerline{%
\includegraphics[trim=0 0 0 0, clip,width=0.999\textwidth] {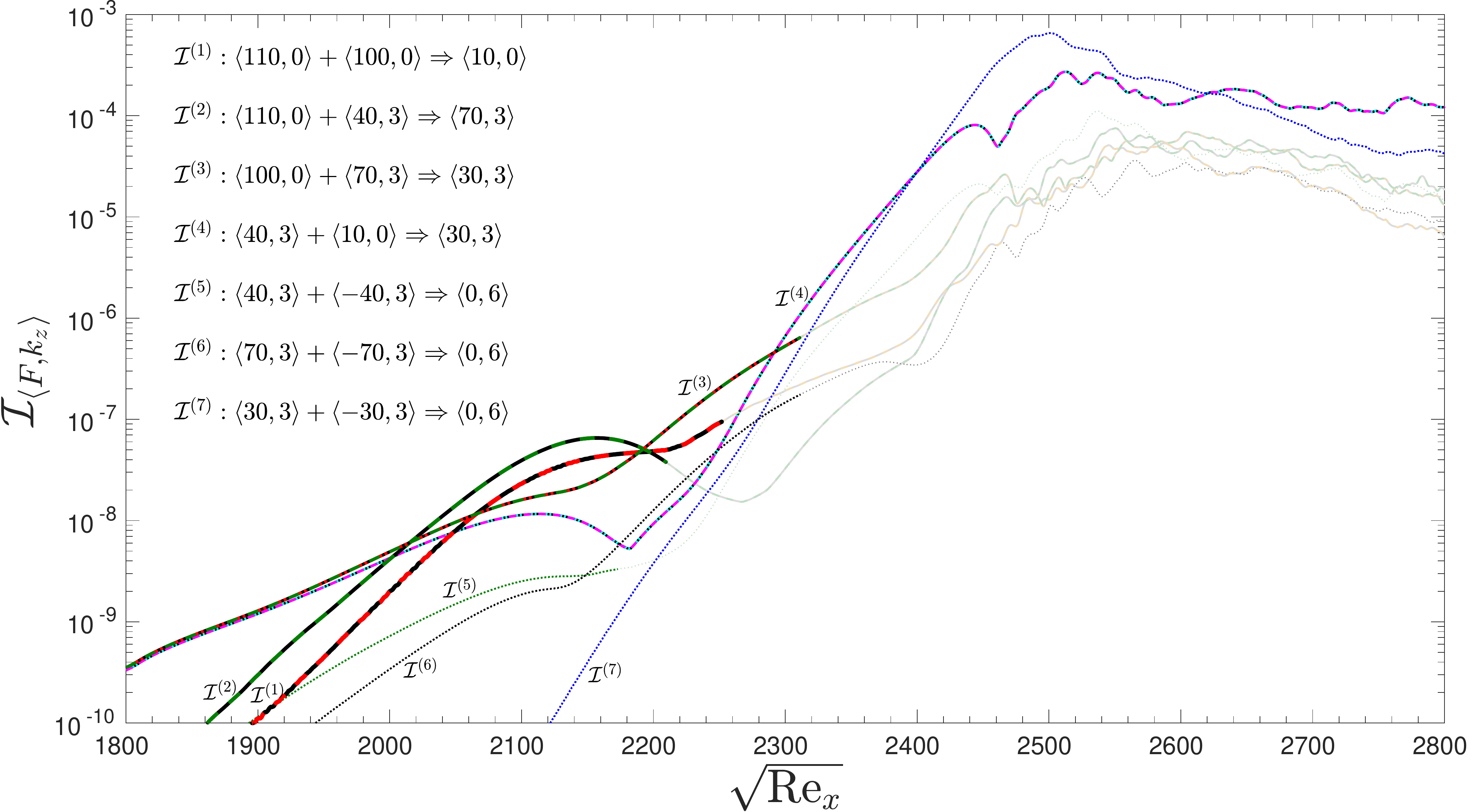}%
}%
\caption{The modal nonlinear energy transfer coefficient, defined in (\ref{Eq:NonLinearEnergyTransfer}), computed for the most important modal interactions corresponding to transition scenario of case E1N.}
\label{FIG:NonLinearEnergyTransfer_NMU_R1E1}
\end{figure}

In case E1N, nonlinear interactions feature prominently and, therefore, $\mathcal{I}_{\left< F , k_z \right>}$ for the seven most important interactions is plotted in figure \ref{FIG:NonLinearEnergyTransfer_NMU_R1E1}.
Note that the designation $\left< F_1 , k_{z,1} \right> + \left< F_{2} , k_{z,2} \right> \Rightarrow \left< F , k_z \right>$ in the figure is only intended to identify the triad, and bears no physical significance.  
However, together with the energy spectra, $\mathcal{I}_{\left< F , k_z \right>}$ can provide a clearer view of the nonlinear interactions that spur new instability waves.
The interaction $\mathcal{I}^{(1)}$ takes place between inflow modes $\left< 100 , 0 \right> $ and $ \left< 110 , 0 \right>$ and generates $\left< 10 , 0 \right>$.  
Similarly, $\mathcal{I}^{(2)}$ takes place between inlet modes $\left< 40 , 3 \right>$ and $\left< 110 , 0 \right>$ and gives rise to $\left<70 , 3 \right>$. 
This emergent mode has a strong interaction $\mathcal{I}^{(3)}$ with $\left< 100 , 0 \right>$ and leads to $\left< 30 , 3 \right> $ which amplifies at the highest rate.  
Note also that near $\sqrt{\textrm{Re}_x} = 2300$, this interaction is overtaken by another triad, $\mathcal{I}^{(4)}$, that involves $\left< 30 , 3 \right> $. 
The source of the streaks $\left< 0 , 6 \right>$ is not a single interaction, but rather three $\mathcal{I}^{(5,6,7)}$, and each of them is dominant in a region of the streamwise length.  
While $\left< 0 , 6 \right>$ shadows $\left<30,3\right>$ in the spectra, $\mathcal{I}^{(7)}$ is only the final interaction that becomes dominant as mode  $\left<30,3\right>$ becomes most energetic.

The role of every interaction, including those that precede the amplification of $\left< 30,3\right>$, can not be discounted.  
We performed exhaustive tests that involved assigning a significant portion of the inflow energy to mode $\left< 30,3\right>$ but transition was delayed.  
Thus, all the events, including nonlinear interactions and base-flow distortion, that precede the formation of this mode must take place for transition to set in as upstream as recorded in our simulations.

Terms in the TKE equation are shown versus downstream Reynolds number in figure \ref{FIG:Disturbances_Spectral_Evolution_R1E1}$e$. 
Three regions can be distinguished, based on the behaviour of the rate of production $P$, which represents the energy exchange between the base state and the instability waves.
(i) In the range $ 2400 < \sqrt{\textrm{Re}_{x}} < 2500$, the production increases faster than dissipation;
In this region, the instability wave $\left< 30,3\right>$ dominates the spectra and reaches saturation.  
(ii) In the region $ 2500 < \sqrt{\textrm{Re}_{x}} < 2550$, the rate of production increases as the secondary instability sets in and leads to the formation of the $\Lambda$-structures and their local breakdown.  
(iii) Finally,  in the range $ 2550 < \sqrt{\textrm{Re}_{x}} < 2650$, the flow starts to approach the statistical state of a fully turbulent boundary layer.

The transition scenario for case E2N is of different ilk. 
Figure \ref{FIG:Disturbances_Spectral_Evolution_R1E2}$a$ captures the fast emergence and amplification of a streamwise elongated streak mode, $\left< 0, 2 \right>$.  
This wave becomes the seat of secondary instability and onset of turbulence.  
Its instantaneous form is evident in the visualization of the streamwise velocity iso-surfaces (panels $b$-$d$).
The spanwise size of the structure is approximately 3 times the thickness of boundary layer at $\sqrt{\textrm{Re}_{x}} \approx 2250$, which is the location where the flow has reached a fully turbulent state.
Similar sizes of streaks were reported in transition at $Ma_{\infty} = 6$ \citep{Franko2013} and also at $Ma_{\infty} = 3$ \citep{Mayer2011}.
The streak $\left< 0 , 2 \right>$ is generated by the mode $\left< 110 , 1 \right>$, which initially amplifies linearly and reaches saturation around $\sqrt{\textrm{Re}_{x}} \approx 2000$\textemdash the location where the $C_f$ curve starts to rise (c.f.\,figure \ref{FIG:Cf_Curves_Cost_Function}$b$).
At this stage, the mode $\left< 50 , 6 \right>$ starts to amplify, and reaches its highest energy level at $\sqrt{\textrm{Re}_{x}} \approx 2100$.
In the instantaneous realization, it appears to effect the secondary instability of the streak mode $\left< 0 , 2 \right>$, which meanders and breaks down to turbulence (figure \ref{FIG:Disturbances_Spectral_Evolution_R1E2}$b$-$d$ and movie \# 2 in the supplementary material).
The early stages of transition in movie \# 2 involve the alternation of light and dark regions near the wall without any visible structures at the edge of the boundary layer.
This process is very different from the appearance of rope-like structures in movie \# 1 for case E1N.
A change in the transition mechanism with inflow disturbance energy is consistent with previous experiments, e.g.\,at high and low enthalpy conditions at Mach 6-7 \citep{Laurence2016}.

We can further examine the transition zone of case E2N by considering the evolution of terms in the TKE equation in various subregions (figure \ref{FIG:Disturbances_Spectral_Evolution_R1E2}$e$). 
In a similar manner to the discussion of case E1N, three regions can be identified: 
(i) In the range $ 2000 < \sqrt{\textrm{Re}_{x}} < 2100$, the increase in the rate of production, $P$, is moderate but $P$ itself exceeds the integrated $\epsilon$.  
The energy of the streaks overtake all other modes at the start of this region, and their secondary instability is amplifying.  
(ii) In the region $ 2100 < \sqrt{\textrm{Re}_{x}} < 2200$, the secondary instability $\left< 50 , 6 \right>$ is nearly saturated, and localized patches of turbulence emerge; here the increase in the rate of production and dissipation become steeper, the latter enhanced by the emergence of small turbulent scales.
Within the localized patches of turbulence, production exceeds the levels of fully turbulent boundary layers \citep{Marxen2018accepted}.
(iii) In the region $ 2200 < \sqrt{\textrm{Re}_{x}} < 2300$, the turbulence spread and fills the domain, and terms in the TKE equation start to relax towards their equilibrium values for fully turbulent boundary layers.

\begin{figure}
\centerline{%
\includegraphics[trim=0 0 0 0, clip,width=0.5\textwidth] {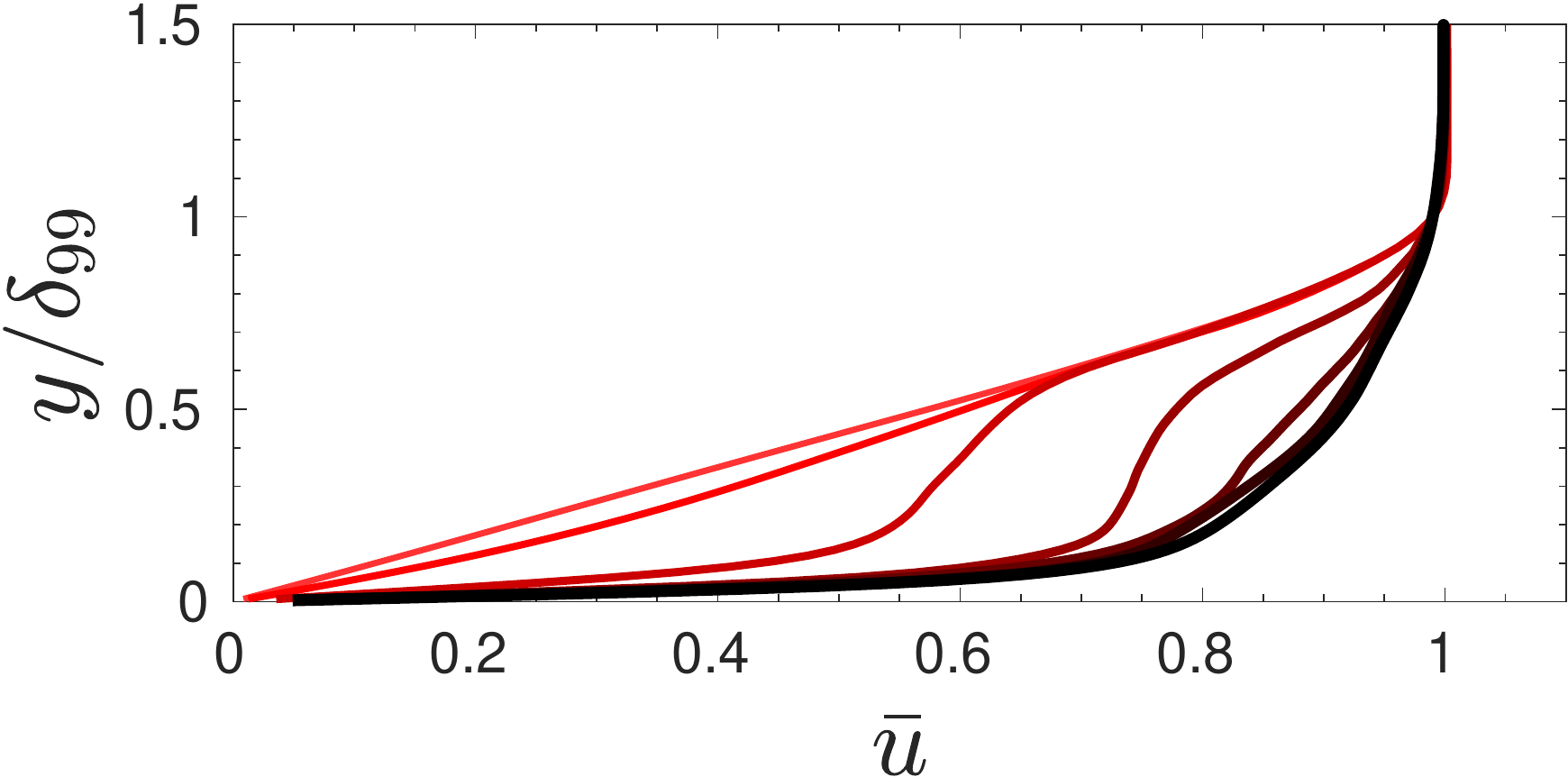}%
\hspace{1.00mm}
\includegraphics[trim=0 0 0 0, clip,width=0.5\textwidth] {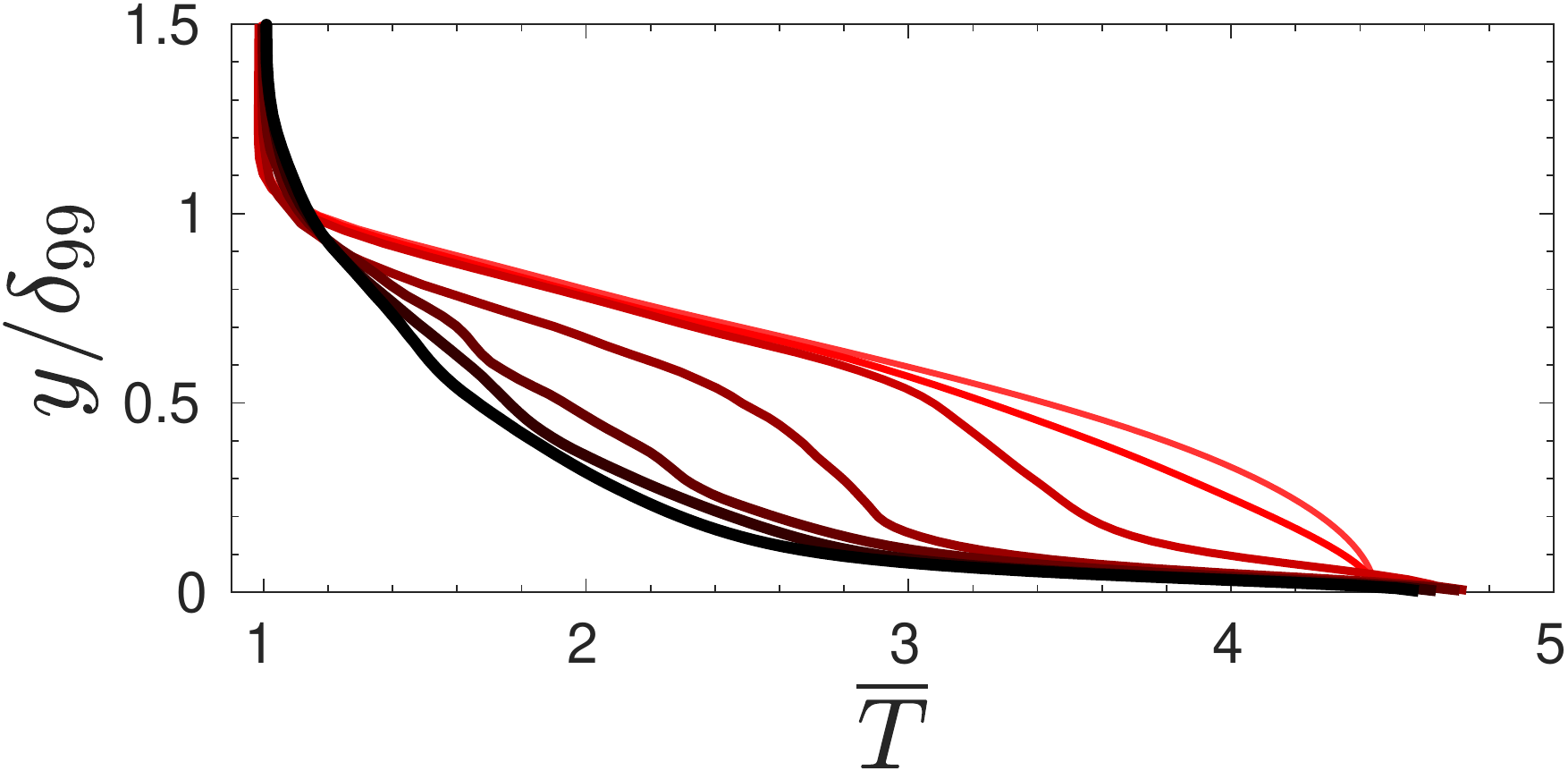}%
\put(-390.0,90.0){$(a)$}
}%
\centerline{%
\includegraphics[trim=0 0 0 0, clip,width=0.5\textwidth] {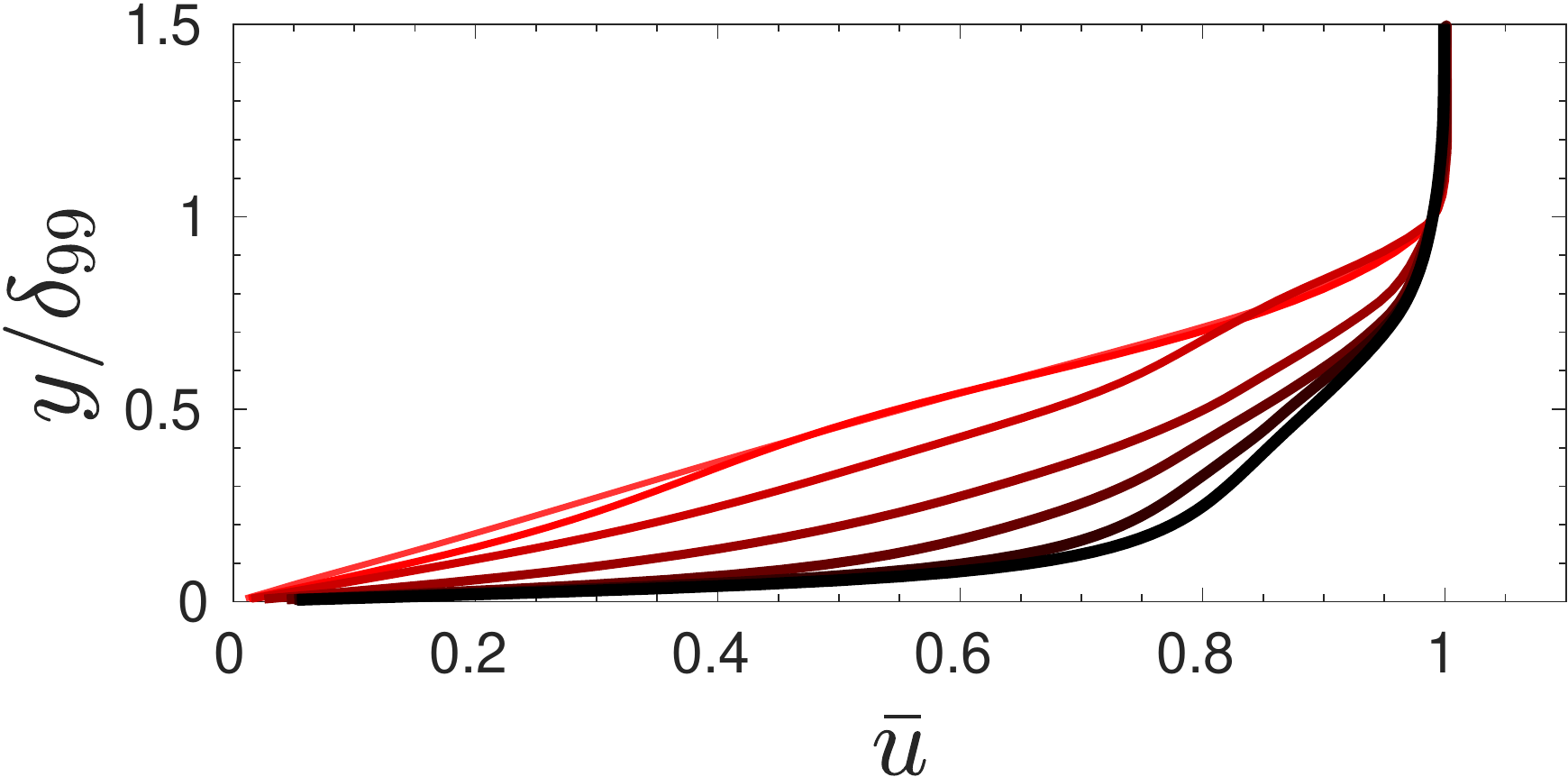}%
\hspace{1.00mm}
\includegraphics[trim=0 0 0 0, clip,width=0.5\textwidth] {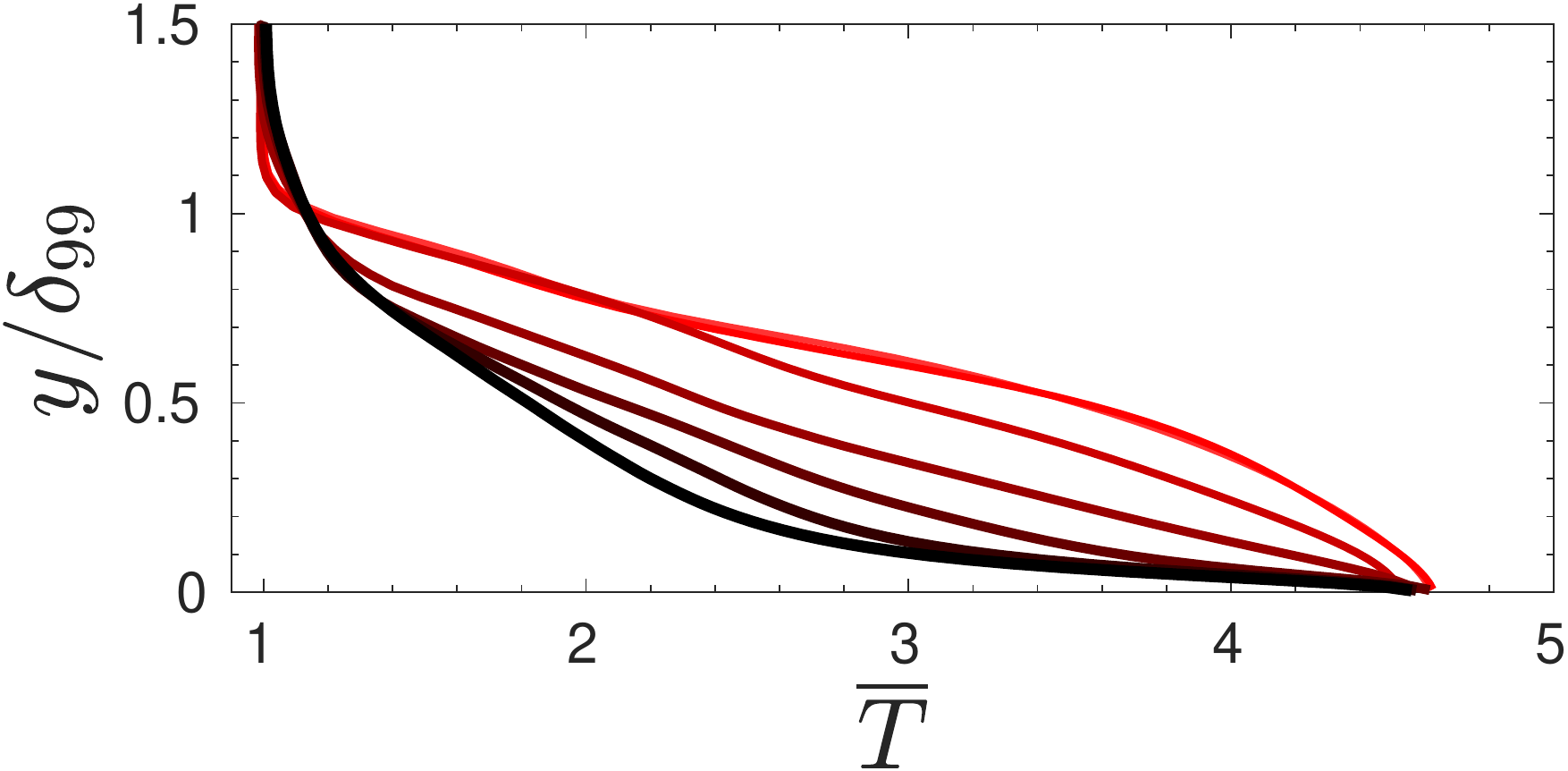}%
\put(-390.0,90.0){$(b)$}
}%
\caption{Wall-normal profiles of time and spanwise averaged streamwise velocity (left plots), $\bar{u}$, and temperature (right plots), $\overline{T}$, at different locations along the transition zone. $(a)$ The profiles of case E1N correspond to $2400 \le\sqrt{\textrm{Re}_x} \le 2700$ with increment of $50$ from light to dark colours (also thin to thick lines). $(b)$ The profiles of case E2N correspond to $2000 \le \sqrt{\textrm{Re}_x} \le 2300$ with increment of $50$ from light to dark colours (also thin to thick lines).}
\label{FIG:R1E1_R1E2_Ubar_Tbar}
\end{figure}

\begin{figure}
\centerline{%
\includegraphics[trim=0 60 0 0, clip,width=0.5\textwidth] {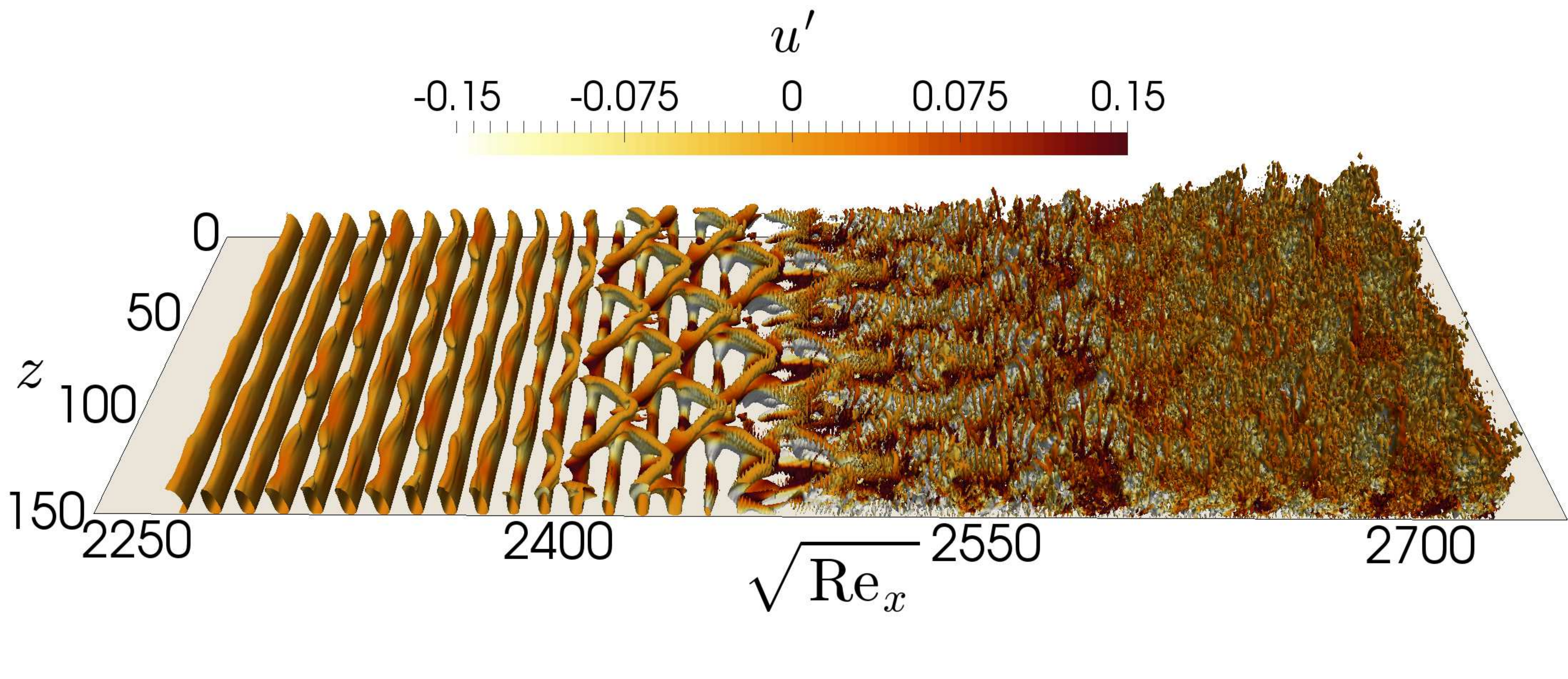}%
\hspace{1.00mm}
\includegraphics[trim=0 60 0 0, clip,width=0.5\textwidth] {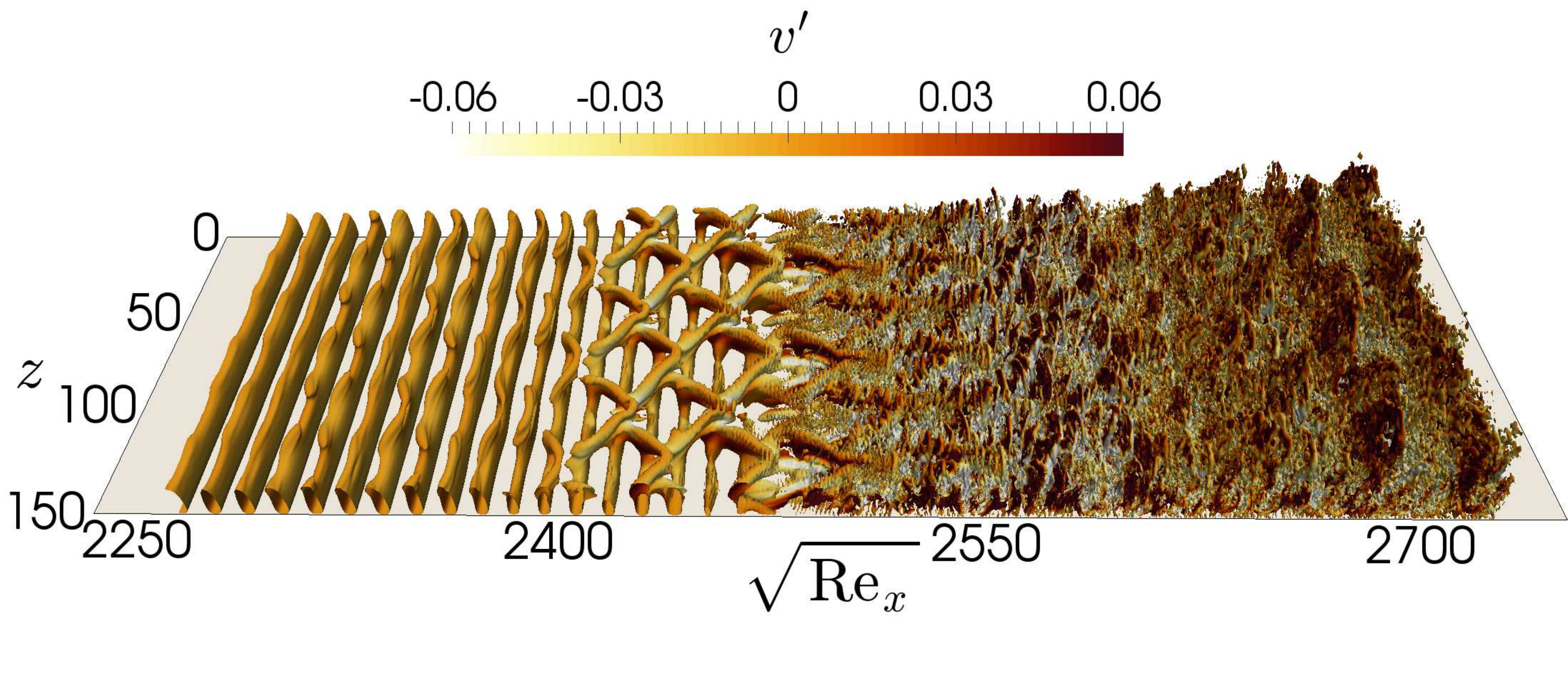}%
\put(-390.0,60.0){$(a)$}
}%
\vspace{1.00mm}
\centerline{%
\includegraphics[trim=0 60 0 0, clip,width=0.5\textwidth] {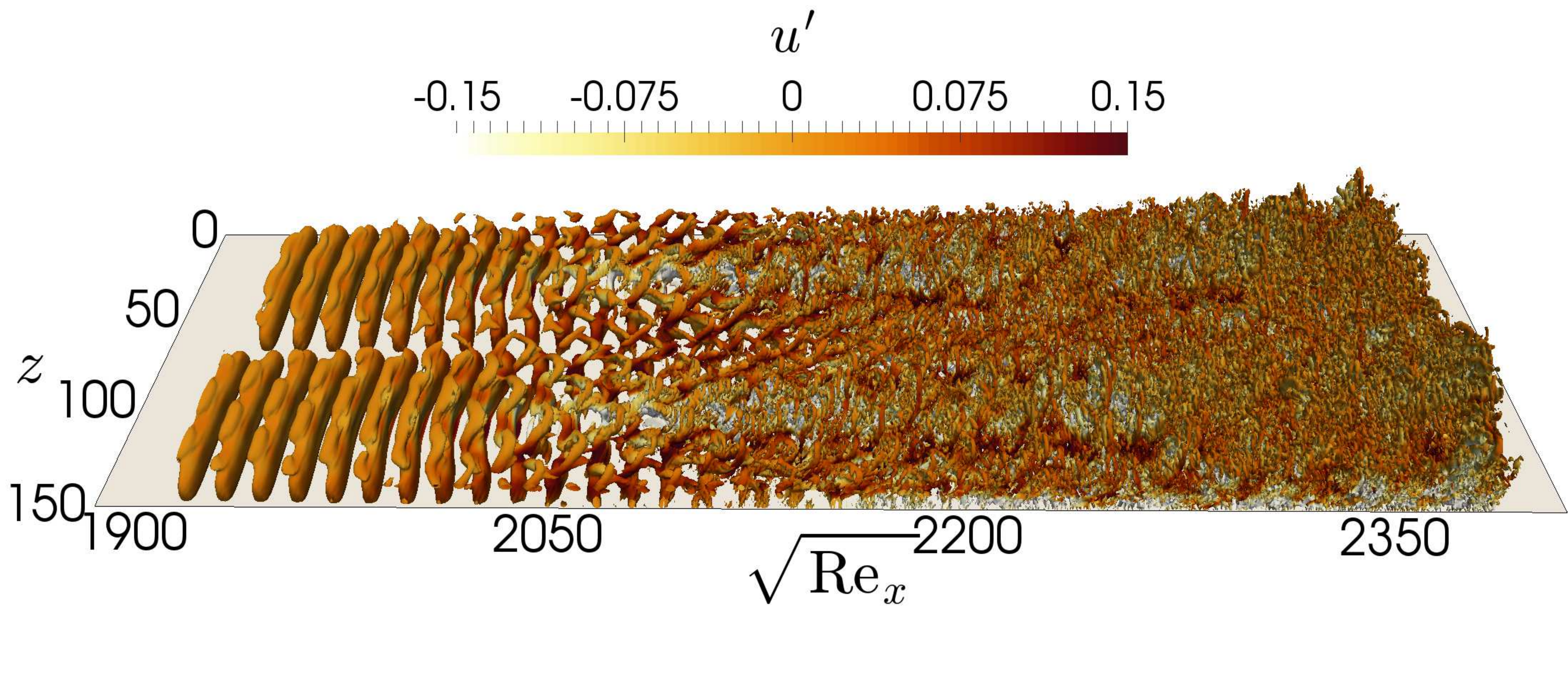}%
\hspace{1.00mm}
\includegraphics[trim=0 60 0 0, clip,width=0.5\textwidth] {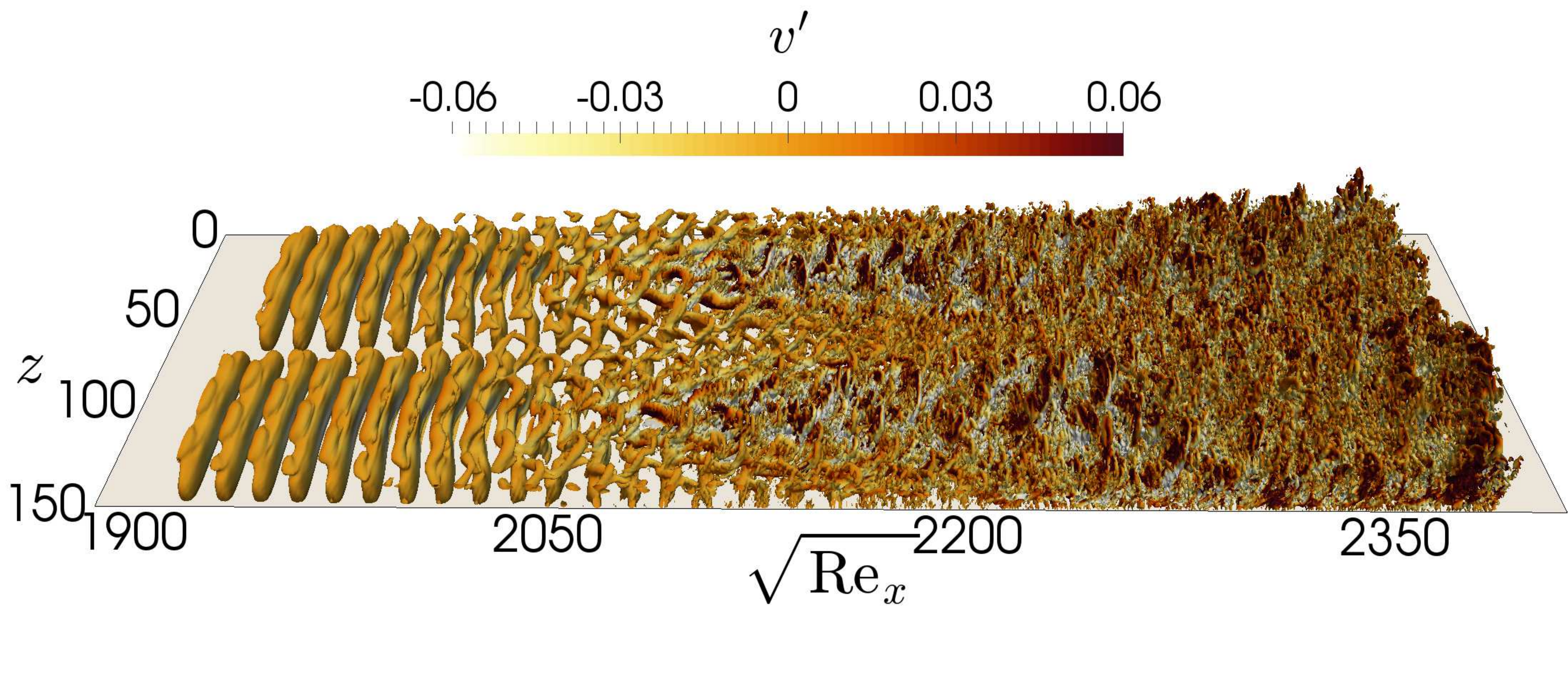}%
\put(-390.0,60.0){$(b)$}
}%
\caption{Iso-surfaces of $Q = 1 \times 10^{-4}$ coloured by $u^{\prime}$ (left plots) and $v^{\prime}$ (right plots) corresponding to cases $(a)$ E1N and $(b)$ E2N.}
\label{FIG:R1E1_R1E2_Q_Contours}
\end{figure}

The transition processes in cases E1N and E2N are contrasted in figures \ref{FIG:R1E1_R1E2_Ubar_Tbar} and \ref{FIG:R1E1_R1E2_Q_Contours}. 
The former figure shows wall-normal profiles of the average streamwise velocity and temperature, evaluated at different streamwise positions within the transition zone.  
And the latter figure shows three-dimensional views of the vortical structures, visualized using the $Q-$criterion, within the transition regions of both cases. 
Transition in case E1N takes place via the formation and breakdown of $\Lambda-$shaped vortices. 
This process takes place over a relatively short streamwise distance as shown qualitatively in figure \ref{FIG:R1E1_R1E2_Q_Contours} and demonstrated by the skin-friction profile (figure \ref{FIG:Cf_Curves_Cost_Function}).
The relatively abrupt transition is accompanied by a distortion of the mean streamwise velocity profile in the near-wall region, followed by a subsequent relaxation of the outer part of the profile towards the fully turbulent curve. 
The same trend is captured in the mean-temperature profile.  
In case E2N, transition is due to the secondary instability of streamwise-elongated streaks.  
While the streaks themselves are not captured in visualization of the $Q$-criterion, their secondary instability is seen in the figure.  
The transition length is longer in this case, and therefore the mean-velocity and temperature profiles gradually approaches the turbulent curve.

For case E1N, the left and right panels of figure \ref{FIG:R1E1_R1E2_Q_Contours}$a$ show that the $\Lambda$-structures, which are generated as a result of the amplification of mode $\left< 30 , 3 \right>$, are located close to the edge of the boundary layer.
They lie above the near-wall streaks shown in figure \ref{FIG:Disturbances_Spectral_Evolution_R1E1}$d$, and successive rows of $\Lambda$'s are staggered in the spanwise direction.  
Typical of hairpin structures in wall-bounded flows \citep[see e.g.][]{Adrian2007,Farano2015}, their legs lead to ejections in the plane of symmetry while the head generates a strong sweep motion.
The wall-normal and streamwise velocity fluctuations therefore have opposite signs in those regions. 
Ejections due to positive wall-normal velocity fluctuations are accompanied by transport of low streamwise momentum upward and a negative streamwise fluctuation;
sweep due to negative $v'$ leads to downward transport of high-momentum fluid and a positive $u'$ perturbation. 
The late-stages in the evolution of these $\Lambda$-structures resemble natural transition to turbulence, where trains of hairpin vortices are formed and break down over a short streamwise distance.  
Since the structures are staggered in the span, a fully turbulent flow is established quickly downstream thus leading to a short transition length.

Figure \ref{FIG:R1E1_R1E2_Q_Contours}$b$ for case E2N shows that $\Lambda$-structures form in this case as well, and successive rows are similarly staggered in the spanwise direction.  
However, breakdown to turbulence does not commence at the tips of individual structures; instead it is initiated due to the instability of the underlying steady streaks.  
These streaks are clearly captured in figure \ref{FIG:Disturbances_Spectral_Evolution_R1E2}$d$, and at the same locations the breakdown pattern is clear in figure \ref{FIG:R1E1_R1E2_Q_Contours}$b$. 
In this region, the streaks are elevated low-speed structures (negative $u^{\prime}$) straddled by $\Lambda$'s that induce their ejection (positive $v^{\prime}$).  
Since breakdown takes place on the low-speed streaks, it directly impacts every other row of the $\Lambda$'s and subsequently spreads to fill the domain; the transition length is therefore longer in this case relative to E1N.

\subsection{Prediction from Linear Stability Theory}
\label{sec:LST}

\begin{figure}
\centerline{%
\includegraphics[trim=0 0 0 0, clip,width=0.9999\textwidth] {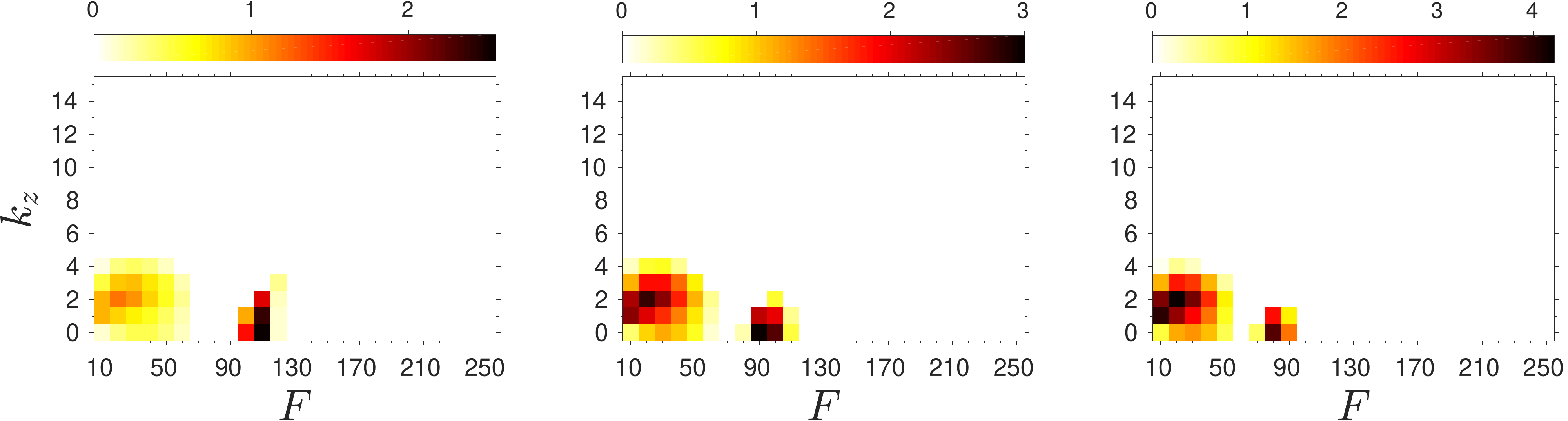}%
\put(-390.0,90.0){$(a)$}
}%
\centerline{%
\includegraphics[trim=0 0 30 0, clip,width=0.5\textwidth] {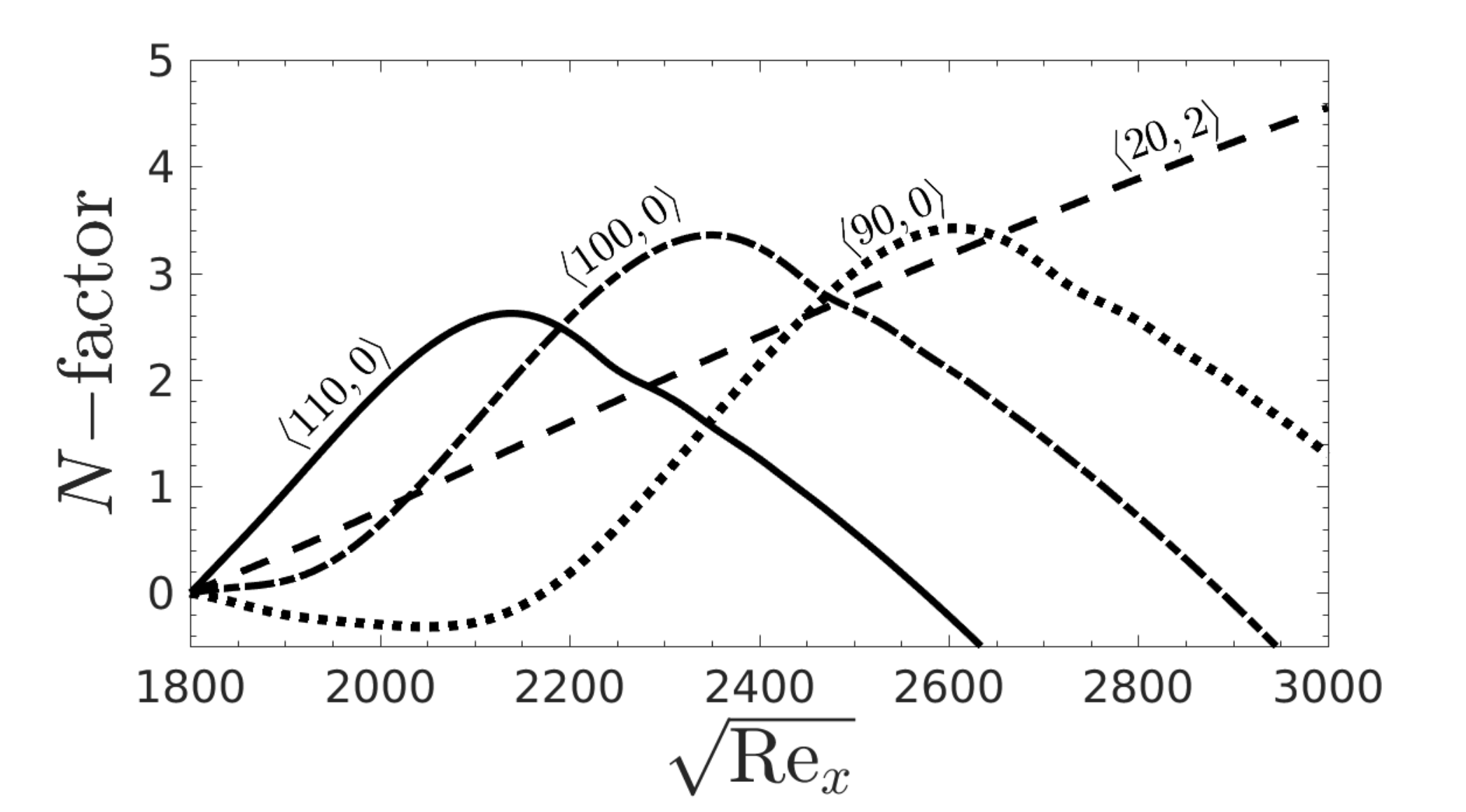}%
\includegraphics[trim=0 0 30 0, clip,width=0.5\textwidth] {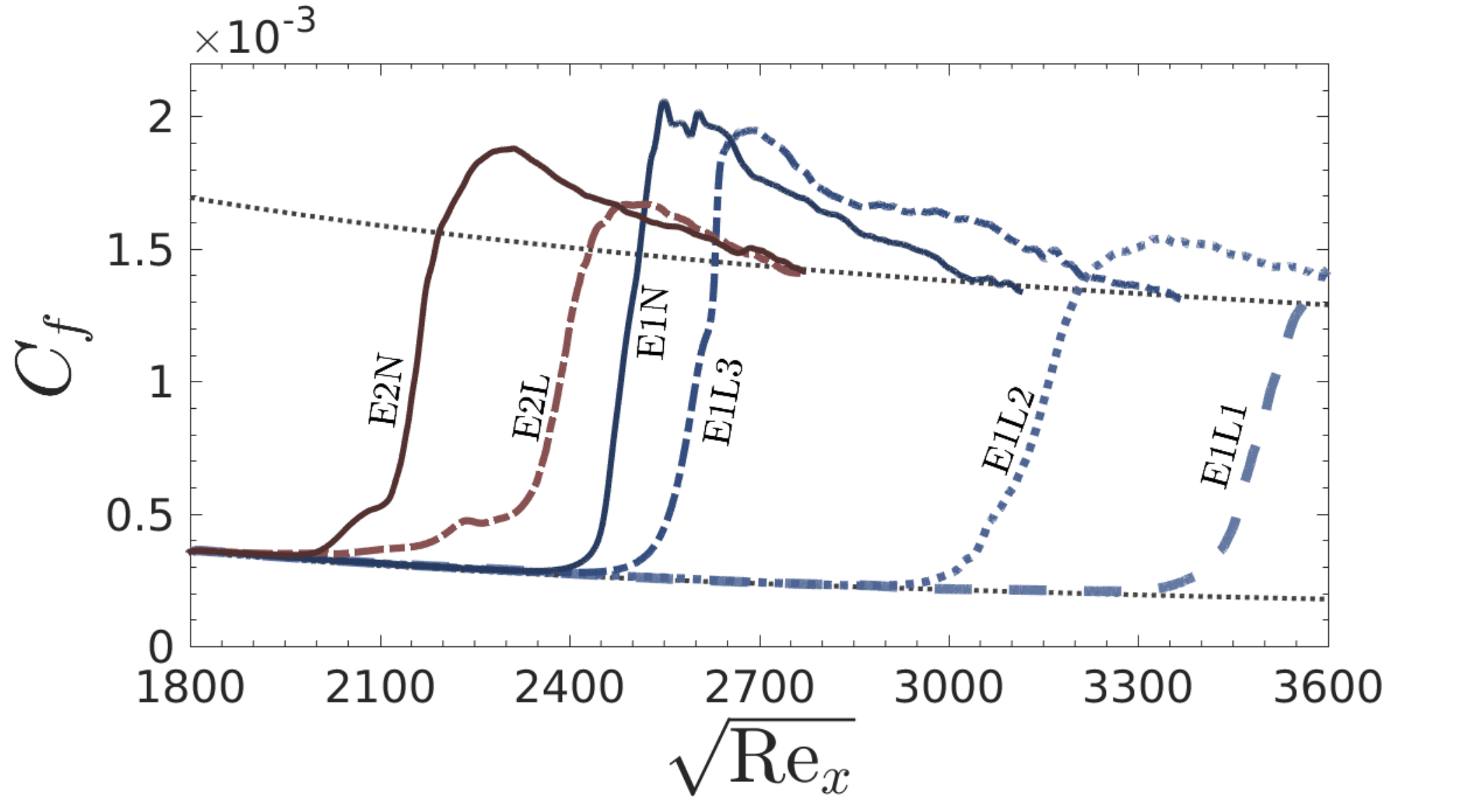}%
\put(-390.0,100.0){$(b)$}
\put(-190.0,100.0){$(c)$}
}%
\caption{$(a)$ Contours of $N-$factor computed from equation \ref{Eq:N_factor} from left to right for $\sqrt{\textrm{Re}_{x_f}} = 2100$, $2500$ and $2900$ respectively. $(b)$ The $N-$factor of the important modes versus streamwise location. $(c)$ Skin friction curves corresponding to the nonlinearly most unstable cases (E1N and E2N), and the linearly most unstable cases (E1L1, E1L2, E1L3 and E2L) presented in table \ref{TABLE:NonLinearly_Most_Unstable}.}
\label{FIG:N_factor}
\end{figure}

In order to highlight the importance of the present nonlinear approach in determining the inflow disturbance, we examine other possible inflow conditions that are selected based on linear theory alone.
The starting point is to evaluate the linear evolution of all the Orr-Sommerfeld and Squire instability waves that are part of the inlet condition, using the linear parabolized instability equations \citep{Junho2018accepted}.
The $N$-factor associated with every $\left< F , k_z \right>$ wave was obtained, 
\begin{equation}
	N\textrm{-factor} = \frac{1}{2} \ln \left( \frac{\mathcal{E}_{\left< F , k_z \right>}}{\mathcal{E}_{0}} \right) ,
\label{Eq:N_factor}
\end{equation}
where $\mathcal{E}_{0}$ is the energy of the mode at the inlet.
The results are shown for three downstream locations in figure \ref{FIG:N_factor}$(a)$.
Depending on the point at which the growth rates are computed, different instability waves attain highest $N-$factor value.
Figure \ref{FIG:N_factor}$(b)$ shows the change of $N-$factor versus streamwise location for the four instability waves that are each most amplified within a sub-region of the computational domain; mode $\left< 110 , 0 \right>$ for $ \sqrt{\textrm{Re}_x} < 2190$; mode $\left< 100 , 0 \right>$ along $2190 < \sqrt{\textrm{Re}_x} < 2470$; mode $\left< 90 , 0 \right>$ along $2470 < \sqrt{\textrm{Re}_x} < 2650$, and mode $\left< 20 , 2 \right>$ for $ \sqrt{\textrm{Re}_x} > 2650$.
In order to make use of these results in selecting the inflow condition, prior knowledge of transition location is needed; linear theory does not provide this information.

We examined four inflow conditions that were selected based on the linear results, and computed their evolution using direct numerical simulations in order to compare the outcome to the nonlinearly most unstable disturbance.  
These new cases are designated E1L1, E1L2, E1L3, and E2L, and their parameters are reported in table \ref{TABLE:NonLinearly_Most_Unstable}.
The first three are at the lower inflow energy level and, since transition is expected in the second half of the domain, $99\%$ of the inflow energy is allocated to modes $\left<100,0\right>$, $\left<90,0\right>$ and $\left<20, 2\right>$, respectively.  
The fourth case, E2L, is at the higher inflow energy level and, therefore, $99\%$ of the inflow energy is assigned to mode $\left<110,0\right>$. 
For all four cases, the remaining $1\%$ of the inflow energy was in the form of broadband forcing, in order to enable possible nonlinear interactions and secondary instabilities.
The outcome of these simulations is contrasted to the nonlinearly most unstable cases in figure \ref{FIG:N_factor}$c$, where $C_f$ is plotted versus downstream distance. 
Note that, for some of these simulations, the domain was extended in the streamwise direction in order to capture the delayed transition to turbulence, in particular in E1L1 and E1L2.  

The transition scenarios of the new simulations are also very different from the nonlinearly obtained inflow condition (not shown here). 
Cases E1L1 and E1L2 are classical fundamental second-mode transition, with a detuned secondary instability regions along which streaks are generated and break down.
Case E1L3 is a typical first-mode oblique breakdown, in which the primary three-dimensional wave develops oblique instabilities, $\Lambda$-structures and ultimately hairpin vortices that become the seat for the onset of turbulence.
Similar to cases E1L1 and E1L2, the transition scenario of case E2L is a typical second-mode fundamental mechanism.
However, the incited secondary instability has the same frequency as the primary mode.

In all four cases, the inflow primary instability is the key determinant of the path to turbulence. 
And while the linearly most unstable waves guarantee fastest exponential growth based on linear theory, they do not necessarily guarantee largest energy in the nonlinear regime or earliest secondary instability and transition. 
In contrast, when the nonlinearly most unstable inflow condition was adopted, the resulting transition mechanism was not due to any one particular inflow wave; 
instead, the nonlinear interactions of the inflow instabilities led to the generation of new waves, the distortion of the base-flow and the earliest possible transition location. 


\section{Summary and conclusions}
\label{sec:Conclusions}

An ensemble-based variational (EnVar) algorithm is introduced to evaluate the nonlinearly most unstable inflow disturbance that results in the earliest location of laminar-to-turbulence transition in a hypersonic boundary layer. 
This disturbance (i) has a prescribed amount of total energy, (ii) satisfies the full Navies-Stokes equations, (iii) and undergoes the fastest breakdown to turbulence.
The first two elements are constraints and the third is the objective of the constrained optimization; the latter was modelled using a cost function that is proportional to the skin friction along the plate.  
During each iteration of the algorithm, an ensemble of possible solutions are advanced, and the associated outcomes are used to evaluate the gradient of the cost function.
New candidate solutions are then formed and the process is repeated until convergence.

The algorithm was applied in the case of zero-pressure-gradient boundary-layer flow at free-stream Mach number $M_\infty = 4.5$, and for two levels of the inflow disturbance energy. 
The level of energy of the first case is of the same order as the kinetic energy of stratospheric turbulent layers, measured during a recent experimental campaign in an attempt to replicate the environmental condition of high-altitude flight tests.
The second case had fifty-times larger energy, in order to highlight the effect of this parameter on the outcome of the algorithm. 
The laminar-to-turbulence transition scenarios due to the nonlinearly most dangerous disturbances from each case were examined in detail.  
While at the higher energy level, transition displays symptoms of second-mode oblique breakdown, the lower energy case can not be ascribed to any classical route.  
Instead, transition is initiated due to nonlinear interactions of a couple of normal acoustic (second-mode) disturbances and an oblique vorticity (first-mode) instability wave. 
Their nonlinear interactions spurs new instabilities waves and distort the mean flow, ultimately leading to very rapid growth of three-dimensional waves that form $\Lambda$-structures and break down to turbulence. 

A number of other inflow conditions, all selected based on the amplification rates from linear theory, were also examined using direct numerical simulations. 
These efforts invariably led to classical breakdown scenarios and, for the same level of inflow energy, delayed transition onset relative to the nonlinearly most unstable disturbance.  
The results thus highlight that a nonlinear approach is required to provide a strict, minimum bound on transition Reynolds number.

Despite the present use of direct numerical simulations, the EnVar technique is applicable for any numerical-design tool.
It is specially advantageous relative to adjoint-based techniques which become unstable over long optimization time-horizons in chaotic transitional and turbulent flows.
The EnVar approach can also be adopted with any choice of the cost function, for example based on the disturbance energy or wall temperature. 
The norms associated with these two quantities were evaluated when the cost function was based on skin friction.  
The results demonstrated that the optimal inflow disturbance is likely to differ, in particular if the cost function is based on wall temperature.
 
The nonlinearly most dangerous disturbance provides an objective measure for evaluating design performance, in particular when environmental disturbances are uncertain.  
If a design is modified, for instance for the purpose of delaying transition, the spectral content of what constitutes the most dangerous disturbance will, however, change and must be re-evaluated.
If transition is delayed in the new configuration, design surety is guaranteed.

\section*{Acknowledgement}
This work was supported in part by the US Air Force Office of Scientific Research (grant FA9550-16-1-0103) and by the Office of Naval Research (grant N00014-17-1-2339).


\appendix

\section{Ensemble generation algorithm}
\label{Appendix_A}

In this appendix, the algorithm for constructing the ensemble members that are used in the optimization procedure (\ref{Eq:Optimization_Problem_Final}) is described.
The starting point is the mean control vector, $\textbf{c}^{(e)}$, which is either the initial guess or the outcome of the optimization at the end of the previous iteration, and satisfies the energy constraint.  
The objective is to generate an ensemble of control vectors, $\textbf{c}^{(r)}$ for $r=1$ to $N_{en}$, around this mean; 
the relationship between $\textbf{c}^{(e)}$ and $\textbf{c}^{(r)}$ is given by equation (\ref{Eq:Ensemble_Mean}), and therefore $\textbf{c}^{(e)}$ is not the arithmetic average of the members.  
The latter must not deviate appreciably from the mean, in a sense that will be quantified using the variance of the ensemble, and must each satisfy the energy constraint as well. 
Another important consideration is that, for a particular ensemble size $N_{en}$, the members must span the control vector subspace as best as possible and be well-conditioned. 
The ensemble generation algorithm has two main steps: first, a very large random ensemble is formed that satisfies specific criteria; second, a smaller ensemble is determined from the larger one, and whose members are better conditioned.  A detailed description of the first step is presented here, while for the second step we only quote the procedure and refer the reader to \citet[][chapter 11]{Evensen2009} for its theoretical foundation and proof of concept.

In the first step of the algorithm, our concern is to enforce the constraints that the ensemble members should satisfy rather than how well the ensemble is conditioned.
Therefore, in order to ensure that the members span the control-vector subspace, we generate a very large ensemble.
Since the members will be randomly generated, and assuming a very large size of the ensemble, it is very likely that any possible control vector can be expressed as a linear superposition of the ensemble members.
There are three constraints that must be satisfied, related to the mean, covariance and energy.
As discussed in \S\ref{sec:4DVarEns}, satisfying the energy constraint by the mean and ensemble members is not possible if the arithmetic average is adopted.
In order to address this challenge, we performed two change of variables during the algorithm, referred to as $\textbf{e}$ and $\hat{\textbf{e}}$ later in this section.
And the covariance constraint is to ensure that the members of the ensemble are close to their mean, which is an essential condition of the EnVar optimization procedure.

We start by constructing the very large ensemble of size $\Upsilon \times N_{en}$.  
In analogy to (\ref{Eq:Ensemble_Mean}), the ensemble members are related to the mean by, 
\begin{subequations}
\begin{equation}
	\textbf{A}_1 \textbf{c}^{(e)} \circ \textbf{A}_1 \textbf{c}^{(e)} = \frac{1}{\Upsilon N_{en}} \sum_{r=1}^{ \Upsilon N_{en}} \textbf{A}_1 \textbf{c}^{(r)} \circ \textbf{A}_1 \textbf{c}^{(r)} ,
\end{equation}
and
\begin{equation}
	\textbf{A}_2 \textbf{c}^{(e)} = \frac{1}{\Upsilon N_{en}} \sum_{r=1}^{\Upsilon N_{en}} \textbf{A}_2 \textbf{c}^{(r)}, 
\end{equation}
where $\textbf{A}_{1}$ and $\textbf{A}_2$ are defined in equations (\ref{Eq:dummy_Matrix_1}) and (\ref{Eq:dummy_Matrix_3}), respectively.
\label{Eq:Ensemble_Mean_Appendix}
\end{subequations}
The above two equalities can be encoded into one expression by defining $\textbf{e}^{(r)} \in \mathbb{R}^{2 M \times 1}$, 
\begin{equation}
	\textbf{e}^{(r)} \equiv \textbf{A}_1 \textbf{c}^{(r)} \circ \textbf{A}_1 \textbf{c}^{(r)} + \textbf{A}_2 \textbf{c}^{(r)}, 
\label{Eq:Realization}
\end{equation}
whose arithmetic average is $\textbf{e}^{(e)}$.
A further change of variables is introduced in order to simplify the algorithm, 
\begin{equation}
	\hat{\textbf{e}}^{(r)} \equiv \textbf{S}^{-1} \left( \textbf{e}^{(r)} - \textbf{e}^{(e)} \right) ,
\label{Eq:Realization_Normalization}
\end{equation} 
where the diagonal matrix $\textbf{S}$ is equal to $ \textbf{diag} \left( \boldsymbol{\sigma} \right)$, and $\boldsymbol{\sigma}$ is a vector containing the desired standard deviations of the original ensemble members from their mean.
By construction, $\hat{\textbf{e}}^{(r)}$ has a zero mean and unit variance. 
Furthermore, as long as the members of the original ensemble and their mean, $ \textbf{e}^{(r)} $ and $  \textbf{e}^{(e)} $, satisfy the energy constraint, the energy of $\hat{\textbf{e}}^{(r)}$ is zero.
Thus the ensemble that we are seeking satisfies the following conditions,
\begin{subequations}
\begin{equation}
\hat{\textbf{P}} \textbf{b}_1 = 0 ,
\label{Eq:Mean_Constraint}
\end{equation}
\begin{equation}
 \hat{\textbf{P}} \hat{\textbf{P}}^{tr} = \left( \Upsilon N_{en} -1 \right) \hat{\textbf{C}} ,
\label{Eq:Covariance_Constraint}
\end{equation}
and
\begin{equation}
\textbf{b}_1^{tr} \textbf{A}_1 \hat{\textbf{P}} = 0 ,
\label{Eq:Energy_Constraint}
\end{equation}
where
\label{Eq:Realization_Conditions}
\end{subequations}
\begin{equation}
\hat{\textbf{P}} = \begin{bmatrix}
    \hat{\textbf{e}}^{(1)} & \hat{\textbf{e}}^{(2)} & \dots & \hat{\textbf{e}}^{(\Upsilon N_{en})}
\end{bmatrix}_{2M \times \Upsilon N_{en}} ,
\end{equation}
\begin{equation}
\hat{\textbf{C}} = \begin{bmatrix}
\begin{bmatrix}
    \mathcal{C}_{ij}
\end{bmatrix}_{M \times M} & \textbf{O}_{M \times M} \\
\textbf{O}_{M \times M} & \textbf{I}_{M \times M}
\end{bmatrix}_{2M \times 2M}  ,
\end{equation}
\begin{equation}
\textbf{b}_1 = \begin{bmatrix}
    1 & 1 & \dots & 1
\end{bmatrix}_{1 \times 2M}^{tr}. 
\end{equation}
In the above, $\mathcal{C}_{ij} = \mathcal{C}_{ji}$ are the correlation coefficients and $\mathcal{C}_{ii} = 1$, $\textbf{I}$ is the identity matrix, $\textbf{O}$ is a zero matrix, and $\textbf{A}_1$ is defined in (\ref{Eq:dummy_Matrix_3}).
Equation (\ref{Eq:Mean_Constraint}) ensures that the mean of the ensemble is zero, equation (\ref{Eq:Covariance_Constraint}) is the covariance constraint, and equation (\ref{Eq:Energy_Constraint}) enforces the energy constraint on all members of the ensemble.
Note that the condition $\mathcal{C}_{ii} = 1$ is a consequence of the normalization in equation (\ref{Eq:Realization_Normalization}).


The problem can be simplified further by combing the covariance and the energy constraints.
From equations (\ref{Eq:Covariance_Constraint}) and (\ref{Eq:Energy_Constraint}), we obtain $\textbf{b}_1^{tr} \textbf{A}_1 \hat{\textbf{C}} = 0$, and therefore, the energy constraint is equivalent to $\sum_{j=1}^{M} \mathcal{C}_{ij} = 0 $.
Thus, we can redefine the problem as evaluating a set of $\hat{\textbf{e}}$ whose arithmetic average is zero, and that are correlated with one another such that the components of their covariance matrix satisfy $\sum_{j=1}^{M} \mathcal{C}_{ij} = 0 $.
The former condition is satisfied by randomly generating uncorrelated control vectors, $\textbf{y}^{(r)}$, that have a zero mean, while the latter condition is enforced by performing Cholesky decomposition of the covariance matrix and setting $\hat{\textbf{e}}^{(r)} = \textbf{L} \textbf{y}^{(r)}$, where $\textbf{L}$ is the triangular matrix of Cholesky factorization. By definition, $\hat{\textbf{e}}^{(r)}$ satisfy both conditions.
Using the symmetry of $\hat{\textbf{C}}$, we can obtain $\hat{\textbf{C}} \textbf{A}_1 \textbf{b}_1 = 0$.
Hence, zero is an eigenvalue of $\hat{\textbf{C}}$ and $\textbf{A}_1 \textbf{b}_1$ is an eigenvector; the dimension of the eigenspace corresponding to this eigenvalue is unity, and the rank of $\hat{\textbf{C}}$ is equal to $2 M - 1$.
Since the algorithm involves a Cholesky decomposition of the covariance matrix that can be used to span a linear space $ \mathbb{R}^{2 M \times 2 M}$, a new covariance matrix is defined whose rank is equal to $2 M$, 
\begin{equation}
\hat{\hat{\textbf{C}}} = \begin{bmatrix}
    1 & \textbf{o}^{tr} \\
    \textbf{o} & \hat{\textbf{C}}
\end{bmatrix}_{(2 M + 1) \times (2 M + 1)}, 
\label{Eq:New_Covariance}
\end{equation}
where $\textbf{o} \in \mathbb{R}^{2 M \times 1}$ is a vector of all zeros.
Note that zero is still an eigenvalue of this matrix.

\begin{algorithm}[bt]
\SetAlgoLined
 \textbf{Step 1} \;
 \Indp
 \textbullet~Generate $\Upsilon N_{en}$ random vectors $\textbf{y}^{(r)}$ whose mean and covariance are 0 and $\textbf{I}$ \; 
 \textbullet~Find $ -1 \leq \mathcal{C}_{ij} \leq 1 $ that satisfies $\sum_{j=1}^{M} \mathcal{C}_{ij} = 0$ for all $1 \leq i \leq M$ \;
 \textbullet~Perform Cholesky decomposition of $\hat{\hat{\textbf{C}}}$ in (\ref{Eq:New_Covariance}), and obtain $\textbf{L} \textbf{L}^{tr}$, $\textbf{L} \in \mathbb{R}^{2 M \times 2 M}$\;
 \textbullet~Compute $\hat{\textbf{e}}^{(r)} \in \mathbb{R}^{2 M \times 1}$ as $\hat{\textbf{e}}^{(r)} = \textbf{L} \textbf{y}^{(r)}$, and store them in $\hat{\textbf{P}} \in \mathbb{R}^{2 M \times \Upsilon N_{en}}$ \;
 \Indm
 \textbf{Step 2} \;
 \Indp
 \textbullet~Compute the singular value decomposition $\hat{\textbf{P}} = \hat{\mathtt{U}} \hat{\Sigma} \hat{\mathtt{V}}^{tr}$  \; 
 \textbullet~Retain the first $N_{en}$ singular vectors in $\hat{\mathtt{U}}$, and store them in $\mathtt{U} \in \mathbb{R}^{2 M \times N_{en}}$ \;
 \textbullet~Retain the first $N_{en} \times N_{en}$ quadrant of $\hat{\Sigma}$, and store it in $\Sigma \in \mathbb{R}^{N_{en} \times N_{en}}$ \;
 \textbullet~Retain the first $N_{en} \times N_{en}$ quadrant of $\hat{\mathtt{V}}$, and store it in $\mathtt{V} \in \mathbb{R}^{N_{en} \times N_{en}}$ \;
 \textbullet~Generate the new ensemble perturbations matrix, $\textbf{P} =  \mathtt{U} \frac{1}{\sqrt{\Upsilon}} \Sigma \mathtt{V}^{tr}$.
 Normalization of $\Sigma$ by $\sqrt{\Upsilon}$ ensures the new ensemble has the correct variance\;
 \textbullet~Compute $\textbf{e}^{(r)} = \textbf{S} \textbf{p}^{(r)} + \textbf{e}^{(e)} $, where $\textbf{p}^{(r)}$ are the columns of $\textbf{P} \in \mathbb{R}^{2 M \times N_{en}}$ \;
\textbullet~Compute the control vectors $\textbf{c}^{(r)}$ from the set of $\textbf{e}^{(r)}$ using equation (\ref{Eq:Realization})\;
 \caption{Generation of the ensemble.}
 \label{Algorithm:Esnemble_Gen}
\end{algorithm}

The above procedure completes the first step of the ensemble generation, and provides a set of $\hat{\textbf{e}}^{(r)} \in \mathbb{R}^{2 M \times 1}$ which satisfy (\ref{Eq:Realization_Conditions}).
The second step of the algorithm is to identify, from this large set of members, the best possible ensemble with size $N_{en}$.
This task involves singular-value-decomposition (SVD) of the large ensemble and choosing the $N_{en}$ dominant singular vectors and values as the new ensemble.
In the limit of $\Upsilon N_{en} \rightarrow \infty$, the singular vectors and values of the large ensemble will converge to eigenvectors and eigenvalues of the full-rank ensemble realizations (an ensemble that represents the physical problem exactly).
Using the SVD of the larger-sized ensemble to approximate its smaller-sized representation, therefore, ensures that for a given size the new ensemble provides the best possible rendition.
The complete procedure for the ensemble generation is summarized in algorithm \ref{Algorithm:Esnemble_Gen}. 

\section{Nonlinear energy transfer}
\label{Appendix_B}

In this appendix, the derivation of the nonlinear energy transfer term, 
\begin{equation}
\tag{\ref{Eq:NonLinearEnergyTransfer}}
	\mathcal{I}_{\left< F , k_{z} \right>} = \int_{0}^{L_y} \left\vert \boldsymbol{\hat{\mathcal{A}}}^*_{\left< F_1 , k_{z,1} \right>} \boldsymbol{\hat{\mathcal{F}}}_{\left< F_2 , k_{z,2} \right>} \right\vert dy, 
\end{equation}
is presented. 
The starting point is the transport term $\mathcal{T}_1$ in the TKE equation (\ref{Eq:TKE_Transport}), which is a nonlinear redistribution of energy between different perturbations.  
While all the other contribution in the energy equation originate from linear terms of the momentum balance, $\mathcal{T}_1$ arises from the nonlinear advection term. 

By integrating $\mathcal{T}_1$ in wall-normal direction, we obtain, 
\begin{equation}
	\int_{y} \mathcal{T}_1 d y = - \frac{1}{2} \overline{ \frac{\partial }{\partial x} \left( \underbrace{ \int_{0}^{L_y} \left( \boldsymbol{\mathcal{A}}^{\prime} \cdot \boldsymbol{ \mathcal{F} }^{\prime} \right) dy }_{\mathcal{I}} \right) } ,
\end{equation}
where the vector quantities are defined as $\mathcal{A}_i^{\prime} = \rho u^{\prime\prime} u^{\prime\prime}_i $ and $\mathcal{F}_i^{\prime} = u^{\prime\prime}_i$, and the variable $\mathcal{I}$ represents the nonlinear interaction.
In the above expression, it was assumed that the fluctuations vanish at the top and bottom boundaries of the domain and, as such, only the streamwise advection is retained.
Using the two-dimensional Fourier representations of the vectors $\boldsymbol{\mathcal{A}}^{\prime}$ and $\boldsymbol{\mathcal{F}}^{\prime}$ in temporal frequency and spanwise wavenumber, $\mathcal{I}$ can be rewritten as
\begin{equation}
	\mathcal{I} = \int_{0}^{L_y} \left( \sum_{F_1 , k_{z,1}} \sum_{F_2 , k_{z,2}} \left( \boldsymbol{\hat{\mathcal{A}}}_{\left< F_1 , k_{z,1} \right>} \cdot \boldsymbol{\hat{\mathcal{F}}}_{\left< F_2 , k_{z,2} \right>} \right) e^{-i\left( \frac{2 \pi (k_{z,1}+k_{z,2})}{L_z} z + \frac{\sqrt{\textrm{Re}_{x_0}} (F_1+F_2)}{10^6} t \right) } \right) dy .
\label{eq:app_I}
\end{equation}
A new variable is defined $\hat{\mathcal{I}}_{\left< F , k_{z} \right>} = \sum_{F_1 , k_{z,1}} \sum_{F_2 , k_{z,2}} \boldsymbol{\hat{\mathcal{A}}}_{\left< F_1 , k_{z,1} \right>} \cdot  \boldsymbol{\hat{\mathcal{F}}}_{\left< F_2 , k_{z,2} \right>}$, where the summation is performed over all $\left< F_1 , k_{z,1} \right>$ and $\left< F_2 , k_{z,2} \right>$ that satisfy $F = F_1 \pm F_2$ and $k_z = k_{z,1} \pm k_{z,2}$.
In terms of $\hat{\mathcal{I}}_{\left< F , k_{z} \right>}$, equation (\ref{eq:app_I}) becomes, 
\begin{equation}
	\mathcal{I} = \sum_{F , k_{z}} \left( \int_{0}^{L_y} \hat{\mathcal{I}}_{\left< F , k_{z} \right>} dy \right) e^{-i\left( \frac{2 \pi k_{z}}{L_z} z + \frac{\sqrt{\textrm{Re}_{x_0}} F}{10^6} t \right) } .
\end{equation}
Therefore the quantity $\hat{\mathcal{I}}_{\left< F , k_{z} \right>}$, referred to as the modal energy transfer coefficient \citep{Cheung2010}, represents the amount of energy transferred to mode $\left< F , k_{z} \right>$ via nonlinear interaction of modes $\left< F_1 , k_{z,1} \right>$ and $\left< F_2 , k_{z,2} \right>$.
The nonlinear modal energy transfer coefficient is defined to include the contribution from all four quadrants $\left< \pm F, \pm k_z \right>$, 
\begin{equation}
\mathcal{I}_{\left< F , k_{z} \right>} = \int_{0}^{L_y} \left\vert \boldsymbol{\hat{\mathcal{A}}}^*_{\left< F_1 , k_{z,1} \right>} \boldsymbol{\hat{\mathcal{F}}}_{\left< F_2 , k_{z,2} \right>} \right\vert dy
\end{equation}
where $*$ denotes complex conjugate transpose and $\left\vert . \right\vert$ is the absolute value.


\bibliographystyle{jfm}
\bibliography{Myrefs}

\end{document}